{}

\documentclass[11pt,a4paper]{article}

\newif\ifpublic\publictrue

\ifpublic\else\usepackage{showkeys}\fi
\usepackage[bookmarks=true,hyperfigures=true,colorlinks=true,linkcolor=black,citecolor=black,urlcolor=black,linktoc=all,bookmarksnumbered]{hyperref}
\usepackage[nosort]{cite}
\usepackage[parsep]{collref}
\usepackage[english]{babel}
\usepackage[T1]{fontenc}
\usepackage[latin1]{inputenc}
\usepackage{amsfonts,amsbsy,dsfont,euscript,mathrsfs,fixmath}
\usepackage{amsmath,amssymb}
\usepackage{enumerate}

\def\showkeysrefformat#1{{\normalfont\tiny\ttfamily#1}}
\makeatletter
\def\SK@@ref#1>#2\SK@{{\@inlabelfalse\leavevmode\vbox to\z@{\vss\SK@refcolor\rlap{\vrule\raise .75em \hbox{\showkeysrefformat{#2}}}}}}
\makeatother

\usepackage[a4paper,text={160mm,247mm},centering]{geometry}
\allowdisplaybreaks[3]
\numberwithin{equation}{section}
\def\[{\begin{equation}\begin{aligned}}
\def\]{\end{aligned}\end{equation}}
\newcommand{\nn}{\nonumber}

\usepackage[font=small,labelfont=bf,width=0.85\textwidth]{caption}

\expandafter\def\expandafter\bfseries\expandafter{\bfseries\ifmmode\else\boldmath\fi}
\expandafter\def\expandafter\mdseries\expandafter{\mdseries\ifmmode\else\unboldmath\fi}
\expandafter\def\expandafter\normalfont\expandafter{\normalfont\ifmmode\else\unboldmath\fi}

\RequirePackage{verbatim}

\makeatletter
\newwrite\bibinl@out
\newenvironment{bibtex}[1][\jobname]{%
\immediate\openout\bibinl@out #1.bib%
\immediate\write\bibinl@out{\@percentchar generated from `\jobname' starting line \the\inputlineno^^J}%
\def\verbatim@processline{\immediate\write\bibinl@out{\the\verbatim@line}}%
\@bsphack\let\do\@makeother\dospecials\catcode`\^^M\active\verbatim@start%
}
{\immediate\closeout\bibinl@out\@esphack}
\makeatother

\let\barefrac=\frac
\renewcommand{\frac}[2]{\mathinner{\barefrac{#1}{#2}}}

\let\baresqrt=\sqrt
\makeatletter
\renewcommand{\sqrt}{\@ifnextchar[\@sqrt@space@a\@sqrt@space@b}
\def\@sqrt@space@a[#1]#2{\mathinner{\mathchoice{\mkern-3mu}{\mkern-3mu}{}{}\baresqrt[#1]{#2}}}
\def\@sqrt@space@b#1{\mathinner{\mathchoice{\mkern-3mu}{\mkern-3mu}{}{}\baresqrt{#1}}}
\makeatother

\makeatletter
\let\per@dot@old=\.
\def\.{\ifmmode\def\per@dot@sel{\mkern3mu}\else\def\per@dot@sel{\per@dot@old}\fi\per@dot@sel}
\makeatother

\let\barefootnote=\footnote
\renewcommand{\footnote}[1]{\barefootnote{#1\vspace{3pt}}}

\newcommand{\sfrac}[2]{{\textstyle\frac{#1}{#2}}}
\newcommand{\half}{\sfrac{1}{2}}

\newcommand{\vfrac}[2]{\ifmmode\mathinner{\textstyle^{#1}\!/\!_{#2}}\else$^{#1}\!/\!_{#2}$\fi}

\DeclareMathOperator{\diag}{diag}
\DeclareMathOperator{\tr}{tr}
\DeclareMathOperator{\Tr}{Tr}
\DeclareMathOperator{\str}{str}
\DeclareMathOperator{\STr}{STr}
\DeclareMathOperator{\tSTr}{\widetilde{STr}}
\DeclareMathOperator{\tTr}{\widetilde{Tr}}

\newcommand{\Integer}{\mathds{Z}}

\newcommand{\ind}[1]{{\scriptscriptstyle{#1}}}

\newcommand{\alg}[1]{\mathfrak{#1}}
\newcommand{\grp}[1]{\mathrm{#1}}

\DeclareMathOperator{\Lie}{Lie}

\DeclareMathOperator{\Ad}{Ad}

\newcommand{\com}[2]{[#1,#2]}

\def\<{\big\langle}
\def\>{\big\rangle}

\newcommand{\geom}[1]{\mathrm{#1}}

\newcommand{\AdS}{\geom{AdS}}

\newcommand{\Sp}{\geom{S}}

\newcommand{\To}{\geom{T}}

\newcommand{\extder}{\mathrm{d}}

\newcommand{\Lag}{\mathcal{L}}
\newcommand{\Act}{\mathcal{S}}

\def\g{\gamma}

\def\l{\lambda}

\def\kk{{k}}
\def\h{{\rm h}}
\def\cG{c_{\ind{\grp{G}}}}
\def\cF{c_{\ind{\grp{F}}}}
\def\ddt{\frac{d}{dt}}
\def\go{g_\ind{0}}

\def\Qc{\mathcal{Q}}
\def\hQc{\widehat{\mathcal{Q}}}
\def\del{\partial}
\def\cel{(c_\ind{L} \eta_\ind{L})}
\def\cer{(c_\ind{R} \eta_\ind{R})}
\def\lb{(\!(}
\def\rb{)\!)}
\def\ls{\langle\!\langle}
\def\rs{\rangle\!\rangle}

\newcommand\dA{\mathbb A}
\newcommand\cA{\mathcal A}

\providecommand{\href}[2]{#2}

\makeatletter
\def\mr@ignsp#1 {\ifx\:#1\@empty\else #1\expandafter\mr@ignsp\fi}
\newcommand{\multiref}[1]{\begingroup%
\xdef\mr@no@sparg{\expandafter\mr@ignsp#1 \: }%
\def\mr@comma{}\def\mr@dash{-}%
\@for\mr@refs:=\mr@no@sparg\do{%
\ifx\mr@refs\mr@dash\def\mr@comma{}--\else%
\mr@comma\def\mr@comma{,}\ref{\mr@refs}\fi}%
\endgroup}
\renewcommand{\eqref}[1]{(\multiref{#1})}
\makeatother

\makeatletter
\newcommand{\namedref}[2]{\hyperref[#2]{#1~\ref*{#2}}}
\newcommand{\secref}{\@ifstar{\namedref{Section}}{\namedref{sec.}}}
\newcommand{\appref}{\@ifstar{\namedref{Appendix}}{\namedref{app.}}}
\providecommand{\footref}{}
\renewcommand{\footref}{\@ifstar{\namedref{Footnote}}{\namedref{foot.}}}
\newcommand{\tabref}{\@ifstar{\namedref{Table}}{\namedref{tab.}}}
\newcommand{\figref}{\@ifstar{\namedref{Figure}}{\namedref{fig.}}}
\makeatother

\let\oldbib=\thebibliography
\def\thebibliography{\phantomsection\addcontentsline{toc}{section}{\refname}\oldbib}

\let\oldtoc=\tableofcontents
\def\tableofcontents{\phantomsection\addcontentsline{toc}{section}{\contentsname}\oldtoc}

\providecommand{\hypersetup}[1]{}
\providecommand{\texorpdfstring}[2]{#1}
\hypersetup{plainpages=false}
\hypersetup{pdfpagemode=UseNone}
\hypersetup{bookmarksnumbered=true}
\hypersetup{pdfstartview=FitH}
\hypersetup{colorlinks=false}
\hypersetup{citebordercolor={1 1 1}}
\hypersetup{urlbordercolor={1 1 1}}
\hypersetup{linkbordercolor={1 1 1}}

\makeatletter
\let\@keywords\@empty
\let\@subject\@empty
\providecommand{\keywords}[1]{\gdef\@keywords{#1}}
\providecommand{\subject}[1]{\gdef\@subject{#1}}
\def\thetitle{\@title}
\def\theauthor{\@author}
\def\thesubject{\@subject}
\def\thedate{\@date}
\def\thekeywords{\@keywords}
\makeatother
\AtBeginDocument{
\hypersetup{pdftitle={\thetitle}}
\hypersetup{pdfauthor={\theauthor}}
\hypersetup{pdfsubject={\thesubject}}
\hypersetup{pdfkeywords={\thekeywords}}}

\newif\ifshownote
\shownotetrue

\ifpublic\shownotefalse\fi

\ifshownote

\ifpdf\else\RequirePackage[active]{srcltx}\fi
\RequirePackage{xcolor}
\RequirePackage{ulem}

\newcommand{\remark}[2][]{{\normalfont\sffamily\hspace{1ex}
\def\emph{\textsl}\def\textbullet{$\bullet$}
\def\tmparga{#1}
\def\tmpargb{BH}\ifx\tmparga\tmpargb\color[rgb]{0.5,0,0}\fi
\def\tmpargb{FS}\ifx\tmparga\tmpargb\color[rgb]{0,0.5,0}\fi
\def\tmpargb{NL}\ifx\tmparga\tmpargb\color[rgb]{0,0,0.5}\fi
\def\tmpargb{}\ifx\tmparga\tmpargb\color{red}\fi
\def\tmpargb{}\ifx\tmparga\tmpargb\else \textbf{#1:}\fi
#2\hspace{1ex}}}

\else
\newcommand{\remark}[2][]{\ignorespaces}

\fi

\title{Bi-\texorpdfstring{$\eta$}{eta} and bi-\texorpdfstring{$\lambda$}{lambda} deformations of \texorpdfstring{$\Integer_4$}{Z4} permutation supercosets}
\author{Ben Hoare, Nat Levine and Fiona K. Seibold}

\begin{document}

\pdfbookmark[1]{Title Page}{title}
\thispagestyle{empty}


\vspace*{2cm}
\begin{center}
\begingroup\Large\bfseries\thetitle\par\endgroup
\vspace{1cm}

\begingroup
Ben Hoare,$\,^{a,}$\footnote{ben.hoare@durham.ac.uk}
Nat Levine$\,^{b,c,}$\footnote{nat.levine@phys.ens.fr}
and
Fiona K. Seibold$\,^{d,}$\footnote{f.seibold21@imperial.ac.uk}
\par\endgroup
\vspace{1cm}

$^a${\it Department of Mathematical Sciences, Durham University, Durham DH1 3LE, UK}

\vspace{0.3cm}

$^{b}${\it Laboratoire de Physique,} \qquad $^c${\it Institut Philippe Meyer,\\
{\'E}cole Normale Sup{\'e}rieure, Universit{\'e} PSL, CNRS, Sorbonne Universit{\'e}, Universit{\'e} Paris Cit{\'e}, \\
24 rue Lhomond, F-75005 Paris, France}

\vspace{0.3cm}

$^d${\it Blackett Laboratory, Imperial College, London SW7 2AZ, UK}

\vfill

\textbf{Abstract}\vspace{5mm}

\begin{minipage}{12.5cm}\small
Integrable string sigma models on $\AdS_3$ backgrounds with 16 supersymmetries have the distinguishing feature that their superisometry group is a direct product.
As a result the deformation theory of these models is particularly rich since the two supergroups in the product can be deformed independently.
We construct bi-$\eta$ and bi-$\lambda$ deformations of two classes of $\Integer_4$ permutation supercoset sigma models, which describe sectors of the Green-Schwarz and pure-spinor string worldsheet theories on type II $\AdS_3$ backgrounds with pure R-R flux.
We discuss an important limit of these models when one supergroup is undeformed.
The associated deformed supergravity background should preserve 8 supersymmetries and is expected to have better properties than the full bi-deformation.
As a step towards investigating the quantum properties of these models, we study the two-loop RG flow of the bosonic truncation of the bi-$\lambda$ deformation.
\end{minipage}

\vspace*{2cm}

\end{center}

\newpage

\setcounter{footnote}{0}

\tableofcontents


\section{Introduction}\label{sec:int}

Integrable string sigma models on $\AdS_3$ backgrounds with 16 supersymmetries and supported by pure R-R flux have received considerable attention in recent years
(for some recent developments and further references, see \cite{Frolov:2021fmj,Seibold:2022mgg,Frolov:2021bwp,Cavaglia:2021eqr,Ekhammar:2021pys,Cavaglia:2022xld}).
One of their distinguishing features is the direct product structure of their superisometry group and, as a result, their deformation theory is particularly rich.
For type II superstrings on $\AdS_3 \times \Sp^3 \times \To^4$ and $\AdS_3 \times \Sp^3 \times \Sp^3 \times \Sp^1$ supported by R-R flux,
the superisometry groups are $\grp{PSU}(1,1|2) \times \grp{PSU}(1,1|2) \times \grp{U}(1)^4$ and $\grp{D}(2,1;\alpha) \times \grp{D}(2,1;\alpha) \times \grp{U}(1)$ respectively.
For the curved part of the geometry, the associated worldsheet
theories in the Green-Schwarz (GS)~\cite{Rahmfeld:1998zn,Park:1998un,Metsaev:2000mv,Babichenko:2009dk} and pure-spinor (PS)~\cite{Berkovits:1999im,Berkovits:1999du} formalisms contain sectors described by sigma models on $\Integer_4$ permutation supercosets
\unskip\footnote{In the $\AdS_3 \times \Sp^3 \times \To^4$ and $\AdS_3 \times \Sp^3 \times \Sp^3 \times \Sp^1$ cases, the relevant $\Integer_4$ supercosets capturing the curved part of the geometry are \label{foot1}
\begin{equation*}
\frac{\grp{PSU}(1,1|2) \times \grp{PSU}(1,1|2)}{\grp{SU}(1,1) \times \grp{SU}(2)} ~,
\qquad
\frac{\grp{D}(2,1;\alpha) \times \grp{D}(2,1;\alpha)}{\grp{SU}(1,1) \times \grp{SU}(2) \times \grp{SU}(2)} ~.
\end{equation*}.}
\begin{equation}
\frac{\grp{G} \times \grp{G}}{\grp{G}_0} ~,
\end{equation}
where $\grp{G}$ is a Lie supergroup
and $\grp{G}_0$ is the diagonal even subgroup of the direct product.

In both formalisms, the $\Integer_4$ supercoset sigma models take the form
\begin{equation}\label{eq:z4coset}
\Act = -\frac{T}{2} \int \extder^2x \STr\big(g^{-1}\partial_+ g \, P_- \, g^{-1} \partial_- g\big) ~,
\end{equation}
where $T$ is the string tension, $g(x^\pm) \in \grp{G} \times \grp{G}$ is a supergroup-valued field and $\STr$ is an invariant bilinear form.
The linear operator $P_-$ is a sum of projectors onto the $\Integer_4$ graded subspaces of the Lie superalgebra $\alg{g} \oplus \alg{g} = \Lie(\grp{G}\times \grp{G})$.
In the Green-Schwarz (GS) formalism, $P_- = P_1 + 2 P_2 - P_3$~\cite{Metsaev:1998it,Henneaux:1984mh}, while in the pure-spinor (PS) formalism, $P_- = P_1 + 2 P_2 + 3 P_3$~\cite{Berkovits:1999zq,Berkovits:2000fe} (henceforth referred to as the GS and PS cases).
For both choices of $P_-$ the action~\eqref{eq:z4coset} is classically integrable -- the equations of motion can be written as the zero-curvature of a Lax connection~\cite{Bena:2003wd,Vallilo:2003nx} and the Poisson bracket of the Lax matrix is a Maillet bracket of twist form~\cite{Mikhailov:2007eg,Magro:2008dv,Vicedo:2010qd}, ensuring that the conserved charges extracted from the monodromy of the Lax matrix are in involution~\cite{Maillet:1985ek,Maillet:1985ec,Sevostyanov:1995hd,Lacroix:2018njs}.
Recalling that $P_0$ and $P_2$ project onto Grassmann-even subspaces of $\alg{g} \oplus\alg{g}$, and $P_1$ and $P_3$ onto Grassmann-odd subspaces, the bosonic truncation of
both the GS and PS sigma models is the symmetric space sigma model on the $\Integer_2$ permutation coset
\begin{equation}
\frac{\grp{G} \times \grp{G}}{\grp{G}} ~,
\end{equation}
which is equivalent, upon gauge fixing, to the principal chiral model (PCM) on the group $\grp{G}$ (where $\grp{G}$ is now an ordinary Lie group).

In this paper we explore integrable deformations of $\Integer_4$ permutation supercoset sigma models.
Integrable deformations are typically associated with deformations of the underlying symmetry group.
In this case the direct product structure of the superisometry group allows us to deform each copy of $\grp{G}$ independently.
In particular, our goal will be to construct bi-deformations of these models, with the two copies of $\grp{G}$ deformed with different strengths.
Constructing the bi-deformed models is important since it allows us to take the limit where one copy of $\grp{G}$ is undeformed.
The resulting model
still
has half the supersymmetry of the original model, so can have
``nicer''
properties than if the symmetry is fully deformed.
One such example was recently studied in detail in \cite{Hoare:2022asa}; starting from the bi-$\eta$ deformation of the $\AdS_3 \times \Sp^3 \times \To^4$ superstring~\cite{Hoare:2014oua,Seibold:2019dvf}, in the limit where only one copy of $\grp{PSU}(1,1|2)$ is deformed, the geometry becomes smooth and the dilaton is constant.

\bigskip

The first type of bi-deformations that we discuss are the bi-$\eta$ deformations.
The $\eta$ deformation, or Yang-Baxter deformation, was introduced by Klim\v{c}\'{i}k as an integrable deformation of the PCM~\cite{Klimcik:2002zj}, and later generalised to the symmetric space~\cite{Delduc:2013fga} and $\Integer_4$ supercoset sigma models~\cite{Delduc:2013qra,Kawaguchi:2014qwa,Benitez:2018xnh,Hoare:2021dix}.
There are three classes of $\eta$ deformations: homogeneous, split inhomogeneous or non-split inhomogeneous, depending on whether the operator $R$ defining the deformation solves the unmodified, split modified or non-split modified classical Yang-Baxter equation.
For a given model, the symmetry algebra determines which of these are possible and the full space of $\eta$ deformations.
The bi-$\eta$, or bi-Yang-Baxter deformation, of the PCM~\cite{Klimcik:2008eq} was an early example of a bi-deformation, which coincides with the two-parameter deformation of the $\grp{O}(4)$ sigma model~\cite{Fateev:1996ea} for $\grp{G} = \grp{SU}(2)$~\cite{Hoare:2014pna}.
It was subsequently generalised to the GS sigma model on $\Integer_4$ permutation supercosets in \cite{Hoare:2014oua}.
In \secref{sec1} we review and further generalise this construction, and derive the bi-$\eta$ deformation of the PS sigma model on $\Integer_4$ permutation supercosets.
In particular, we will allow the two copies of $\grp{G}$ to be deformed in different ways, which will be useful when we discuss Poisson-Lie duality in \secref{sec:pl}.
\unskip\footnote{It is known that WZ terms can be added to the PCM~\cite{Novikov:1982ei,Witten:1983tw,Witten:1983ar,Veselov:1984,Abdalla:1984gm} and the GS $\Integer_4$ permutation supercoset sigma model~\cite{Cagnazzo:2012se} while preserving their classical integrability, and this should also be possible in the PS case too as suggested in \cite{Berkovits:1999im}.
Doing so corresponds to supporting the $\AdS_3$ backgrounds by a mix of R-R and NS-NS flux.
Bi-$\eta$ deformations in the presence of these WZ terms can still be constructed~\cite{Delduc:2017fib,Delduc:2018xug,Klimcik:2019kkf}, however, the operator $R$ needs to satisfy an additional compatibility condition~\cite{Hoare:2020mpv}.
On the other hand, Poisson-Lie duality in the presence of a WZ term is more subtle.
While it is still possible to construct an $\mathcal E$ model and integrate out degrees of freedom to obtain the (bi-)$\eta$ deformation (with WZ term), it appears that there is no isotropic subalgebra of the Drinfel'd double that gives a generalisation of the (bi-)$\lambda$ model.}

In \secref{sec2} we construct the bi-$\lambda$ deformations of the GS and PS $\Integer_4$ permutation supercoset sigma models.
The $\lambda$ deformation of the PCM and the symmetric space sigma model was first constructed in~\cite{Sfetsos:2013wia,Hollowood:2014rla} generalising the $\grp{G} = \grp{SU}(2)$ model of \cite{Balog:1993es}.
The deformed model interpolates between the non-abelian T-dual of the original model~\cite{delaOssa:1992vci} and the (gauged) Wess-Zumino-Witten model~\cite{Novikov:1982ei,Witten:1983tw,Witten:1983ar,Witten:1991mm}.
It was later generalised to both the GS $\Integer_4$ case~\cite{Hollowood:2014qma} and the PS $\Integer_4$ case~\cite{Hoare:2021dix}.
As mentioned above, the bosonic truncation of both the GS and PS $\Integer_4$ permutation supercoset sigma models is the PCM.
The bi-$\lambda$ deformation for this model was introduced in~\cite{Sfetsos:2014cea} and a Lax connection was constructed in~\cite{Sfetsos:2017sep}.
Therefore, the models we construct should give an embedding of this bosonic model into string theory for $\grp{G} = \grp{PSU}(1,1|2)$ or $\grp{D}(2,1;\alpha)$.
A potentially important limit, which we discuss in some detail, is when one copy of $\grp{G}$ becomes undeformed.
In this limit, undoing the non-abelian T-duality in the undeformed copy of $\grp{G}$, the resulting model is expected to describe an embedding of the $\lambda$ deformation of the PCM into string theory.
Therefore, in the limit $\lambda \to 0$ the bosonic truncation is just the WZW model.
This deformation still has half the supersymmetry of the undeformed model and, just as for the bi-$\eta$ deformation, the associated supergravity background may
thus have ``nicer'' properties.

The split $\eta$ deformation is known to be the Poisson-Lie dual~\cite{Klimcik:1995ux,Klimcik:1995jn} of the $\lambda$ deformation~\cite{Vicedo:2015pna,Hoare:2021dix}, while the non-split $\eta$ deformation is also dual up to analytic continuation~\cite{Hoare:2015gda,Sfetsos:2015nya,Hoare:2017ukq,Klimcik:2015gba}.
In \secref{sec:pl} we show that the bi-$\eta$ and bi-$\lambda$ models that we construct in secs.~\ref{sec1} and~\ref{sec2} are similarly related by Poisson-Lie duality.
This is achieved by showing that both models follow from a first-order model, the $\mathcal{E}$ model~\cite{Klimcik:1995dy,Klimcik:1996nq,Klimcik:2015gba,Klimcik:2019kkf}, on the Drinfel'd double, generalising the duality-invariant action of~\cite{Tseytlin:1990nb,Tseytlin:1990va} underlying abelian T-duality.
Starting from this $\mathcal{E}$ model it is then possible to construct further bi-deformations.
This includes the $\eta$-$\lambda$ deformation, where the $\eta$ deformation is associated to one copy of $\grp{G}$ and the $\lambda$ to the other.
The bosonic truncation of this model was earlier constructed via Poisson-Lie duality and analytic continuation in~\cite{Sfetsos:2015nya}.
Again, the $\Integer_4$ generalisation is expected to define an embedding into string theory.

Having constructed classical integrable bi-deformations, it is interesting to explore the quantum properties of these models.
In the context of string theory, a key question is whether or not the deformations preserve Weyl invariance.
Typically, the $\lambda$ deformation of $\Integer_4$ supercosets leads to Weyl invariant string sigma models, while this is only the case for the $\eta$ deformation when the operator $R$ is unimodular~\cite{Borsato:2016ose}.
Examples of such unimodular operators for string sigma models have been studied for homogeneous~\cite{Borsato:2016ose,vanTongeren:2019dlq} and non-split modified~\cite{Hoare:2018ngg,Seibold:2019dvf} deformations.
Weaker conditions that can be investigated are renormalisability~\cite{Balog:1993es,Fateev:1996ea,Itsios:2014lca,Sfetsos:2015nya} and scale invariance~\cite{Appadu:2015nfa,Arutyunov:2015mqj}, both of which are generically preserved by these deformations.
Much is known about the one-loop properties of the bi-deformations -- the bi-$\eta$ and bi-$\lambda$ deformations of the PCM are renormalisable~\cite{Sfetsos:2015nya,Sfetsos:2017sep}, while the bi-deformations of the $\Integer_4$ permutation supercoset string sigma models are expected to be scale and Weyl invariant (assuming unimodularity in the bi-$\eta$ case)~\cite{Seibold:2019dvf}.
At higher loops, less is known about the bi-deformations.
Therefore, in~\secref{sec:rg} we study the renormalisation group flow of the bi-$\lambda$ deformation of the PCM using the ``tripled'' formulation introduced in~\cite{Hoare:2019mcc,Levine:2021fof}.
We show that, in this formulation, the model is renormalisable to all orders due to its manifest symmetries and the decoupling of certain fields.
We explicitly compute the two-loop beta function in a standard minimal scheme \cite{Metsaev:1987zx}.

We conclude in~\secref{sec6} with comments on our results and future directions.

\section{Bi-\texorpdfstring{$\eta$}{eta} models}\label{sec1}

In this section we review the integrable bi-$\eta$ deformation of the GS sigma model on $\Integer_4$ permutation supercosets
\cite{Hoare:2014oua}.
We also write down an integrable action for the bi-$\eta$ deformation of the PS sigma model, generalising the one-parameter deformation of~\cite{Benitez:2018xnh,Hoare:2021dix}.

As outlined in the \hyperref[sec:int]{Introduction}, $\Integer_4$ permutation supercosets take the form
\begin{equation}\label{eq:psc}
\frac{\grp{G}\times\grp{G}}{\grp{G}_0} ~,
\end{equation}
where $\grp{G}$ is a Lie supergroup and $\grp{G}_0$ is the diagonal
even
subgroup of $\grp{F} \equiv \grp{G} \times \grp{G}$.
The Lie superalgebra $\alg{f} \equiv \alg{g} \oplus \alg{g} = \Lie(\grp{G} \times \grp{G})$ admits a $\Integer_4$ automorphism
\begin{equation}
\sigma(X_\ind{L},X_\ind{R}) = (X_\ind{R},
(p_0-p_1)X_\ind{L}) ~, \qquad X_{{\ind{L},\ind{R}}}\in\alg{g} ~,
\end{equation}
where $p_0$ and $p_1$ project onto the Grassmann-even and Grassmann-odd subspaces of $\alg{g}$ respectively.
This leads to a $\Integer_4$ grading of $\alg{g} \oplus \alg{g}$ with the projectors $P_i$ onto the grade-$i$ subspaces given by
\unskip\footnote{Note that, strictly speaking, for this to be a $\Integer_4$ grading of the real form, one needs different matrix realisations of the superalgebra for each copy.
Equivalently, we can modify the reality conditions obeyed by the Grassmann-odd fields accordingly.}
\begin{equation} \label{eq:proj} \begin{split}
P_0(X_\ind{L},X_\ind{R}) & = \frac12(p_0(X_\ind{L}+X_\ind{R}),p_0(X_\ind{L}+X_\ind{R})) ~,
\\
P_1(X_\ind{L},X_\ind{R}) & = \frac12(p_1(X_\ind{L}-iX_\ind{R}),p_1(X_\ind{R}+iX_\ind{L})) ~,
\\
P_2(X_\ind{L},X_\ind{R}) & = \frac12(p_0(X_\ind{L}-X_\ind{R}),p_0(X_\ind{R}-X_\ind{L})) ~,
\\
P_3(X_\ind{L},X_\ind{R}) & = \frac12(p_1(X_\ind{L}+iX_\ind{R}),p_1(X_\ind{R}-iX_\ind{L})) ~.
\end{split}\end{equation}

To write down the deformed action it will be useful to introduce the operator
\begin{equation}
W = P_\ind{L} - P_\ind{R}~, \qquad W P_{0,2} = P_{2,0} W ~, \qquad W P_{1,3} = P_{3,1} W ~,
\end{equation}
where $P_\ind{L}$ and $P_\ind{R}$ project onto the left (first) and right (second) copies of $\alg{g}$,
along with the $\Integer_2$-symmetric bilinear form
\begin{equation}\label{eq:str}
\STr\big((X_\ind{L},X_\ind{R}) (Y_\ind{L},Y_\ind{R}) \big) = \str(X_\ind{L}Y_\ind{L}) + \str(X_\ind{R}Y_\ind{R}) ~,
\end{equation}
with $\str$ an ad-invariant bilinear form on $\alg{g}$.
For linear operators $\mathcal{O}$ on $\alg{g}\oplus\alg{g}$ we denote their transpose with respect to this bilinear form as
\begin{equation}
\STr\big(u\, \mathcal{O}v\big) = \STr\big((\mathcal{O}^t u )\, v\big)~.
\end{equation}
Note that we have
\begin{equation}
P_i^t = P_{4-i \mod 4} ~, \qquad P_{\ind{L},\ind{R}}^t = P_{\ind{L},\ind{R}} ~, \qquad W^t = W ~.
\end{equation}

\paragraph{Action and equations of motion.}
The action of the bi-$\eta$ deformation of the $\mathbb{Z}_4$ supercoset sigma model~\eqref{eq:z4coset} is of the form
\begin{equation} \label{eq:acteta}
\Act_{\eta_\ind{L},\eta_\ind{R}} = -\frac{T}{2} \int \extder^2 x \, \STr \Big( g^{-1} \partial_+ g \, \mathcal{P}_- \frac{1}{1-R_g(\eta_\ind{L} P_\ind{L} + \eta_\ind{R} P_\ind{R}) \mathcal{P}_-} g^{-1} \partial_- g \Big)~.
\end{equation}
The two-dimensional base manifold is parametrised by $x^0 \equiv \tau$ and $x^1 \equiv \sigma$ and we use the light-cone coordinates $x^\pm = \half (x^0 \pm x^1)$ and $\partial_\pm = \partial_0 \pm \partial_1$.
The action is for the supergroup-valued field $g(x^\pm) \in \grp{G} \times \grp{G}$ and depends on three real parameters: $T$ is an overall constant (the string tension in the context of
string theory), while $\eta_\ind{L}$ and $\eta_\ind{R}$ parametrise the strength of deformation of the left and right copies of $\grp{G}$ respectively.
The dressed operator $R_g = \Ad_g^{-1} R \Ad_g$ is defined in terms of the deforming linear operator $R: \alg{g} \oplus \alg{g} \rightarrow \alg{g} \oplus \alg{g}$.
Note that while $P_\ind{L}$ and $P_\ind{R}$ commute with $\Ad_g$, they do not necessarily commute with $R$.
\unskip\footnote{In \cite{Hoare:2014oua}, $R$ was taken to be of factorised form, i.e., $R = R_\ind{LL} \oplus R_\ind{RR}$, in which case $P_\ind{L}$ and $P_\ind{R}$ commute with $R$.}
We take the operator $R$ to have the following symmetry property with respect to the bilinear form $\STr$
\begin{equation}
R^t = -(\eta_\ind{L}^{-1} P_\ind{L} + \eta_\ind{R}^{-1} P_\ind{R}) R (\eta_\ind{L} P_\ind{L} + \eta_\ind{R} P_\ind{R}) ~,
\end{equation}
and to satisfy the (modified) classical Yang-Baxter equation
\begin{equation} \label{eq:cYBE}
\com{ RX}{ RY} - R(\com{ RX}{Y}+ \com{X}{ RY}) = - \big(c_\ind{L}^2 P_\ind{L} + c_\ind{R}^2 P_\ind{R}\big) \com{X}{Y}~, \qquad X,Y \in \alg{g}\oplus\alg{g}~.
\end{equation}
Without loss of generality, the constants $c_\ind{L}$ and $c_\ind{R}$ can be either $0$ (homogeneous), $1$ (split) or $i$ (non-split).
We treat all these cases on an equal footing, in particular allowing for different classes of deformation for the two copies of $\grp{G}$.
Finally, the constant linear operator $\mathcal{P}_-$ (as well as its transpose $\mathcal{P}_+ = \mathcal{P}_-^t$) depends on the projectors $P_j$ defined in \eqref{eq:proj}.
Its explicit form, fixed by requiring the classical integrability of \eqref{eq:acteta}, is discussed below.

In terms of the auxiliary currents
\begin{equation}
A_\pm := \frac{1}{1\pm R_g (\eta_\ind{L} P_\ind{L} + \eta_\ind{R} P_\ind{R}) \mathcal{P}_\pm} g^{-1} \partial_\pm g~,
\end{equation}
the equations of motion following from the action~\eqref{eq:acteta} and the zero-curvature equation for $g^{-1}\partial_\pm g$ take the form
\begin{gather}
\label{eq:eom1}
\partial_+ (\mathcal{P}_- A_-) + \partial_- (\mathcal{P}_+ A_+) + \com{A_+}{\mathcal{P}_- A_-} + \com{A_-}{\mathcal{P}_+ A_+} = 0 ~, \\
\label{eq:eom2}
\partial_+ A_- - \partial_- A_+ + \com{A_+}{A_-} - (\cel^2 P_\ind{L} + \cer^2 P_\ind{R}) \com{\mathcal{P}_+ A_+}{\mathcal{P}_- A_-} = 0~.
\end{gather}
It is also insightful to define the quantities
\begin{equation}
B_\pm = \Ad_g \mathcal{P}_\pm A_\pm~,
\end{equation}
in terms of which the equations of motion take the manifestly Poisson-Lie symmetric form
\unskip\footnote{From~\eqref{eq:cYBE} it follows that $\widehat{R} = R (\eta_\ind{L} P_\ind{L} + \eta_\ind{R} P_\ind{R} )$ solves the (modified) classical Yang-Baxter equation
\begin{equation*}
\com{ \widehat RX}{ \widehat RY} - \widehat R(\com{ \widehat RX}{Y}+ \com{X}{ \widehat RY}) = - \big(c_\ind{L}^2 \eta_\ind{L}^2 P_\ind{L}+ c_\ind{R}^2\eta_\ind{R}^2 P_\ind{R}\big) \com{X}{Y}~, \qquad X,Y \in \alg{g}\oplus\alg{g}~.
\end{equation*}
}
\begin{equation} \label{eq:eom3}
\partial_+ B_- + \partial_- B_+ = (\com{R(\eta_\ind{L} P_\ind{L} + \eta_\ind{R} P_\ind{R} ) B_+}{B_-} + \com{B_+}{R(\eta_\ind{L} P_\ind{L} + \eta_\ind{R} P_\ind{R} ) B_-})~.
\end{equation}
In \secref{sec:pl} we will show that the Poisson-Lie duals of the bi-$\eta$ deformations constructed here are the bi-$\lambda$ deformations constructed in \secref{sec2}.

\paragraph{Integrability and solutions for $\mathcal{P}_\pm$.}
The $\grp{G}_0$ gauge invariance of the model requires that
\begin{equation} \label{eq:cP}
\mathcal{P}_\pm = \rho P_2 + \alpha_\pm P_1 + \alpha_\mp P_3 + \beta_+ W P_1 + \beta_- W P_3~.
\end{equation}
We would now like to find constant parameters $\alpha_\pm$, $\beta_\pm$ and $\rho$ such that the equations \eqref{eq:eom1}, \eqref{eq:eom2} can be recast as the zero-curvature condition of a Lax connection, which gives a strong indication that the model is classically integrable.
\unskip\footnote{The final step to prove Hamiltonian integrability would be to demonstrate that there are infinitely many local conserved charges in involution.
One way to do this is to show that the Poisson bracket of the Lax matrix takes the form of a Maillet bracket governed by a twist function \cite{Maillet:1985ek,Maillet:1985ec,Sevostyanov:1995hd,Lacroix:2018njs}.}
A way to ensure this is if, upon redefining the currents as $J_\pm = \mathcal O_\pm A_\pm$ where $\mathcal O_\pm$ are constant invertible linear operators, the equations take the following form
\begin{equation} \label{eq:eomundef} \begin{gathered}
\partial_+(P_- J_-) + \partial_- (P_+ J_+) + [J_+,P_- J_-] + [J_-, P_+ J_+] = 0 ~,
\\
\partial_+ J_- - \partial_- J_+ + [J_+,J_-] = 0 ~,
\end{gathered}\end{equation}
where $P_+ = P_-^t$ with $P_- = P_1 + 2P_2 - P_3$ in the GS case and $P_- = P_1 + 2 P_2 + 3 P_3$ in the PS case.
In both cases, the equations \eqref{eq:eomundef} follow from a Lax connection.
In the GS case the Lax connection is given by
\begin{equation}\label{LaxGS}
L_\pm = J_\pm^{(0)} + z J_\pm^{(1)} + z^{\mp 2} J_\pm^{(2)} + z^{-1} J_\pm^{(3)}~,
\end{equation}
where $J_\pm^{(j)} = P_j J_\pm$ for $j=0,1,2,3$, while in the PS case it is
\begin{equation}\label{LaxPS}
L_\pm = J_\pm^{(0)} + z^{-1\mp2} J_\pm^{(1)} + z^{\mp 2} J_\pm^{(2)} + z^{1\mp 2} J_\pm^{(3)}~.
\end{equation}

In terms of the auxiliary currents $A_\pm$, the equations \eqref{eq:eom1}, \eqref{eq:eom2} depend only on the combinations $\cel^2$ and $\cer^2$, so this will also be true for $\mathcal{P}_\pm$.
For brevity, we define
\begin{equation}
a_\ind{L} = \frac{1}{\sqrt{1+\cel^2}} ~, \qquad
a_\ind{R} = \frac{1}{\sqrt{1+\cer^2}} ~.
\end{equation}

Let us start with the GS case, for which it is known~\cite{Hoare:2014oua} that the solution is given by eq.~\eqref{eq:cP} with
\begin{equation} \begin{split}
\rho = 2 a_\ind{L} a_\ind{R}~, \quad \alpha_\pm = \mp 1~, \quad \beta_\pm=0~.
\label{etaGS}
\end{split}\end{equation}
The currents appearing in the Lax connection are
\begin{equation} \label{eq:JGS}\begin{aligned}
&\begin{aligned}
J_\pm^{(0)} &= A_\pm^{(0)} + (a_\ind{L}^2 - a_\ind{R}^2) W A_\pm^{(2)}~, &
J_\pm^{(2)} &= (a_\ind{L}^2 + a_\ind{R}^2 - 1) A_\pm^{(2)}~, \\
J_\pm^{(1)} &= \xi^{1/2}(A_\pm^{(1)} + \omega W A_\pm^{(3)})~, &
J_\pm^{(3)} &= \xi^{1/2}(A_\pm^{(3)} + \omega W A_\pm^{(1)})~, \\
\end{aligned}
\\
&\omega = \frac{a_\ind{L} - a_\ind{R}}{a_\ind{L} + a_\ind{R}} ~, \qquad
\xi = \frac{(a_\ind{L}^{-1} + a_\ind{R}^{-1})^2 (a_\ind{L}^2 + a_\ind{R}^2 - 1)}{4} ~.
\end{aligned}
\end{equation}
In the PS case, we find the coefficients determining $\mathcal{P}_\pm$ to be
\begin{equation} \label{etaPS} \begin{aligned}
\rho= 2 a_\ind{L} a_\ind{R} ~, \qquad
\alpha_\pm = \frac{(a_\ind{L}\pm a_\ind{R})^2}{a_\ind{L}a_\ind{R}(3-a_\ind{L}^2-a_\ind{R}^2)} \mp 1 ~,
\qquad \beta_\pm = \frac{a_\ind{L}^2-a_\ind{R}^2}{a_\ind{L}a_\ind{R}(3-a_\ind{L}^2-a_\ind{R}^2)} ~.
\end{aligned}
\end{equation}
The bosonic currents $J_\pm^{(0)}$, $J_\pm^{(2)}$ are the same as the GS case~\eqref{eq:JGS}.
For the fermions, we have
\begin{equation}
J_\pm^{(1)} = \xi_\pm^{1/2} (A_\pm^{(1)} + \omega_\pm W A_\pm^{(3)})~, \qquad
J_\pm^{(3)} = \xi_\mp^{1/2} (A_\pm^{(3)} + \omega_\mp W A_\pm^{(1)})~,
\end{equation}
with
\begin{equation}\begin{aligned}
\omega_+ &= \frac{a_\ind{L} - a_\ind{R}}{a_\ind{L} + a_\ind{R}} ~, &
\omega_- &= \frac{a_\ind{L} - a_\ind{R}}{a_\ind{L} + a_\ind{R}} \frac{3 + a_\ind{L}^2 +4 a_\ind{L}a_\ind{R} + a_\ind{R}^2}{3+a_\ind{L}^2-4a_\ind{L}a_\ind{R} + a_\ind{R}^2} ~,\\
\xi_+ &= \frac{(a_\ind{L}^{-1} + a_\ind{R}^{-1})^2(a_\ind{L}^2 + a_\ind{R}^2 -1)^3}{4(3-a_\ind{L}^2-a_\ind{R}^2)^2} ~, &
\xi_- &= \frac{(a_\ind{L}^{-1} + a_\ind{R}^{-1})^2(a_\ind{L}^2 + a_\ind{R}^2 -1)(3+a_\ind{L}^2-4a_\ind{L}a_\ind{R} + a_\ind{R}^2)^2}{4(3-a_\ind{L}^2-a_\ind{R}^2)^2} ~.
\end{aligned}\end{equation}

\paragraph{Limits and truncations.}
In both the GS and PS cases the bosonic truncation of the bi-$\eta$ deformation gives the bi-$\eta$ (or bi-Yang-Baxter) deformation of the PCM \cite{Klimcik:2008eq}.
When both $\eta_\ind{L} \rightarrow 0 $ and $\eta_\ind{R} \rightarrow 0$ we find the undeformed sigma model \eqref{eq:z4coset} with $\mathcal{P}_- \to P_- = P_1 + 2 P_2 - P_3$ in the GS case and $\mathcal{P}_- \to P_- = P_1 + 2 P_2 + 3 P_3$ in the PS case as expected.
The symmetric deformation $c_\ind{L}\eta_\ind{L}=c_\ind{R}\eta_\ind{R} = c\eta$
corresponds to the standard $\eta$ deformation of the $\Integer_4$ supercoset, with
the known result $\mathcal{P}_- = P_1 + \frac{2}{1+c^2\eta^2} P_2 - P_3$ in the GS case~\cite{Delduc:2013qra,Kawaguchi:2014qwa} and $\mathcal{P}_- = P_1 + \frac{2}{1+c^2\eta^2} + \frac{3+c^2\eta^2}{1+3c^2 \eta^2} P_3$ in the PS case~\cite{Benitez:2018xnh,Hoare:2021dix}.

Another interesting limit is when one deformation parameter is set to vanish, e.g.~$\eta_\ind{R} = 0$.
In this limit one copy of $\grp{G}$ is undeformed and the deformation preserves half the supersymmetries of the original model.
This limit (also including a WZ term) was studied in detail for the $\AdS_3 \times \Sp^3 \times \To^4$ background~\cite{Hoare:2022asa}, where it was observed that the deformed background has particularly ``nice'' properties, including a smooth geometry and constant dilaton.

\section{Bi-\texorpdfstring{$\lambda$}{lambda} models}\label{sec2}

In this section we present the construction of the integrable bi-$\lambda$ models on $\Integer_4$ permutation supercosets and their Lax connections.

\paragraph{Action and equations of motion.}
Recalling the construction of the standard $\lambda$-models \cite{Sfetsos:2013wia,Hollowood:2014qma}, the action of the integrable bi-$\lambda$ models on $\Integer_4$ permutation supercosets is expected to take the form
\begin{equation}\begin{split} \label{act1}
k_\ind{L} \Act_{\grp{G}/\grp{G}}(g_\ind{L},A_\ind{L};\str) + k_\ind{R} \Act_{\grp{G}/\grp{G}}(g_\ind{R},A_\ind{R};\str) + \text{bilinear}\left( (A_{\ind{L}+},A_{\ind{R}+}), (A_{\ind{L}-},A_{\ind{R}-})\right) ~,
\end{split}\end{equation}
where $g_{\ind{L},\ind{R}}$ are fields valued in the supergroup $\grp{G}$, $A_{\ind{L},\ind{R}\pm}$ are valued in the superalgebra $\alg{g}$, $4\pi k_{\ind{L},\ind{R}}$ are (integer-quantized) levels and $\Act_{\grp{G}/\grp{G}} (g,A;\str)$ denotes the action of the gauged WZW model
\begin{align}\nn
\Act_{\grp{G}/\grp{G}} (g,A;\str) & = -\frac{1}{2} \int \extder^2x \, \str(g^{-1}\partial_+ g g^{-1}\partial_- g) +
\Act_{\textrm{WZ}}(g;\str)
\\ \label{gWZW} & \quad \qquad + \int \extder^2x \, \str( A_+ g^{-1}\partial_- g - \partial_+ g g^{-1} A_- + A_+ g^{-1} A_- g - A_+ A_- ) ~.
\end{align}
Here $\Act_{\textrm{WZ}}(g;\str)$ denotes the Wess-Zumino term
\begin{equation}
\Act_{\textrm{WZ}}(g;\str) = \frac16 \int d^3 x \, \epsilon^{ijk} \str(g^{-1}\partial_i g [g^{-1}\partial_j g, g^{-1} \partial_k g]) ~.
\end{equation}
Given the form of \eqref{act1} it is convenient to introduce a second bilinear form on $\alg{f}$ (in addition to the one defined in \eqref{eq:str}), which takes account of the different levels:
\begin{equation}\begin{split}\label{eq:modbil}
\tSTr \big((X_\ind{L},X_\ind{R})(Y_\ind{L},Y_\ind{R})\big) & = k_\ind{L} \str(X_\ind{L}Y_\ind{L}) + k_\ind{R} \str(X_\ind{R}Y_\ind{R})
\\ & = \STr\big((X_\ind{L},X_\ind{R}) (k_\ind{L} P_\ind{L} + k_\ind{R} P_\ind{R}) (Y_\ind{L},Y_\ind{R}) \big) ~.
\end{split}\end{equation}
We denote the transposes with respect to this new ad-invariant bilinear form as
\begin{equation}
\tSTr\big(u\, \mathcal{O}v\big) = \tSTr\big((\mathcal{O}^T u )\, v\big) ~,
\end{equation}
for linear operators $\mathcal{O}$ on $\alg{g}\oplus\alg{g}$.
Note that we have
\begin{equation}
\mathcal{O}^T = (k_\ind{L}^{-1} P_\ind{L} + k_\ind{R}^{-1} P_\ind{R}) \mathcal{O}^t (k_\ind{L} P_\ind{L} + k_\ind{R} P_\ind{R}) ~.
\end{equation}
Written using the bilinear form~\eqref{eq:modbil}, our ansatz
\eqref{act1}
for the actions of the bi-$\lambda$ models is
\begin{equation}\label{eq:ansatz}
\Act(g,A) = \Act_{\frac{\grp{G}\times\grp{G}}{\grp{G}\times\grp{G}}}(g,A;\tSTr) - \int \extder^2x \, \tSTr(A_+ (\Qc -1) A_-) ~,
\end{equation}
where $g = (g_\ind{L},g_\ind{R}) \in \grp{G}\times\grp{G}$, $A_\pm = (A_{\ind{L}\pm},A_{\ind{R}\pm}) \in \alg{g} \oplus\alg{g}$ and $\Qc$ is a constant linear operator on $\alg{g} \oplus \alg{g}$.
$\Act_{\frac{\grp{G}\times\grp{G}}{\grp{G}\times\grp{G}}}(g,A;\tSTr)$ denotes the action of the $\frac{\grp{G}\times\grp{G}}{\grp{G}\times\grp{G}}$ gauged WZW model, which takes the form~\eqref{gWZW}, with the bilinear form~\eqref{eq:modbil}.
Starting from the action~\eqref{eq:ansatz}, the equations of motion for the gauge fields take the simple form
\begin{equation} \label{eomA} \begin{gathered}
g^{-1}\partial_- g + g^{-1} A_- g = \Qc A_- ~,
\qquad
-\partial_+g g^{-1} + g A_+ g^{-1} = \Qc^T A_+ ~,
\end{gathered}\end{equation}
while the equation of motion for $g=(g_\ind{L}, g_\ind{R}) \in \grp{G} \times \grp{G}$ is
\begin{equation}\begin{gathered}\label{eomg}
\partial_+(g^{-1}\partial_- g + g^{-1} A_- g) - \partial_- A_+ + [A_+,g^{-1} \partial_- g + g^{-1} A_- g] = 0 ~,
\\
\iff
\\
\partial_+ A_- - \partial_-(-\partial_+g g^{-1} + g A_+ g^{-1}) + [-\partial_+g g^{-1} + g A_+ g^{-1}, A_-] = 0 ~.
\end{gathered}\end{equation}

If we integrate out the auxiliary field $A_\pm$ in the action~\eqref{eq:ansatz} we find the sigma model action
\begin{equation}\label{qact}
\begin{aligned}
\Act(g) = - \frac{1}{2} \int \extder^2 x \, \tSTr \Big(g^{-1} \partial_+ g\frac{\Qc+\Ad_g^{-1} }{\Qc-\Ad_g^{-1} } g^{-1} \partial_- g\Big) + \Act_{\textrm{WZ}}(g;\tSTr) ~.
\end{aligned}
\end{equation}
This action is invariant under the formal $\Integer_2$ transformation
\begin{equation}\label{eq:z2trans}
g\to g^{-1} ~, \qquad k_{\ind{L},\ind{R}} \to - k_{\ind{L},\ind{R}} ~, \qquad \Qc \to \Qc^{-1} ~.
\end{equation}

\paragraph{Integrability and solutions for $\Qc$.}

Since we are interested in constructing models on the $\Integer_4$ permutation supercoset~\eqref{eq:psc}, we require that the action~\eqref{eq:ansatz} is invariant under the $\grp{G}_0$ gauge symmetry
\begin{equation}\begin{split}\label{eq:gauge}
(g_\ind{L}, \, g_\ind{R}) & \to (\go^{-1} g_\ind{L} \go , \, \go^{-1} g_\ind{R} \go) ~, \qquad \go(x) \in \grp{G}_0 ~,
\\
(A_{\ind{L}\pm}, \, A_{\ind{R}\pm}) & \to (\go^{-1} A_{\ind{L}\pm} \go + \go^{-1} \partial_\pm \go , \, \go^{-1} A_{\ind{R}\pm} \go + \go^{-1} \partial_\pm \go) ~.
\end{split}\end{equation}
The most general $\Qc$ built from $P_{0,1,2,3}$ and $W$ consistent with gauge invariance is
\begin{equation}\label{Qparam}\begin{gathered}
\Qc = 1 + (k_\ind{L}^{-1} P_\ind{L} + k_\ind{R}^{-1} P_\ind{R}) \big((\alpha_1 + \beta_1 W) P_1 + \alpha_2 P_2 + (\alpha_3 + \beta_3 W) P_3\big) ~,
\\
\Qc^T = 1 + (k_\ind{L}^{-1} P_\ind{L} + k_\ind{R}^{-1} P_\ind{R}) \big( (\alpha_3 + \beta_1 W) P_1 + \alpha_2 P_2 + (\alpha_1 + \beta_3 W) P_3 \big) ~.
\end{gathered}\end{equation}

Substituting the equations of motion for $A_\pm$~\eqref{eomA} into the equations of motion for $g$~\eqref{eomg} we find
\begin{equation}\begin{gathered}
\partial_+(\Qc A_-) - \partial_- A_+ + [A_+,\Qc A_-] = 0 ~,
\\
\partial_+ A_- - \partial_+(\Qc^T A_+) + [\Qc^T A_+ , A_-] = 0 ~.
\end{gathered}
\end{equation}
Making the ansatz
$J_\pm = \mathcal{O}_\pm A_\pm$ where $\mathcal{O}_\pm$ are constant invertible linear operators, we would now like to find for which parameters $\alpha_j$, $\beta_j$ in \eqref{Qparam}
these equations are equivalent to the zero-curvature ones \eqref{eq:eomundef}.
The resulting models can be understood as bi-$\lambda$ models with both copies of $\alg{g}$ being $\lambda$ deformed with different strengths.

The models we are constructing will depend on three parameters.
In the following discussion we use two different sets of parameters $\{k_\ind{L}, k_\ind{R}, \gamma\}$ and $\{k, \lambda, \chi\}$, related to each other as
\begin{equation}\label{eq:klamchi}
k_\ind{L} = k\frac{1-\lambda}{1+\lambda}\frac{\chi+\lambda}{1-\chi\lambda} ~,
\qquad
k_\ind{R} = k\frac{1-\lambda}{1+\lambda}\frac{\chi^{-1}+\lambda}{1-\chi^{-1}\lambda} ~,
\qquad
\gamma = k\frac{1-\lambda}{1+\lambda} ~,
\end{equation}
and
\begin{equation}\begin{split}\label{eq:klkrgam}
\lambda & = \frac{\gamma (k_\ind{L} + k_\ind{R}) - \sqrt{(\gamma^2 + k_\ind{L}^2) (\gamma^2 + k_\ind{R}^2)}}{\gamma^2 - k_\ind{L} k_\ind{R}} ~, \qquad
\\
\chi & = \frac{\gamma (k_\ind{L} - k_\ind{R}) + \sqrt{(\gamma^2 + k_\ind{L}^2) (\gamma^2 + k_\ind{R}^2)}}{\gamma^2 + k_\ind{L} k_\ind{R}} ~, \qquad
\\
k & = \frac{k_\ind{L} k_\ind{R}-\gamma^2 + \sqrt{(\gamma^2 + k_\ind{L}^2) (\gamma^2 + k_\ind{R}^2)}}{k_\ind{L} + k_\ind{R}} ~.
\end{split}\end{equation}
Requiring $k_\ind{L}$, $k_\ind{R}$ and $k$ to be positive leads to the following two regimes
\begin{equation}\begin{aligned}
|\lambda| < 1 ~, \qquad && |\lambda| < \chi < |\lambda|^{-1} ~, \qquad \gamma > 0 ~,
\\
|\lambda| > 1 ~, \qquad && |\lambda|^{-1} < \chi < |\lambda| ~, \qquad \gamma < 0 ~.
\end{aligned}\end{equation}
The coupling $k$ only appears as an overall coefficient rescaling the action, so drops out of the equations of motion.
The classically integrable choices for the linear operator $\Qc$, i.e., the values of the coefficients in \eqref{Qparam}, and the operators $\mathcal{O}_\pm$ defining the Lax connection will then be determined in terms of the remaining couplings $\lambda$ and $\chi$.

We find the following solution
\unskip\footnote{We have checked that \eqref{GSsol} is the unique solution in the GS case perturbatively around $\chi = 1$.
This solution can be found assuming the ansatz~\eqref{Qparam} for $\Qc$, along with a similar one for $\mathcal{O}_\pm$ and solving the resulting equations.
In \secref{sec:pl} we show that it also follows from PL dualising the bi-$\eta$ model of \cite{Hoare:2014oua}.}
corresponding to the GS case
\begin{equation} \label{GSsol}\begin{split}
\Qc & = \frac{2\chi}{1+\chi^2}\Big(
P_0 W_\chi^2 + \lambda^{-2} W_\chi^2 P_2
+\lambda^{-1} W_\chi P_1 W_\chi^{-1}
+\lambda W_\chi^{-1} P_3 W_\chi \Big) ~,
\\
\Qc^T & = \frac{2\chi}{1+\chi^2}\Big(
P_0 W_\chi^2 + \lambda^{-2} W_\chi^2 P_2
+\lambda W_\chi^{-1} P_1 W_\chi
+\lambda^{-1} W_\chi P_3 W_\chi^{-1} \Big) ~,
\end{split}\end{equation}
with
\begin{equation}\begin{split}
\mathcal{O}_+ & = \frac{2\chi}{1+\chi^2} \Big(
P_0 W_\chi^2 + \lambda^{-1}P_2
+ \lambda^{\frac12} P_1 W_\chi
+ \lambda^{-\frac12} P_3 W_\chi^{-1}
\Big) ~,
\\
\mathcal{O}_- & = \frac{2\chi}{1+\chi^2} \Big(
P_0 W_\chi^2 + \lambda^{-1}P_2
+ \lambda^{-\frac12} P_1 W_\chi^{-1}
+ \lambda^{\frac12} P_3 W_\chi
\Big) ~.
\end{split}\end{equation}
Here we have defined the operator
\begin{equation}
W_\chi := \frac{1+\chi + (1-\chi) W}{2\sqrt{\chi}} =
\frac{1}{\sqrt{\chi}}P_\ind{L} + \sqrt{\chi}P_\ind{R} ~, \qquad W_{\chi}^{-1} = W_{\chi^{-1}} ~,
\qquad W_\chi^t = W_\chi^T = W_\chi ~.
\end{equation}
We note the relation
\begin{equation}\label{eq:relation}
\Qc(\lambda,\chi)^{-1} = \Qc(\lambda^{-1},\chi^{-1}) ~,
\end{equation}
meaning, in combination with \eqref{eq:z2trans}, that the resulting sigma model action is invariant under the following $\Integer_2$ transformation acting on fields and parameters
\begin{equation}\label{eq:z21}
g \to g^{-1} ~, \qquad k \to - k ~, \qquad \lambda \to \lambda^{-1} ~, \qquad \chi \to \chi^{-1} ~,
\end{equation}
or equivalently, in terms of the parameters $\{k_\ind{L}, k_\ind{R}, \gamma\}$~\eqref{eq:klkrgam},
\begin{equation}\label{eq:z22}
g \to g^{-1} ~, \qquad k_\ind{L} \to - k_\ind{L} ~, \qquad k_\ind{R} \to -k_\ind{R} ~, \qquad \gamma \to \gamma ~.
\end{equation}
In terms of the parameters $\{k_\ind{L}, k_\ind{R}, \gamma\}$ we find that $\Qc$ in \eqref{GSsol} has the form~\eqref{Qparam} as required, with the parameters $\alpha_{1,2,3}$ and $\beta_{1,3}$ given by
\begin{equation}\begin{aligned}
& \alpha_1 = \frac{\gamma^2(k_\ind{L}+k_\ind{R}) + 2 \gamma k_\ind{L} k_\ind{R}}{k_\ind{L} k_\ind{R} - \gamma^2 } ~,
\qquad
\alpha_3 = \frac{\gamma^2(k_\ind{L}+k_\ind{R})-2\gamma k_\ind{L} k_\ind{R}}{k_\ind{L} k_\ind{R} - \gamma^2} ~, \label{sol}\\
&\alpha_2=
\frac{4 \gamma k_\ind{L} k_\ind{R}\big(\gamma(k_\ind{L}+k_\ind{R}) +\sqrt{(\gamma^2 + k_\ind{L}^2) (\gamma^2 + k_\ind{R}^2)}\big)}{(k_\ind{L} k_\ind{R} - \gamma^2)^2} ~, \qquad
\beta_1 = \beta_3 = \frac{\gamma^2(k_\ind{L}-k_\ind{R})}{k_\ind{L} k_\ind{R} - \gamma^2} ~.
\end{aligned}\end{equation}

In the PS case we find the following solution
\unskip\footnote{Again we have checked that this is the unique solution perturbatively around $\chi = 1$.}
for $\Qc$
\begin{equation}\begin{split}\label{eq:PSsol}
\Qc & = \frac{2\chi}{1+\chi^2}\Big(
P_0 W_\chi^2 + \lambda^{-2} W_\chi^2 P_2
+\lambda^{-1} W_\chi P_1 W_\chi^3
+\lambda^{-3} W_\chi^3 P_3 W_\chi \Big) ~,
\\
\Qc^T & = \frac{2\chi}{1+\chi^2}\Big(
P_0 W_\chi^2 + \lambda^{-2} W_\chi^2 P_2
+\lambda^{-3} W_\chi^{3} P_1 W_\chi
+\lambda^{-1} W_\chi P_3 W_\chi^{3} \Big) ~,
\end{split}\end{equation}
with
\begin{equation}\begin{split}
\mathcal{O}_+ & = \frac{2\chi}{1+\chi^2} \Big(
P_0 W_\chi^2 + \lambda^{-1}P_2
+ \lambda^{-\frac32}P_1 W_\chi
+ \lambda^{-\frac12}P_3 W_\chi^3\Big) ~,
\\
\mathcal{O}_- & = \frac{2\chi}{1+\chi^2} \Big(
P_0 W_\chi^2 + \lambda^{-1}P_2
+ \lambda^{-\frac12}P_1 W_\chi^3
+ \lambda^{-\frac32}P_3 W_\chi\Big) ~.
\end{split}\end{equation}
Again this solution satisfies the relation~\eqref{eq:relation} implying that the resulting sigma model action is invariant under the $\Integer_2$ transformation~\eqref{eq:z21}, or equivalently~\eqref{eq:z22}.
Furthermore, again writing in terms of the parameters $\{k_\ind{L},k_\ind{R},\gamma\}$ we find that it takes the form~\eqref{Qparam} as required.

\paragraph{Bosonic truncation.}
The bosonic truncations of the GS and PS cases both give the same bi-$\lambda$ model on the $\Integer_2$ permutation coset
\begin{equation}
\frac{\grp{G} \times \grp{G}}{\grp{G}} ~,
\end{equation}
where $\grp{G}$ is now an ordinary Lie group.
This model was introduced in \cite{Sfetsos:2014cea} and shown to be classically integrable in \cite{Sfetsos:2017sep} -- its Lax connection follows from the bosonic truncation of \eqref{LaxGS} or \eqref{LaxPS}.
Explicitly, the action is given by
\begin{equation}\label{bmr}
\Act(g,A) = \Act_{\frac{\grp{G}\times\grp{G}}{\grp{G}\times\grp{G}}}(g,A;\tTr)
- \h\int \extder^2x \, \Tr(A_+ P_2 A_-) ~,
\end{equation}
where the coupling $\h$ is given by
\begin{equation}\label{eq:couph}
\h =
\frac{4 \gamma k_\ind{L} k_\ind{R}\big(\gamma(k_\ind{L}+k_\ind{R}) +\sqrt{(\gamma^2 + k_\ind{L}^2) (\gamma^2 + k_\ind{R}^2)}\big)}{(k_\ind{L} k_\ind{R} - \gamma^2)^2} ~,
\end{equation}
$g = (g_\ind{L},g_\ind{R}) \in \grp{G} \times \grp{G}$, $A_\pm = (A_{\ind{L}\pm},A_{\ind{R}\pm}) \in \alg{g} \oplus\alg{g}$ and $P_2(X_\ind{L},X_\ind{R}) = \frac12(X_\ind{L}-X_\ind{R},X_\ind{R}-X_\ind{L})$.
The bilinear forms $\Tr$ and $\tTr$ are defined analogously to the superalgebra counterparts
$\STr$ and $\tSTr$ in \eqref{eq:str}, \eqref{eq:modbil},
\begin{equation}\begin{split}
&\Tr\big( (X_\ind{L},X_\ind{R})(Y_\ind{L},Y_\ind{R})\big) = \tr (X_\ind{L} Y_\ind{L}) + \tr(X_\ind{R} Y_\ind{R}) ~, \\
&\tTr\big( (X_\ind{L},X_\ind{R})(Y_\ind{L},Y_\ind{R})\big) = k_\ind{L} \tr (X_\ind{L} Y_\ind{L}) + k_\ind{R} \tr(X_\ind{R} Y_\ind{R}) ~,
\end{split}\end{equation}
in terms an ad-invariant non-degenerate bilinear form $\tr$ on $\alg{g}$, and $k_\ind{L}$ and $k_\ind{R}$.

The action~\eqref{bmr} has a $\grp{G}$ gauge symmetry acting as in eq.~\eqref{eq:gauge} with $\go \in \grp{G}$, and is also invariant under the $\Integer_2$ transformation~\eqref{eq:z22}, under which the coupling $\h$ transforms as
\begin{equation}
\h \to \frac{2 \h k_\ind{L}k_\ind{R}}{2k_\ind{L}k_\ind{R}+ \h (k_\ind{L}+k_\ind{R})} ~.
\label{Z2}
\end{equation}
We will investigate this bosonic model further in \secref{sec:rg} when we discuss its two-loop RG flow.

\subsection{Limits}

\paragraph{Symmetric $\lambda$ model.}
The symmetric limit $\chi \to 1$, or equivalently $k_\ind{L} \to k_\ind{R} = k$,
corresponds to the standard $\l$ deformation of the $\Integer_4$ supercoset
(with deformation parameter
$\lambda = \tfrac{k-\g}{k+\g}$ and WZ level $k$).
Since the two levels are equal in this limit, the left and right symmetries are deformed in the same way and with the same deformation parameter.
The resulting action takes the form
\begin{equation}\begin{split}\label{331}
\Act(g,A) & = k \Act_{\frac{\grp{G}\times\grp{G}}{\grp{G}\times\grp{G}}}(g,A;\STr) - k \int \extder^2x \STr\big( A_+ (\Qc -1) A_-\big) ~.
\end{split}\end{equation}

In this limit the formulae \eqref{GSsol}, \eqref{eq:PSsol} above for the operator $\Qc$ reproduce the known ones for $\l$ deformed $\Integer_4$ cosets in the GS and PS formalisms.
We obtain in the GS case \cite{Hollowood:2014qma}
\begin{equation}\begin{split}\label{332}
\Qc & = P_0 + \lambda^{-2} P_2 + \lambda^{-1} P_1 + \lambda P_3 ~,
\end{split}\end{equation}
and
in the PS case
\cite{Hoare:2021dix}
\begin{equation}\label{333}
\Qc = P_0 + \lambda^{-2} P_2 + \lambda^{-1} P_1 + \lambda^{-3} P_3 ~.
\end{equation}

\paragraph{NATD-$\lambda$ model.}
A second interesting limit is to take $\chi \to \lambda$, or equivalently $k_\ind{R} \to \infty$, while zooming in on $g_\ind{R} = 1$
according to
\begin{equation}\label{eq:grzoom}
g_\ind{R} = \exp\big(\frac{v_\ind{R}}{k_\ind{R}}\big) ~, \qquad v_\ind{R} \in \alg{g} ~,
\end{equation}
which, as we will argue, gives the non-abelian T-dual (NATD) of the single-sided $\l$-deformation.
Under the $\grp{G}_0$ gauge symmetry~\eqref{eq:gauge} $v_\ind{R}$ transforms as
\begin{equation}
v_\ind{R} \to g_\ind{0}^{-1} v_\ind{R} g_\ind{0} ~.
\end{equation}
After taking the limit, the resulting gauge-invariant action is
\begin{equation}\begin{aligned}
&\Act(g_{\ind{L}},v_\ind{R},A_{\ind{L},\ind{R}})
\\ & \qquad = k_\ind{L} \Act_{\grp{G}/\grp{G}}(g_\ind{L},A_\ind{L};\str)
+ \int \extder^2x \, \str\big(v_\ind{R} F_{+-}(A_\ind{R}) \big)
- \int \extder^2x \, \STr\big(A_+ \hQc A_-\big) ~,
\label{actrnatd}
\end{aligned}\end{equation}
where $F_{+-}(A_\ind{R}) = \partial_+ A_{\ind{R}-} - \partial_- A_{\ind{R}+} + [A_{\ind{R}+},A_{\ind{R}-}]$ and
$\hQc = \lim_{k_\ind{R}\to\infty} \big((k_\ind{L} P_\ind{L} + k_\ind{R} P_\ind{R}) (\Qc-1)\big)$.
In the limit $k_\ind{R} \to \infty$ we have~\eqref{eq:klkrgam}
\begin{equation}
\gamma = k_\ind{L}\frac{1-\lambda^2}{2\lambda} ~,
\end{equation}
and in terms of the parameters $k_\ind{L}$ and $\lambda$ we find in the GS case
\begin{equation}\begin{split}\label{qhatgs}
\hQc = k_\ind{L}\frac{1-\lambda^2}{\lambda}\Big( 2 \lambda^{-1} P_2 + P_1 - P_3 + \frac{1-\lambda^2}{2\lambda} P_\ind{R} (P_1+P_3)\Big)~,
\end{split}\end{equation}
and in the PS case
\begin{equation}\begin{split}\label{qhatps}
\hQc = k_\ind{L} \frac{1-\lambda^2}{\lambda} \Big( 2 \lambda^{-1} P_2 - \lambda^{-2}(P_1 - P_3) + \frac{3+\lambda^2}{2\lambda} P_\ind{R} (P_1+P_3) + \frac{1+\lambda^2}{\lambda^3} P_\ind{L} (P_1+P_3)\Big)~.
\end{split}\end{equation}

From the action~\eqref{actrnatd} we can obtain two integrable sigma models.
The first is given by integrating out the auxiliary fields $A_{\ind{L}\pm}$ and $A_{\ind{R}\pm}$.
This is the same procedure that gives the sigma model~\eqref{qact} from~\eqref{eq:ansatz}, hence it follows that the resulting action is a limit of~\eqref{qact}.
Moreover, this model is the non-abelian T-dual, with respect to $\grp{G}_\ind{R}$, of the second sigma model, which is obtained by instead integrating out the auxiliary field $A_{\ind{L}\pm}$ and the Lagrange multiplier $v_\ind{R}$, i.e., imposing $A_{\ind{R}\pm}=\tilde g^{-1}\partial_\pm \tilde g$ where $\tilde g\in \grp{G}$.
Under the $\grp{G}_0$ gauge symmetry~\eqref{eq:gauge} the field $\tilde g$ transforms as
\begin{equation}\label{eq:gsymtg}
\tilde g \to \tilde g g_\ind{0} ~.
\end{equation}

To interpret this second model let us just integrate out the Lagrange multiplier $v_\ind{R}$ to give
\begin{equation}\begin{split} \label{single}
\Act(g_{\ind{L}},A_{\ind{L}},\tilde g) & = k_\ind{L}
\Act_{\grp{G}/\grp{G}}(g_\ind{L},A_\ind{L};\str)
- \int \extder^2x \, \STr\big( (A_{\ind{L}+},\tilde g^{-1}\partial_+ \tilde g) \widehat{\mathcal{Q}} (A_{\ind{L}-},\tilde g^{-1} \partial_- \tilde g)\big) ~.
\end{split}\end{equation}
As discussed above, the bosonic truncations of the GS and PS cases both give the same model.
Since in the truncated model $\grp{G}_0 = \grp{G}$, we can use the gauge symmetry \eqref{eq:gsymtg} to fix $\tilde g = 1$, while $\widehat{\mathcal{Q}} = - 2 k_\ind{L}(1-\lambda^{-2})P_2$.
Therefore, the action of the bosonic truncation is
\begin{equation}\begin{split}\label{eq:bostrunc}
\Act(g_{\ind{L}},A_{\ind{L}}) & =
k_\ind{L}
\Act_{\grp{G}/\grp{G}}(g_\ind{L},A_\ind{L};\tr)
+ k_\ind{L}(1-\lambda^{-2})\int \extder^2x \, \tr( A_{\ind{L}+}A_{\ind{L}-}) ~,
\end{split}\end{equation}
which we recognise as the well-known $\lambda$ deformation of the PCM \cite{Sfetsos:2013wia} with level $k_\ind{L}$.
It follows that the action~\eqref{single} can be interpreted as the single-sided $\lambda$ deformation of the $\Integer_4$ supercoset sigma model \eqref{eq:z4coset}.

In contrast with the symmetric $\lambda$~deformation~\eqref{331}--\eqref{333}, which has no global symmetries, the action~\eqref{single} has a global $\grp{G}$ symmetry acting as
\unskip\footnote{After using the gauge symmetry to fix $\tilde g = 1$ in the bosonic truncation, this global symmetry acts as $g_\ind{L} \to \ell g_\ind{L} \ell^{-1}$, $A_{\ind{L}\pm} \to \ell A_{\ind{L}\pm} \ell^{-1}$ in the action~\eqref{eq:bostrunc}.}
\begin{equation}
\tilde g \to \ell \tilde g ~, \qquad \ell \in \grp{G} ~.
\end{equation}
Therefore, for the $\AdS_3$ $\Integer_4$ permutation supercosets in \footref{foot1}, the corresponding supergravity backgrounds are expected to preserve 8 supersymmetries.
As discussed in the \hyperref[sec:int]{Introduction}, it is natural to expect that as a result they will have ``nicer'' properties than the generic deformations that preserve no supersymmetries~\cite{Hoare:2022asa}.
As we will discuss in \secref{sec:pl}, it is also possible to $\eta$ deform the left copy of $\grp{G}$, instead of $\lambda$ deforming, to give an $\eta$-$\lambda$ model generalising that of \cite{Sfetsos:2015nya}.

\paragraph{Bi-NATD model.}
Starting from either of the above limits it is then possible to take a further limit to give the bi-NATD model.
In the symmetric $\lambda$ model~\eqref{331} we take $k\to \infty$ and $\lambda \to 1$, while zooming on $g =1$.
On the other hand, starting from the NATD-$\lambda$ model~\eqref{actrnatd}, we take $k_\ind{L} \to \infty$, while zooming in on $g_\ind{L} = 1$.
The resulting model is
\begin{equation}\begin{split} \label{actnatd}
\Act(v,A) = & \int \extder^2x \, \STr\big(
v F_{+-}(A)
- 2\gamma A_+ P_- A_-\big) ~,
\\
F_{+-}(A) & = \partial_+ A_{-} - \partial_- A_{+} + [A_{+},A_{-}] ~,
\end{split}\end{equation}
where $v = (v_\ind{L},v_\ind{R}) \in \alg{g} \oplus \alg{g}$, $A_\pm = (A_{\ind{L}\pm},A_{\ind{R}\pm}) \in \alg{g} \oplus\alg{g}$, and we recall that $P_- = P_1 + 2 P_2 - P_3$ for the GS case and $P_- = P_1 + 2P_2 + 3P_3$ for the PS case.
The bi-NATD model is given by integrating out the auxiliary field $A_\pm$ in the action~\eqref{actnatd}, while instead integrating out the Lagrange multiplier $v$ gives $A_\pm = g^{-1} \partial_\pm g$ and we recover the $\Integer_4$ supercoset sigma models~\eqref{eq:z4coset} if we set $\gamma = \frac{T}{4}$.

\paragraph{$\lambda \to 0$ limit.}
The final limit we discuss is $\lambda \to 0$, which is equivalent to $\gamma \to \sqrt{k_\ind{L}k_\ind{R}}$~\eqref{eq:klamchi} or $\h \to \infty$~\eqref{eq:couph}.
In this limit we are left with the parameters $\{\chi, k\}$ or $\{k_\ind{L},k_\ind{R}\}$ related as
\begin{equation}
\chi = \sqrt{\frac{k_\ind{L}}{k_\ind{R}}}~, \qquad k = \sqrt{k_\ind{L} k_\ind{R}}~.
\end{equation}

We can see the importance of this limit by taking it in the bosonic truncation~\eqref{bmr}.
Doing so, the coefficient of the final term diverges, hence the equation of motion for $P_2 A_\pm$ simply becomes $P_2 A_\pm = 0$, which we can solve by setting $A_{\ind{L}\pm} = A_{\ind{R}\pm} = B_\pm$.
Substituting this into~\eqref{bmr} gives the action of the $(\grp{G}_{k_\ind{L}} \times \grp{G}_{k_\ind{R}})/\grp{G}_{k_\ind{L}+k_\ind{R}}$ gauged WZW model
\begin{equation}\label{eq:glgrb}
\Act(g_\ind{L},g_\ind{R},B) = k_\ind{L} \Act_{\grp{G}/\grp{G}}(g_\ind{L},B;\str) + k_\ind{R} \Act_{\grp{G}/\grp{G}}(g_\ind{R},B;\str)~,
\end{equation}
which, as we discuss in \secref{sec:rg}, is a fixed point of the RG flow.
Further taking $k_\ind{R} \to\infty$ while zooming in on $g_\ind{R} = 1$ according to~\eqref{eq:grzoom}, the second gauged WZW action in~\eqref{eq:glgrb} becomes a flatness constraint on the gauge field $B_\pm$ and we can use the $\grp{G}$ gauge symmetry to fix $B_\pm=0$.
Therefore, we find the WZW action with level $k_\ind{L}$, which we can also recover as the $\lambda\to 0$ limit of~\eqref{eq:bostrunc}, i.e., the two limits commute.

Taking the $\lambda \to 0$ limit in the $\Integer_4$ supercoset bi-$\lambda$ model~\eqref{eq:ansatz}, we similarly observe that the operators $\mathcal Q$ and $\mathcal Q^T$ diverge, due to the negative powers of $\lambda$ in their expressions~\eqref{GSsol}, \eqref{eq:PSsol}.
It again follows that the equations of motion~\eqref{eomA} will set certain components of the gauge fields to vanish.
In the GS case, however, due to the positive powers of $\lambda$ in~\eqref{GSsol} it is not clear if the $\lambda \to 0$ limit will be well-defined.
The PS case~\eqref{eq:PSsol} behaves more straightforwardly like the bosonic truncation and we find
\begin{equation} \label{eq:psconstraints}
P_2 A_\pm=0~, \qquad P_1 A_\pm =0~, \qquad P_3 A_\pm =0 ~.
\end{equation}
We thus obtain the action of the $(\grp{G}_{k_\ind{L}} \times \grp{G}_{k_\ind{R}})/(\grp{G}_0)_{k_\ind{L}+k_\ind{R}}$ gauged WZW model.
It would then be interesting to investigate the further limit $k_\ind{R} \to \infty$ and whether it agrees with the $\lambda \to 0$ limit of~\eqref{actrnatd} in the PS case \eqref{qhatps}
(and also in the GS case~\eqref{qhatgs} assuming the $\lambda \to 0$ limit exists).

\subsection{\texorpdfstring{$\kappa$}{kappa}-symmetry in the GS case}

$\Integer_4$ permutation supercoset sigma models are of interest in the context of string theory \cite{Babichenko:2009dk}, e.g., those mentioned in \footref{foot1}.
In this context, the GS string sigma model should be invariant under a local fermionic $\kappa$-symmetry to ensure that the theory describes the correct number of fermionic degrees of freedom~\cite{Witten:1985nt,Grisaru:1985fv,Howe:1983sra}.
The $\kappa$-symmetry of the model ensures that the deformed background satisfies a set of generalised supergravity equations of motion~\cite{Wulff:2016tju}, which should also imply scale invariance~\cite{Arutyunov:2015mqj}.
Moreover, due to the lack of isometries we expect the bi-$\lambda$ deformations to be Weyl invariant, similarly to the symmetric $\lambda$ deformation limit~\cite{Borsato:2016ose}.

Here we show that the bi-$\lambda$ deformation in the GS case \eqref{eq:ansatz}, \eqref{GSsol} has a local fermionic $\kappa$-symmetry.
\unskip\footnote{Note that this is not the full $\kappa$-symmetry of the GS string sigma model on $\AdS_3 \times \Sp^3 \times \To^4$ and $\AdS_3 \times \Sp^3 \times \Sp^3 \times \Sp^1$ since the $\Integer_4$ permutation supercoset sigma model only describes a sector of the theory. Nevertheless, the undeformed supercoset model has a $\kappa$-symmetry in the GS case and we expect any deformation consistent with string theory to preserve this.}
To do so, we follow the construction given in \cite{Hollowood:2014qma} for the symmetric $\lambda$ deformation.
We start by considering a local $\grp{G}_\ind{L} \times \grp{G}_\ind{R}$ symmetry acting infinitesimally on the fields as
\begin{equation}\label{eq:infin}
\delta g = \varepsilon_\ind{L} g - g \varepsilon_\ind{R} ~, \qquad \delta A_+ = \com{\varepsilon_\ind{L}}{A_+} - \partial_+ \varepsilon_\ind{L}~, \qquad \delta A_- = \com{\varepsilon_\ind{R}}{A_-} - \partial_- \varepsilon_\ind{R} ~, \qquad \varepsilon_\ind{L}, \varepsilon_\ind{R}\in\alg{g}~.
\end{equation}
The action \eqref{eq:ansatz} transforms as
\begin{equation}
\delta \Act = \int \extder^2 x \, \tSTr \big((\Qc^T \varepsilon_\ind{L}-\varepsilon_\ind{R}) \partial_+ A_- - (\varepsilon_\ind{L}-\Qc \varepsilon_\ind{R}) \partial_- A_+ + \varepsilon_\ind{L} \com{A_+}{\Qc A_-} - \varepsilon_\ind{R} \com{\Qc^T A_+}{A_-} \big)~.
\end{equation}
Requiring that the derivative terms vanish we have
\begin{equation}\label{eq:alphabeta}
\varepsilon_\ind{R} = \Qc^T \varepsilon_\ind{L}~, \qquad \varepsilon_\ind{L} = \Qc \varepsilon_\ind{R}~.
\end{equation}
In the GS case
\begin{equation}
\Qc \Qc^T = \Qc^T \Qc = 1 + (\lambda^{-4}-1) P_2 + (1+\lambda^{-2})^2 \frac{1-\chi^2}{1+\chi^2} W P_2 ~,
\end{equation}
hence the compatibility of the two equations~\eqref{eq:alphabeta} implies that $P_2 \varepsilon_\ind{L}=P_2\varepsilon_\ind{R} = 0$.

We are left with the variations
\begin{equation}\begin{aligned}
\delta g & = \varepsilon g - g \Qc^T \varepsilon ~, \qquad \delta A_+ = \com{\varepsilon}{A_+} - \partial_+ \varepsilon \qquad \delta A_- = \com{\Qc^T\varepsilon}{A_-} - \partial_- \Qc^T \varepsilon ~,
\\
\delta \Act & = \int \extder^2 x \tSTr \big(\varepsilon (\com{A_+}{\Qc A_-} - \Qc \com{\Qc^T A_+}{A_-} )\big) ~, \qquad \varepsilon \in\alg{g} ~, \quad P_2 \varepsilon = 0~.
\end{aligned}\end{equation}
When $\varepsilon=P_0 \varepsilon$ the variation of the action vanishes demonstrating the invariance of the action under the $\grp{G}_0$ gauge symmetry~\eqref{eq:gauge}.
\unskip\footnote{In the PS case, we also have that $P_1 \varepsilon = P_3 \varepsilon = 0$ follow from the compatibility of the two equations~\eqref{eq:alphabeta}.
Therefore, the only local symmetry with infinitesimal action~\eqref{eq:infin} is the $\grp{G}_0$ gauge symmetry~\eqref{eq:gauge}.}

For $\kappa$-symmetry we instead require that the variation vanishes for some Grassmann-odd quantity $\varepsilon = (P_1+P_3) \varepsilon = \varepsilon^{(1)} + \varepsilon^{(3)}$.
Using the explicit form of the operators $\mathcal Q$ and $\mathcal Q^T$ in the GS case~\eqref{GSsol}, we find that the variation of the action is proportional to
\begin{equation}\begin{aligned}\label{eq:fermvar}
\delta \Act & \sim \int \extder^2 x \STr \big(\varepsilon( W_\chi \com{J_+^{(1)}}{J_-^{(2)}} - \lambda^{-1} W_\chi^{-1} \com{J_+^{(2)}}{J_-^{(3)}})\big)
\\
&\sim \int \extder^2 x \STr \big(\epsilon( \com{J_+^{(1)}}{J_-^{(2)}} - \com{J_+^{(2)}}{J_-^{(3)}})\big) ~,
\end{aligned}\end{equation}
where $\epsilon = \frac{2\chi}{1+\chi^2} (P_1 W_\chi + \lambda^{-1} P_3 W_\chi^{-1})\varepsilon$.
On the other hand, considering the sigma model \eqref{eq:ansatz} on a curved background, we find that the non-vanishing components of the energy-momentum tensor are proportional to
\begin{equation}
T_{\pm \pm} \sim \tSTr[A_\pm (\Qc \Qc^T - 1) A_\pm] \sim \STr[ J_\pm^{(2)} J_\pm^{(2)}]~,
\end{equation}
in the GS case.
Therefore, after writing in terms of the auxiliary currents $J_\pm$ and redefining $\varepsilon$, we find that both the variation of the action and the energy-momentum tensor are independent of $\lambda$ and $\chi$ (up to constants of proportionality).
It follows that the variation of the action~\eqref{eq:fermvar} vanishes on the Virasoro constraints $T_{\pm\pm} = 0$ if either $\epsilon^{(1)} = \com{J_-^{(2)}}{\kappa^{(1)}}_+$
with $\kappa^{(1)} =P_1 \kappa^{(1)}$ or
$\epsilon^{(3)} = \com{J_+^{(2)}}{\kappa^{(3)}}_+ $ with $\kappa^{(3)} =P_3 \kappa^{(3)}$ \cite{Metsaev:1998it,Arutyunov:2009ga}.
\unskip\footnote{Here $\com{\cdot}{\cdot}_+$ denotes the anticommutator.}
In the first case we have $\epsilon^{(3)} = 0$, which implies $\varepsilon = W_\chi \epsilon^{(1)}$ and the infinitesimal transformations of the fields are given by
\begin{equation} \begin{gathered}\label{eq:ks1}
\delta g = W_\chi \epsilon^{(1)} g - g \lambda W_\chi^{-1} \epsilon^{(1)} ~,
\qquad \epsilon^{(1)} = \com{J_-^{(2)}}{\kappa^{(1)}}_+ ~,
\qquad \kappa^{(1)} =P_1 \kappa^{(1)} ~,
\\ \delta A_+ = W_\chi (\com{ \epsilon^{(1)}}{A_+} - \partial_+ \epsilon^{(1)})~, \qquad \delta A_- = \lambda W_\chi^{-1} (\com{ \epsilon^{(1)}}{A_-} - \partial_- \epsilon^{(1)})~.
\end{gathered}
\end{equation}
In the second case, $\epsilon^{(1)} = 0$, hence $\varepsilon = \lambda W_\chi^{-1} \epsilon^{(3)}$ and the infinitesimal transformations of the fields are given by
\begin{equation} \begin{gathered}\label{eq:ks2}
\delta g = \lambda W_\chi^{-1} \epsilon^{(3)} g - g W_\chi \epsilon^{(3)} ~,
\qquad \epsilon^{(3)} = \com{J_+^{(2)}}{\kappa^{(3)}}_+ ~, \qquad \kappa^{(3)} =P_3 \kappa^{(3)} ~,
\\ \delta A_+ = \lambda W_\chi^{-1} (\com{ \epsilon^{(3)}}{A_+} - \partial_+ \epsilon^{(3)})~,\qquad \delta A_- = W_\chi (\com{ \epsilon^{(3)}}{A_-} - \partial_- \epsilon^{(3)})~.
\end{gathered}
\end{equation}

\section{Poisson-Lie duality}\label{sec:pl}

In this section we show that the bi-$\lambda$ models introduced in \secref{sec2} are the Poisson-Lie duals of the bi-$\eta$ deformed GS and PS models defined in \secref{sec1}, with $R$ solving the split modified classical Yang-Baxter equation,
\unskip\footnote{They are also the Poisson-Lie duals of the bi-$\eta$ deformed models with non-split operator $R$ upon analytic continuation.}
and parameters related by $(T,\eta_\ind{L}, \eta_\ind{R}) = (4\gamma, \tfrac{\gamma}{k_\ind{L}}, \tfrac{\gamma}{k_\ind{R}})$.
This provides an explicit construction of the bi-$\lambda$ models.

Poisson-Lie duality is a generalisation of abelian and non-abelian T-duality to sigma models that do not necessarily have global symmetries, but whose currents $K_\pm \in \alg{f}$ obey the Poisson-Lie symmetric equation of motion
\begin{equation}
\label{eq:PLsym}
\partial_+ K_- + \partial_- K_+ + \com{K_+}{K_-}_{\tilde{\alg{f}}} =0~,
\end{equation}
where $\com{\cdot}{\cdot}_{\tilde{\alg{f}}}$ denotes the Lie bracket on a dual Lie algebra $\tilde{\alg{f}}$.
The presence of the two algebraic structures $\alg{f}$ and $\tilde{\alg{f}}$ makes it possible to construct an $\mathcal E$ model, with degrees of freedom in the Drinfel'd double $\Lie(\grp{D}) = \alg{d}= \alg{f}+\tilde{\alg{f}}$.
Provided there exists an ad-invariant bilinear form on $\alg{d}$ with respect to which the two Lie algebras are isotropic, one can construct two sigma models by integrating out the degrees of freedom associated to the dual Lie algebra: integrating out the degrees of freedom in $\alg{\tilde{f}}$ one gets a sigma model on $\tilde{\grp{F}} \backslash \grp{D} \cong \grp{F}$, while integrating out the degrees of freedom in $\alg{f}$ one obtains a sigma model on $\grp{F} \backslash\grp{D} \cong \tilde{\grp{F}}$.
The two sigma models produced through this procedure are then said to be Poisson-Lie dual to each other.
\unskip\footnote{One can also consider other decompositions $\alg{d}=\alg{f}_1 + \alg{f}_2$, where only $\alg{f}_2$ is an isotropic subalgebra.
Integrating out the associated degrees of freedom generates new Poisson-Lie dual models.} Introducing a gauge field it is also possible to obtain Poisson-Lie dual sigma models on coset spaces.

The $\eta$ deformation of the GS and PS $\mathbb{Z}_4$ cosets are indeed characterised by equations of motion of the form \eqref{eq:PLsym}, see \eqref{eq:eom3}.
Choosing the $\mathcal E$ model and the two subalgebras $\alg{f}$ and $\tilde{\alg{f}}$ appropriately, this will lead to respectively the $\eta$ and $\lambda$ deformations of the GS and PS $\mathbb{Z}_4$ model.
In this section we extend this to the case of the bi-$\eta$ deformations and find their dual bi-$\lambda$ models.
We also construct hybrid deformations, with one copy of the symmetry algebra $\eta$ deformed and the other $\lambda$ deformed.

\subsection{\texorpdfstring{$\mathcal E$}{E} model}

We start by summarising the construction of the $\mathcal E$ model on the Drinfel'd double.
For additional details the reader is referred to the review \cite{Hoare:2021dix}, from which most of the notation is taken.

The action of the $\mathcal E$ model for the group-valued field $l \in \grp{D}$, with gauge field $\dA \in \alg{h} = \Lie(\grp{H})$,
\unskip\footnote{The Lie algebra $\alg{h}$ is required to be isotropic with respect to the invariant bilinear form $\langle \!\langle \cdot, \cdot \rangle \!\rangle $.
For the purposes of recovering the bi$-\eta$ and bi-$\lambda$ deformations, the Lie group $\grp{H}$ will be identified with $\grp{G}_0$ in \eqref{eq:psc}, see eq.~\eqref{eq:defh}.} is
\begin{equation} \begin{aligned}
\label{eq:Emod}
\Act_{\mathcal E} &= \int \extder^2x \, \langle \!\langle l^{-1} \partial_\tau l, l^{-1} \partial_\sigma l \rangle \!\rangle + \frac{1}{6} \int \extder^3 x \, \epsilon^{ijk} \langle \!\langle l^{-1} \partial_i l, \com{l^{-1}\partial_j l}{l^{-1} \partial_k l} \rangle \!\rangle \\
&- 2 \int \extder^2 x \, \langle \!\langle \dA_\tau , l^{-1} \partial_\sigma l \rangle \!\rangle - \int \extder^2 x \, \langle \!\langle (l^{-1} \partial_\sigma l - \dA_\sigma), \mathcal E (l^{-1} \partial_\sigma l - \dA_\sigma)\rangle \!\rangle ~.
\end{aligned}
\end{equation}
The operator $\mathcal E : \alg{d} \rightarrow \alg{d}$ satisfies $\mathcal E^2=1$ and is symmetric with respect to the invariant bilinear form $\langle \!\langle \cdot, \cdot \rangle \!\rangle $ on $\alg{d}$, so that $\ls \mathcal E X, Y \rs = \ls X, \mathcal E Y \rs$ for any two elements $X,Y \in \alg{d}$.
Let us now consider a subalgebra $\alg{b} \subset \alg{d}$ that is isotropic, i.e., $\langle \!\langle X,Y \rangle \!\rangle =0$ for any two elements $X,Y \in \alg{b}$.
We call $\grp{B}$ the associated Lie group.
Sending $l \rightarrow b l$ where $b\in \grp{B}$, and integrating out the degrees of freedom associated to $\alg{b}$,
\unskip\footnote{This is possible as long as $\Ad_l^{-1} \alg{b}$ and $\mathcal E \Ad_l^{-1} \alg{b}$ have trivial intersection.} one gets the following model on $\grp{B}\backslash\grp{D}/\grp{H}$,
\begin{equation} \begin{aligned}
\label{eq:EmodB}
\Act &= \frac{1}{2}\int \extder^2 x \, \langle \!\langle (l^{-1} \partial_+ l - \dA_+), \mathcal E \mathcal P (\mathcal E +1) (l^{-1} \partial_- l - \dA_-) \rangle \!\rangle \\
&\qquad - \frac{1}{2} \int \extder^2 x \, \langle \!\langle (l^{-1} \partial_- l - \dA_-), \mathcal E \mathcal P (\mathcal E -1) (l^{-1} \partial_+ l - \dA_+) \rangle \!\rangle \\
&\qquad + \int \extder^2 x \, \epsilon^{\mu \nu} \langle \!\langle l^{-1} \partial_\mu l, \dA_\nu \rangle \!\rangle + \frac{1}{6} \int \extder^3 x \, \epsilon^{ijk} \langle \!\langle l^{-1} \partial_i l, \com{l^{-1}\partial_j l}{l^{-1} \partial_k l} \rangle \!\rangle ~.
\end{aligned}
\end{equation}
We recall that we use light-cone coordinates $x^\pm = \half (\tau \pm \sigma)$, $\partial_\pm = \partial_\tau \pm \partial_\sigma$ and our convention for the antisymmetric Levi-Civita symbol is $\epsilon^{+-}= - \epsilon^{-+} = -1/2$.
The projector $\mathcal P$ satisfies $\text{im} [\mathcal P] = \mathcal E \Ad_l^{-1} \alg{b}$ ($\text{im}$ denotes the image, not the imaginary part) and $\text{ker} [\mathcal P ]= \Ad_l^{-1} \alg{b}$.
By virtue of the condition $\mathcal E^2=1$ also $\mathcal E \mathcal P (\mathcal E \pm 1)$ are projectors, with
\begin{equation} \label{eq:epe}
\text{im}[\mathcal E \mathcal P (\mathcal E \pm 1)] = \Ad_l^{-1} \alg{b}~, \qquad \text{ker}[\mathcal E \mathcal P (\mathcal E \pm 1)] = \alg{e}_\mp~,
\end{equation}
where $\alg{e}_\mp$ are the eigenspaces of $\mathcal E$ with eigenvalues $\mp 1$.
All this ensures that the action \eqref{eq:EmodB} has a gauge symmetry
\begin{equation}
l \rightarrow b l h~, \qquad \dA_\pm \rightarrow h^{-1} \dA_\pm h + h^{-1} \partial_\pm h~,
\end{equation}
with $b(x) \in \grp{B}$ and $h(x) \in \grp{H}$.

Choosing the operator $\mathcal E$, as well as the Drinfel'd double $\alg{d}$ and its isotropic algebra $\alg{b}$, appropriately, the action \eqref{eq:EmodB} gives the bi-$\eta$, bi-$\lambda$ and $\eta$-$\lambda$ deformations.
We discuss these choices below.

\paragraph{Drinfel'd double.} The $\eta$ deformation is governed by an antisymmetric linear operator $R$ satisfying the (in)homogeneous Yang-Baxter equation \eqref{eq:cYBE}, which for convenience we rewrite in the form
\begin{equation}
(R \pm \hat{c}) \com{X}{Y}_R= \com{(R \pm \hat{c})(X) }{(R \pm \hat{c})(Y) }~, \qquad \com{X}{Y}_R = \com{R(X)}{Y} + \com{X}{R(Y)}~,
\end{equation}
with $\hat{c} = c_\ind{L} P_\ind{L} + c_\ind{R} P_\ind{R}$.
At this point we already see the emergence of a dual Lie algebra $\tilde{\alg{f}}$, which as a vector space is the same as $\alg{f}$, but endowed with the Lie bracket $\com{\cdot}{\cdot}_R$, known as the R-bracket.
In what follows we restrict to the case $c_\ind{L} = c_\ind{R} = c \in \mathbb{R}_{\neq 0}$, so that the operator $R$ satisfies the split inhomogeneous Yang-Baxter equation in both the left and the right copy, and without loss of generality we fix $c=1$. More general cases of $c$ can then be obtained simply through rescalings of $R$.

Then, we define
\begin{equation} \label{eq:fdiag}
\alg{f}_{\text{diag}} = \{(\!(X,X)\!),X \in \alg{f}\}~,
\end{equation}
as well as
\begin{equation} \label{eq:fc}
\tilde{\alg{f}} = \{(\!((R + 1)X,(R - 1)X)\!),X \in \alg{f}\} ~.
\end{equation}
Both $\alg{f}_{\diag}$ and $\tilde{\alg{f}}$ are Lie algebras. This is obvious for the former, while for the latter it is a consequence of the inhomogeneous Yang-Baxter equation. Moreover, they are isotropic with respect to the bilinear form
\begin{equation}
\label{eq:bilin}
\langle \!\langle (\!(X_1,Y_1)\!), (\!(X_2,Y_2)\!) \rangle \!\rangle = \tSTr\left[ X_1 X_2 \right] - \tSTr \left[ Y_1 Y_2 \right]~,
\end{equation}
where $\tSTr$ was defined in \eqref{eq:modbil}.
Again, for $\alg{f}_{\diag}$ this is obvious, while for $\tilde{\alg{f}}$ one needs to use the antisymmetry of $R$ with respect to $\tSTr$. We then construct the Drinfel'd double
\begin{equation}\label{eq:dd}
\alg{d} = \alg{f}_{\diag} + \tilde{\alg{f}}~.
\end{equation}
Notice that the Drinfel'd double is the same as in the one-parameter case, the only modification lies in the bilinear form \eqref{eq:bilin}.

For general $\Integer_4$ supercosets $\grp{F}/\grp{F}_0$, we take the Lie algebra $\alg{h}$ to be
\begin{equation} \label{eq:defh}
\alg{h} = \{(\!(X,X)\!),~X \in \alg{f}^{(0)} = \Lie(\grp{F}_0) \}~,
\end{equation}
and write $\dA_\pm = (\!(\mathcal A_\pm, \mathcal A_\pm)\!)$ with $\mathcal A_\pm \in \alg{f}^{(0)}$, where $\alg{f}^{(0)}$ denotes the grade-0 subalgebra of the $\mathbb{Z}_4$ graded superalgebra $\alg{f}$.
For the $\Integer_4$ permutation supercosets~\eqref{eq:psc} in which we are interested, we have $\alg{f} = \alg{g} + \alg{g}$ and $\alg{f}^{(0)}=\alg{g}_0$ is the diagonal bosonic subalgebra.

\paragraph{Operator $\mathcal E$.} We define
\begin{equation}
\label{eq:PGPB}
P_G = \gamma \hat{k}^{-1} \left( P_0 + \frac{1}{2} (\mathcal P_- + \mathcal P_+) \right)~, \qquad P_B = \frac{\gamma}{2} \hat{k}^{-1} (\mathcal P_- - \mathcal P_+)~, \qquad \hat{k}= k_\ind{L} P_\ind{L} + k_\ind{R} P_\ind{R} ~,
\end{equation}
where $\mathcal P_\pm$ are the quantities appearing in the $\eta$ deformed action, defined in \eqref{eq:cP}, with coefficients given by \eqref{etaGS} for the GS case and in \eqref{etaPS} for the PS case. We also make the identification
\begin{equation} \label{eq:etal}
\eta_\ind{L} = \frac{\gamma}{k_\ind{L}}~, \qquad \eta_\ind{R} = \frac{\gamma}{k_\ind{R}}~,
\end{equation}
so that $P_G$ and $P_B$ depend only on $\gamma,k_\ind{L}, k_\ind{R}$.
We then introduce the $\mathcal E$ operator which acts on $\lb X,Y \rb \in \alg{d}$ as
\unskip\footnote{This is the same type of ansatz as for the one-parameter $\lambda$ deformation.
With respect to the notation in \cite{Hoare:2021dix} we remove the tilde $\tilde{P}^\lambda_\pm \rightarrow P_\pm^\lambda$.
This definition of $\mathcal E$ is such that
\begin{equation*}
\mathcal E \lb X, X \rb = P_G^{-1} P_B (\!(X,X )\!)- (P_G - P_B (P_G^{-1}) P_B) (\!(X,-X )\!)~.
\end{equation*}
}
\begin{equation} \begin{aligned}
\mathcal E \lb X, Y \rb &= \lb - ((P_+^\lambda)^{-1} - P_-^\lambda)^{-1} (((P_+^\lambda)^{-1} + P_-^\lambda) X - 2 Y), \\
&\qquad \qquad ((P_-^\lambda)^{-1} - P_+^\lambda)^{-1} (((P_-^\lambda)^{-1} + P_+^\lambda)Y-2 X) \, \rb~,
\end{aligned}
\end{equation}
where $P_\pm^\lambda$ are defined implicitly through the relations
\begin{equation}
\begin{aligned}
(P_G)^{-1} P_B - P_G + P_B (P_G)^{-1} P_B - (P_G)^{-1} - P_B (P_G)^{-1} &= -2 ((P_+^\lambda)^{-1} - P_-^\lambda)^{-1} ((P_+^\lambda)^{-1} + P_-^\lambda)~, \\
(P_G)^{-1} P_B - P_G + P_B (P_G)^{-1} P_B + (P_G)^{-1} + P_B (P_G)^{-1} &= +4 ((P_+^\lambda)^{-1} - P_-^\lambda)^{-1}~,\\
(P_G)^{-1} P_B + P_G - P_B (P_G)^{-1} P_B - (P_G)^{-1} + P_B (P_G)^{-1} &= -4 ((P_-^\lambda)^{-1} - P_+^\lambda)^{-1}~,\\
(P_G)^{-1} P_B + P_G - P_B (P_G)^{-1} P_B + (P_G)^{-1} - P_B (P_G)^{-1} &= +2 ((P_-^\lambda)^{-1} - P_+^\lambda)^{-1} ((P_-^\lambda)^{-1} + P_+^\lambda)~.
\end{aligned}
\end{equation}
Important identities are $P_+^\lambda = (P_-^\lambda)^T$, where $^T$ denotes the transpose with respect to $\tSTr$ of \eqref{eq:modbil}, as well as
\begin{equation}
\label{eq:idPGPB}
P_E \equiv P_G+P_B = (1+P_-^\lambda)^{-1}(1-P_-^\lambda) = -(1-(P_-^\lambda)^{-1})(1+(P_-^\lambda)^{-1})^{-1}~,
\end{equation}
and
\begin{equation}
\text{ker}\left[ \mathcal E \mathcal P (\mathcal E \pm 1) \right] = \left\{ \left(X,(P_\mp^\lambda)^{\pm 1} X\right)~,~ X \in \alg{f} \right\}.
\end{equation}
It is possible to check that in the symmetric case $k_\ind{L} = k_\ind{R} = k$ one recovers
\begin{equation}
\label{eq:Pmlone}
P_-^\lambda = \lambda P_0 +\lambda^2 P_2 + \lambda P_1 + \lambda^{-1} P_3~, \qquad \lambda = \frac{k-\gamma}{k+\gamma}~,
\end{equation}
in the GS case, and
\begin{equation}
\label{eq:PmlonePS}
P_-^\lambda = \lambda P_0 +\lambda^2 P_2 + \lambda P_1 + \lambda^{3} P_3~,
\end{equation}
in the PS case. These are the usual combination of projectors arising in the action of the (one-parameter) $\lambda$ deformation, with the relation $P_-^\lambda = \Qc^{-1}+(\lambda-1)P_0$.

\subsection{Bi-\texorpdfstring{$\eta$}{eta} models}

To obtain the action of the bi-$\eta$ deformation we take $\alg{b}=\tilde{\alg{f}}$ and fix the gauge $l=\lb g,g \rb \in \grp{F}_{\text{diag}}$.
An ansatz for $\mathcal E \mathcal P (\mathcal E \pm 1)$ with image and kernel satisfying \eqref{eq:epe} is given by
\begin{equation}
\label{eq:epsYB}
\mathcal E \mathcal P (\mathcal E \pm 1)\lb X,Y \rb = \lb (R_g + 1) f_\pm K_\pm , (R_g - 1) f_\pm K_\pm \rb~, \qquad K_\pm = Y - (P_\mp^\lambda)^{\pm 1} X~,
\end{equation}
where $R_g = \Ad_g^{-1} R \Ad_g$.
The functions $f_\pm$ are fixed by requiring that \eqref{eq:epsYB} is a projector, leading to
\begin{equation}
f_\pm = \left((1-(P_\mp^\lambda)^{\pm 1}) R_g - (1+ (P_\mp^\lambda)^{\pm 1})\right)^{-1}~.
\end{equation}
Plugging into \eqref{eq:EmodB} and using the identities \eqref{eq:idPGPB} one arrives at the action
\begin{equation} \label{eq:etaA}
\Act = - 2 \int \extder^2 x \, \tSTr \left[ (g^{-1} \partial_+ g - \mathcal A_+) \gamma \hat{k}^{-1} (P_0+\mathcal P_-) \frac{1}{1- R_g \gamma \hat{k}^{-1} (P_0+\mathcal P_-)} (g^{-1} \partial_- g - \mathcal A_-) \right]~.
\end{equation}

At this point we would like to integrate out the gauge fields $\mathcal A_\pm \in \alg{g}_0$
in order to obtain the action in sigma-model form.
The equation of motion for $\mathcal A_+$ reads
\begin{equation}
P_0 C_- = 0 ~, \qquad C_- = (P_0+\mathcal P_-) \frac{1}{1- R_g \gamma \hat{k}^{-1} (P_0+\mathcal P_-)} (g^{-1} \partial_- g - \mathcal A_-)~,
\end{equation}
from which we can deduce that
\begin{equation}
C_- = \mathcal P_- \frac{1}{1- R_g \gamma \hat{k}^{-1} \mathcal P_-} g^{-1} \partial_- g~.
\end{equation}
Replacing in the action gives
\begin{equation}
\Act = -2 \gamma \int \extder^2 x \, \STr\left[g^{-1} \partial_+ g \mathcal P_- \frac{1}{1- R_g \gamma \hat{k}^{-1} \mathcal P_-} g^{-1} \partial_- g \right]~.
\end{equation}
Therefore, upon identifying the parameters
\begin{equation}
T= 4 \gamma~, \qquad \eta_\ind{L} = \frac{\gamma}{k_\ind{L}}~, \qquad \eta_\ind{R} = \frac{\gamma}{k_\ind{R}}~,
\end{equation}
we recover the two-parameter $\eta$ deformation \eqref{eq:acteta} of the GS and PS sigma model respectively ($\mathcal P_\pm$ take different forms in the two models). Note however that this construction assumed that $R$ satisfies the split inhomogeneous Yang-Baxter equation in both left and right copy.

\subsection{Bi-\texorpdfstring{$\lambda$}{lambda} models}

To obtain the action of the dual $\lambda$-model we take $\alg{b}= \alg{f}_{\text{diag}}$.
An arbitrary element of $\grp{D}$ can be parametrised as $l=\lb g' g,g' \rb= \lb g',g'\rb \lb g,1 \rb$, with $\lb g',g'\rb \in \grp{F}_{\text{diag}}$ and $\lb g,1\rb \in \grp{F}_+$.
The gauge freedom of \eqref{eq:EmodB} then allows to choose $l=\lb g,1\rb$.

From the conditions \eqref{eq:epe} it follows that the image of $\mathcal E \mathcal P (\mathcal E \pm 1 )$ should be $\Ad_l^{-1} \alg{b} = \lb \Ad_g^{-1},1 \rb \alg{f}_{\text{diag}}$, which motivates the ansatz (the kernel remains the same)
\begin{equation}
\mathcal E \mathcal P (\mathcal E \pm 1 ) \lb X,Y \rb = \lb \Ad_g^{-1} f_\pm K_\pm , f_\pm K_\pm \rb~, \qquad K_\pm = Y- (P_\mp^\lambda)^{\pm 1} X~.
\end{equation}
Requiring that these are projectors further selects
\begin{equation}
f_\pm= \frac{1}{1-(P_\mp^\lambda)^{\pm 1} \Ad_g^{-1}}~.
\end{equation}
Plugging this into \eqref{eq:EmodB} gives
\begin{equation} \label{eq:lambdaA} \begin{aligned}
\Act &=- \frac{1}{2}\int \extder^2 x \, \tSTr \left[(g^{-1} \partial_+ g - \mathcal A_+ + \Ad_g^{-1} \mathcal A_+ ) \frac{1+\Ad_g^{-1} P_-^\lambda}{1-\Ad_g^{-1} P_-^\lambda} (g^{-1} \partial_- g - \mathcal A_- + \Ad_g^{-1} \mathcal A_- ) \right] \\
&\quad + \frac{1}{2} \int \extder^2 x \, \tSTr \left[ \mathcal A_+ (g^{-1} \partial_- g + \partial_- g g^{-1}) - \mathcal A_- ( g^{-1} \partial_+ g + \partial_+ g g^{-1} )\right]\\
&\quad + \frac{1}{2}\int \extder^2 x \, \tSTr \left[ \mathcal A_+ g^{-1} \mathcal A_- g - \mathcal A_+ g \mathcal A_- g^{-1}\right]+ \Act_{\textrm{WZ}}(g; \tSTr)~.
\end{aligned}
\end{equation}
The final step to obtain the bi $\lambda$-model consists in integrating out the gauge fields $\mathcal A_\pm \in \alg{g}_0$.
For this we follow but slightly modify the procedure used in the previous subsection for the $\eta$ deformation.
The equation of motion for $\mathcal A_+$ reads
\begin{equation} \label{eq:eomAp2}
P_0 C_- = 0 ~, \quad C_- = (k_\ind{L} P_\ind{L} + k_\ind{R} P_\ind{R}) (1-P_-^\lambda) \frac{1}{1-\Ad_g^{-1} P_-^\lambda} (g^{-1} \partial_- g - \mathcal A_- + \Ad_g^{-1} \mathcal A_- )~.
\end{equation}
A complication arises because of the presence of the $(k_\ind{L} P_\ind{L} + k_\ind{R} P_\ind{R})$ term in $C_-$.
We however observe that the auxiliary operator
\begin{equation}
Q = (1-P_-^\lambda)^{-1} (\Qc^{-1} - P_-^\lambda)~,
\end{equation}
with $\Qc$ defined in \eqref{GSsol} for the GS case and \eqref{eq:PSsol} for the PS case satisfies, in both cases,
\begin{align} \label{eq:cond1}
Q (1-P_-^\lambda)^{-1} (k_\ind{L}^{-1} P_\ind{L} + k_\ind{R}^{-1} P_\ind{R}) P_i=0 ~, \qquad i=1,2,3~.
\end{align}
From \eqref{eq:cond1} if follows that the equation of motion \eqref{eq:eomAp2} can be rewritten
\begin{equation}
Q (P_-^\lambda)^{-1} \hat{C}_-=0~, \qquad \hat{C}_- = P_-^\lambda \frac{1}{1-\Ad_g^{-1} P_-^\lambda} (g^{-1} \partial_- g - \mathcal A_- + \Ad_g^{-1} \mathcal A_- )~,
\end{equation}
from which we deduce that
\begin{equation}
\hat{C}_- = \Qc^{-1} \frac{1}{1-\Ad_g^{-1} \Qc^{-1}} (g^{-1} \partial_- g - \mathcal A_- + \Ad_g^{-1} \mathcal A_-)~.
\end{equation}
Injecting into the action \eqref{eq:lambdaA} leads to
\begin{equation}
\begin{aligned}
\Act &=-\frac{1}{2} \int \extder^2 x \,\tSTr \left[(g^{-1} \partial_+ g)\frac{\Qc+\Ad_g^{-1} }{\Qc-\Ad_g^{-1} } (g^{-1} \partial_- g - \mathcal A_- + \Ad_g^{-1} \mathcal A_- ) \right] \\
&\quad - \frac{1}{2} \int \extder^2 x \, \tSTr \left[ g^{-1} \partial_+ g(1+\Ad_g^{-1})\mathcal A_- \right] + \Act_{\textrm{WZ}}(g;\tSTr)~.
\end{aligned}
\end{equation}
Finally, one can check that the terms involving $\mathcal A_-$ cancel, owing to condition $\Qc^{-1} P_0 = P_0$, and the action of the $\lambda$-model, without gauge fields, becomes
\begin{equation}
\begin{aligned}
\Act = - \frac{1}{2} \int \extder^2 x \,\tSTr \left[(g^{-1} \partial_+ g) \frac{\Qc+\Ad_g^{-1} }{\Qc-\Ad_g^{-1} } (g^{-1} \partial_- g)\right]+ \Act_{\textrm{WZ}}(g; \tSTr)~.
\end{aligned}
\end{equation}
This is precisely the action of the bi-$\lambda$ deformation in sigma model form as obtained in \eqref{qact}.
Therefore, the bi-$\lambda$ deformation is the Poisson-Lie dual of the bi-$\eta$ deformation for operators $R$ satisfying the split inhomogeneous Yang-Baxter equation.

\subsection{\texorpdfstring{$\eta$-$\lambda$}{eta-lambda} deformation}

Let us now take advantage of the $\mathcal E$-model formulation to derive new hybrid deformations, with one copy of the symmetry algebra $\eta$ deformed, and the other $\lambda$ deformed.
We write $\alg{f}=\alg{g}_\ind{L} \oplus \alg{g}_\ind{R}$, where we keep track of the two different copies of $\alg{g}$ with the labels ``L'' and ``R.'' Any element $X \in \alg{f}$ can therefore be written $X=(X_\ind{L}, X_\ind{R})$ where $X_\ind{L} \in \alg{g}_\ind{L}$ and $X_\ind{R} \in \alg{g}_\ind{R}$. Clearly, also the diagonal algebra \eqref{eq:fdiag} takes this direct sum structure,
\begin{equation}
\alg{f}_{\diag} = \left\{\lb (X_\ind{L}, X_\ind{R}), (X_\ind{L}, X_\ind{R}) \rb \right\} = \alg{g}_{\diag,\ind{L}} \oplus \alg{g}_{\diag, \ind{R}}~,
\end{equation}
where
\begin{equation}
\alg{g}_{\diag, \ind{L}} = \{ \lb X_\ind{L}, X_\ind{L} \rb \}~,
\end{equation}
and similarly for the right copy. It is clear that $\alg{g}_{\diag, \ind{L}}$ and $\alg{g}_{\diag, \ind{R}}$ are algebras on their own.
For operators $R$ of the form $R= R_\ind{L} \oplus R_\ind{R}$, with $R_\ind{L,R} : \alg{g}_\ind{L,R} \rightarrow \alg{g}_\ind{L,R}$ (this is in particular true if the operator $R$ is of Drinfel'd-Jimbo type), also the algebra $\tilde{\alg{f}}$ defined in \eqref{eq:fc} takes this direct sum structure, with
\begin{align}
\tilde{\alg{f}} = \lb ((R_\ind{L} + 1) X_\ind{L}, (R_\ind{R}+1) X_\ind{R}), ((R_\ind{L} - 1) X_\ind{L}, (R_\ind{R}-1) X_\ind{R}) \rb= \tilde{\alg{g}}_{\ind{L}} \oplus \tilde{\alg{g}}_{\ind{R}}~,
\end{align}
where
\begin{equation}
\tilde{\alg{g}}_{\ind{L}} = \lb (R_\ind{L} + 1) X_\ind{L}, (R_\ind{L} - 1) X_\ind{L} \rb~,
\end{equation}
and similarly for the right copy.
Due to the requirement on $R$ also $\tilde{\alg{g}}_{\ind{L}}$ and $\tilde{\alg{g}}_{ \ind{R}}$ are algebras. These are then four subalgebras of the Drinfel'd double
\begin{equation}
\alg{d} = \alg{f}_{\diag} \oplus \tilde{\alg{f}} = \alg{g}_{\diag, \ind{L}} \oplus \alg{g}_{\diag, \ind{R}} \oplus \tilde{\alg{g}}_{\ind{L}} \oplus \tilde{\alg{g}}_{ \ind{R}}~.
\end{equation}
They are all isotropic with respect to the bilinear form \eqref{eq:bilin}.

We have seen in the previous two sections that integrating out the degrees of freedom associated to $\tilde{\alg{f}} = \tilde{\alg{g}}_{ \ind{L}} \oplus \tilde{\alg{g}}_{ \ind{R}}$ one gets the (bi-)$\eta$ deformation, while integrating out the degrees of freedom associated to $\alg{f}_{\diag} = \alg{g}_{\diag, \ind{L}} \oplus \alg{g}_{\diag, \ind{R}}$ one gets the (bi-)$\lambda$-model instead.
But one can do more.
In particular, one can integrate out the degrees of freedom associated to $\alg{f}_1 = \alg{g}_{\diag, \ind{L}} \oplus \tilde{\alg{g}}_{\ind{R}}$ or $\alg{f}_2 = \tilde{\alg{g}}_{\ind{L}} \oplus \alg{g}_{\diag, \ind{R}} $. This is possible because both $\alg{f}_1$ and $\alg{f}_2$ are subalgebras of $\alg{d}$ which are isotropic with respect to the bilinear form $\eqref{eq:bilin}$. The resulting models will be hybrid $\eta-\lambda$ deformations.

Without loss of generality (it is always possible to relabel the left and right copies), let us consider the case where we integrate out
\begin{equation} \label{eq:bhybrid}
\alg{f}_2 = \tilde{\alg{g}}_{\ind{L}} \oplus \alg{g}_{\diag, \ind{R}} \equiv \alg{b}_\ind{L} \oplus \alg{b}_\ind{R} \equiv \alg{b}~.
\end{equation}
Naively the resulting model should be $\eta$ deformed in the left copy and $\lambda$ deformed in the right copy, with additional non-trivial coupling terms between the two copies. An arbitrary element $l \in \grp{D}$ can be decomposed into
\begin{equation}
l = b_\ind{L} \lb (g_\ind{L}, g_\ind{R}' g_\ind{R}) , (g_\ind{L}, g_\ind{R}') \rb = b_\ind{L} \lb (1, g_\ind{R}') , (1, g_\ind{R}') \rb \lb (g_\ind{L}, g_\ind{R}) , (g_\ind{L}, 1) \rb~,
\end{equation}
where $b_\ind{L} \in \grp{B}_\ind{L}$ and $\lb (1, g_\ind{R}') , (1, g_\ind{R}') \rb \in \grp{B}_\ind{R}$.
After gauge fixing the left-acting $\grp{B}$ symmetry we are left with the representative
\begin{equation} \label{eq:lhybrid}
l= \lb (g_\ind{L}, g_\ind{R}), (g_\ind{L}, 1) \rb~.
\end{equation}

Let us now turn to the definition of the projectors $\mathcal E \mathcal P (\mathcal E \pm 1)$, satisfying the two constraints \eqref{eq:epe}. In particular, the image should be $\text{im} [\mathcal E \mathcal P (\mathcal E \pm 1)] = \Ad_{l}^{-1} \alg{b}$. From \eqref{eq:lhybrid} it follows that $\Ad_l^{-1} = \lb (\Ad_{g_\ind{L}}^{-1}, \Ad_{g_\ind{R}}^{-1}), (\Ad_{g_\ind{L}}^{-1}, 1) \rb$, and we recall that $\alg{b}$ is defined in \eqref{eq:bhybrid}. An ansatz with the correct image is therefore given by
\begin{equation} \label{eq:epehybrid}
\mathcal E \mathcal P (\mathcal E \pm 1) \lb X,Y \rb = \lb \begin{pmatrix}
R_{\ind{L}, g_\ind{L}} + 1 & 0 \\
0 & \Ad_{g_\ind{R}}^{-1}
\end{pmatrix} f_\pm K_\pm ,
\begin{pmatrix}
R_{\ind{L}, g_\ind{L}} - 1 & 0 \\
0 & 1
\end{pmatrix} f_\pm K_\pm
\rb~,
\end{equation}
and we recall the (unmodified) kernel
\begin{equation}
K_\pm = Y- (P_\mp^\lambda)^{-1} X~, \qquad X=(X_\ind{L}, X_\ind{R})~, \quad Y=(Y_\ind{L}, Y_\ind{R})~.
\end{equation}
We use a vector/matrix notation where the first component is in the left copy and the second component is in the right copy.
The unknown $f_\pm$ can be seen as $2 \times 2$ matrices, and are fixed by requiring that $\mathcal E \mathcal P (\mathcal E \pm 1)$ are projectors. We find
\begin{equation}
f_\pm = \begin{pmatrix}
f_{\pm, \ind{LL}} & f_{\pm,\ind{LR}} \\
f_{\pm, \ind{RL}} & f_{\pm, \ind{RR}}
\end{pmatrix}
= \left(
\begin{pmatrix}
R_{\ind{L},g_{\ind{L}}}-1 & 0 \\
0 & 1
\end{pmatrix}
- (P_\mp^\lambda)^\pm
\begin{pmatrix}
R_{\ind{L},g_{\ind{L}}} + 1 & 0 \\
0 & \Ad_{g_\ind{R}}
\end{pmatrix}
\right)^{-1}~.
\end{equation}

Then, the gauge field $\dA \in \alg{h}$ is as before, $\dA = \lb \mathcal A, \mathcal A \rb$ with $\mathcal A \in \alg{g}_0$. While it is possible to further decompose $\mathcal A=(\mathcal A_\ind{L}, \mathcal A_\ind{R})$, from the definition of the diagonal subalgebra $\alg{g}_0$ it follows that $\cA_\ind{L}=\cA_\ind{R}$.

Using the definition of the inner product $\ls \cdot , \cdot \rs$ of \eqref{eq:bilin}, and after some manipulation explained in \appref{app:hybrid}, the action of the hybrid model can be put in the form
\begin{equation} \begin{aligned}
\Act &= - \frac{1}{2} \int \extder^2 x \, \tSTr \left[ J_+ \left( 2 P_\ind{L} + P_\ind{R} (1+\Ad_{g_\ind{R}}^{-1} P_-^\lambda) (1-P_-^\lambda)^{-1} \right) \mathcal O^{-1} \left(J_- - \mathcal A_- + P_\ind{R} \Ad_{g_\ind{R}}^{-1} \mathcal A_- \right)\right] \\
&\qquad -\frac{1}{2} \int \extder^2 x \, \tSTr \left[ J_+ (1+\Ad_{g_\ind{R}}^{-1}) P_\ind{R} \mathcal A_-\right] + \Act_{\textrm{WZ}}(g_\ind{R}; \tSTr) \\
&\qquad + \int \extder^2 x \, \tSTr \left[ \mathcal A_+ \mathcal O^{-1}\left( J_- - \mathcal A_- + P_\ind{R} \Ad_{g_\ind{R}}^{-1} \mathcal A_- \right) \right]~,
\end{aligned}
\end{equation}
where
\begin{equation}
\mathcal O = \frac{1}{2} P_\ind{L} (P_E^{-1}-R_{L,g_\ind{L}}) + P_\ind{R} (1-\Ad_{g_\ind{R}}^{-1} P_-^\lambda) (1-P_-^\lambda)^{-1} ~,
\end{equation}
and
\begin{equation}
J_\pm = g^{-1} \partial_\pm g~, \qquad g=(g_\ind{L}, g_\ind{R}) \in \grp{G}_\ind{L} \times \grp{G}_\ind{R}~.
\end{equation}
The action is invariant under the gauge symmetry
\begin{equation}
g_\ind{L} \rightarrow g_\ind{L} \go ~, \qquad g_{\ind{R}} \rightarrow \go^{-1} g_\ind{R} \go~, \qquad \mathcal A_\pm \rightarrow \go^{-1} \mathcal A_\pm \go + \go^{-1} \partial_\pm \go~, \qquad \go(x) \in \grp{G}_0~.
\end{equation}

\section{Renormalisability and scale invariance}\label{sec:rg}

In this section we discuss quantum aspects of the bi-deformed models.
Our main focus will be the RG flow for the bosonic bi-$\l$ model, demonstrating its all-loop renormalisability in a ``tripled'' formulation and explicitly computing its two-loop beta function in a particular subtraction scheme.

Before we turn to the bosonic truncation, let us briefly comment on the deformed $\Integer_4$ permutation supercoset models.
For these to define consistent string sigma models, we require that they are Weyl invariant,
\unskip\footnote{In the context of the pure-spinor worldsheet theory in conformal gauge, the action should have conformal symmetry, zero central charge and a nilpotent fermionic operator.}
hence at one-loop the background fields solve the type II supergravity equations.
We will return to this in the \hyperref[sec6]{Conclusions}.
However, it follows from the results of \cite{Levine:2022hpv}, that, in the PS case, the bi-$\eta$ model \eqref{eq:acteta}, \eqref{eq:cP}, \eqref{etaPS} and bi-$\l$ model \eqref{qact}, \eqref{eq:PSsol} are one-loop renormalisable, and if the superalgebra $\alg{g}$ has a vanishing Killing form (as is the case for both $\alg{psu}(1,1|2)$ and $\alg{d}(2,1;\alpha)$) they are one-loop scale invariant, a necessary condition for Weyl invariance.
\unskip\footnote{This follows since the theories have a $\grp{G}_0$ gauge invariance, under which the Lax connection transforms as a connection, and satisfy a ``Bianchi completeness'' condition \cite{Levine:2022hpv}.
A sufficient condition for the latter is that the currents $J_\pm$ appearing in the Lax connection \eqref{LaxPS} are of the form $J_\pm = O_\pm(g) g^{-1}\partial_\pm g$, where the linear operators $O_\pm(g):\alg{g}\to\alg{g}$ are invertible.}

\subsection{RG flow for the bosonic bi-\texorpdfstring{$\lambda$}{lambda} model}

The bosonic bi-$\l$ model \eqref{bmr} on $\tfrac{\grp{G}\times \grp{G}}{\grp{G}}$ is not scale invariant, but it was observed in \cite{Sfetsos:2017sep} to be renormalisable at one loop with only the coupling $\h$ running.
Thus it provides a further example of the general expectation that integrable sigma models are renormalisable, or stable under RG flow \cite{Fateev:1992tk,Hoare:2019ark}.

Here, following the approach used in \cite{Hoare:2019mcc} for the standard $\l$ deformation, we will make a path integral transformation to a ``tripled'' formulation, after which certain fields decouple, leaving a model that is manifestly renormalisable to all orders due to its symmetries.
The key transformation is to exchange the auxiliary fields $A_{\ind{L},\ind{R}\pm}$ for Lorentz scalars
\begin{equation}
A_{i+} = h_i^{-1} \partial_+ h_i ~, \qquad A_{i-} = \bar h_i \partial_- {\bar h}_i ^{-1} ~, \qquad h_i, \, \bar h_i \in \grp{G} ~, \quad i=L,R ~. \label{tra}
\end{equation}
Applying the Polyakov-Wiegmann identity \cite{Polyakov:1983tt}, a certain combination of fields $\tilde g_i = h_i \, g_i \, \bar h_i$ decouples leaving the Lagrangian
\unskip\footnote{Note that here we take $\Act = \frac{1}{4\pi\alpha'} \int \extder^2 x\,\Lag$, hence the couplings $k_{\ind{L},\ind{R}}$ and $\h$ in this section are equal to those in \secref{sec2} and \secref{sec:pl} multiplied by $4\pi$. $\alpha'$ is understood as a loop-counting parameter, which may be set to one for convenience.}
\begin{gather}
\Lag = \kk_i \, \Lag_{\ind{\grp{G}}}(\tilde g_i) + \Lag' ~, \qquad \Lag' = - \kk_i \Big(\Lag_{\ind{\grp{G}}}(h_i) + \Lag_{\ind{\grp{G}}}(\bar h_i)\Big) + \Tr[ a_{ij}\, J_{i+} {\bar K}_{j-} ] ~, \\
a_{ij}
= \begin{pmatrix}
\kk_{\ind{L}} + \half \h & - \half \h \\
- \half \h & \kk_\ind{R} + \half \h
\end{pmatrix}_{ij} ~, \qquad J_{i+} = h_i^{-1} \partial_+ h_i ~, \quad {\bar K}_{i-} = \partial_- \bar h_i {\bar h}_i^{-1} ~.
\end{gather}
Here we are summing over repeated indices $i,j=L,R$ and $\Lag_{\ind{\grp{G}}}$ denotes the Lagrangian of the WZW model for the group $\grp{G}$.
The resulting sigma model has a ``tripled'' target space $(\grp{G}\times \grp{G})^3$.
The decoupled Lagrangian for $\tilde g_i$ is conformal on its own, leaving the ``truncated'' model $\Lag'$ to determine the RG flow of the bi-$\l$ model.

Since the transformation \eqref{tra} is non-local it gives rise to a finite one-loop determinant
\begin{equation}
\Delta \Lag = -2\cG \big( \Lag_{\ind{\grp{G}}} (h_i \bar h_i) + q \Tr[ J_{+i} {\bar K}_{i-} ] \big)~,
\end{equation}
where the local ambiguity parametrised by $q$ should be fixed to $q=0$ to preserve the $\grp{G}$ gauge symmetry
\begin{equation}
h_i \to h_i g ~, \qquad \bar h_i \to g^{-1} \bar h_i ~, \qquad \quad g(x)\in \grp{G} ~. \label{Gg}
\end{equation}
This has the effect of shifting the WZ levels $k_i \to \tilde k_i = k_i+2\cG$ in $\Lag'$
\begin{align}
\Lag' = -\tilde \kk_i \Big(\Lag_{\ind{\grp{G}}}(h_i) + \Lag_{\ind{\grp{G}}}(\bar h_i)\Big) + \Tr[ \tilde a_{ij}\, J_{i+}(h) \bar K_{j-}(\bar h) ] ~, \label{tr'}
\qquad
\tilde a_{ij} = \begin{pmatrix}
\tilde \kk_\ind{L} + \half \h & - \half \h \\
- \half \h & \tilde \kk_\ind{R} + \half \h
\end{pmatrix}_{ij} ~.
\end{align}
The Lagrangian $\Lag'$~\eqref{tr'} can be viewed as a degenerate gauge-invariant limit of a coupled $\grp{G}^4$ model of the type studied in \cite{Delduc:2018hty}.

In addition to the $\grp{G}$ gauge symmetry~\eqref{Gg}, the Lagrangian $\Lag'$~\eqref{tr'} is invariant under a $(\grp{G}(x^-) \times \grp{G}(x^+))^2$ chiral gauge symmetry, which is an artefact of the change of variables \eqref{tra}
\begin{equation}
h_i \to u_i(x^-) \, h_i ~, \quad \bar h_i \to \bar h_i \, v_i(x^+) ~, \qquad \big((u_{\ind{L}},v_{\ind{L}}),(u_{\ind{R}},v_{\ind{R}})\big)\in \big(\grp{G}(x^-) \times \grp{G}(x^+)\big)^2 ~.
\end{equation}
Crucially, up to the definition of the WZ levels and the coupling $\h$, the theory $\Lag'$ is the unique one with these symmetries.
As such, it must be renormalisable to all orders with only $\h$ running (since the WZ levels do not run).

\bigskip

We shall explicitly demonstrate the two-loop renormalisability of the truncated model \eqref{tr'} using a particular ``GB subtraction scheme'' \cite{Metsaev:1987zx} (see also the discussion in \cite{Levine:2021fof} and references therein), in which a general bosonic sigma model
\begin{equation}\begin{aligned}
\Act &= - \frac{1}{4\pi\alpha'} \int \extder^2x \ (G_{mn}(\varphi) \eta^{ab} + B_{mn}(\varphi) \varepsilon^{ab}) \ \partial_a \varphi^m \partial_b \varphi^n \\
&= \frac{1}{4\pi\alpha'} \int \extder^2x \ (G(\varphi) + B(\varphi) )_{mn} \ \partial_+ \varphi^m \partial_- \varphi^n ~, \label{smodel}
\end{aligned} \end{equation}
has the two-loop beta function
\unskip\footnote{In the beta function \eqref{genb} we have dropped possible diffeomorphism terms $L_{X}(G+B)_{mn}$ and exact terms $\del_{[m}Y_{n]}$, since here they are fixed to zero by global symmetry.}
\begin{align}
\ddt(G_{mn} + B_{mn}) &= \alpha'\, \beta^{(1)}_{mn} + \alpha'^2 \, \beta^{(2)}_{mn} + \ldots \label{genb}\\
&= \alpha' \, \widehat R_{mn} + \alpha'^2 \, \tfrac{1}{2} \Big(\widehat R^{klp}{}_n \widehat R_{mklp} - \tfrac12 \widehat R^{ l p k}{}_n \widehat R_{mklp}
+\tfrac12 \widehat R_{k mn l} H^{k pq} H^ {l}_{\ pq} \Big) + \ldots \ . \nonumber
\end{align}
Here $H_{mnk}=3 \del_{[m} B_{nk]}$ and $\widehat R$ is the curvature of the generalized connection $\widehat\Gamma^k{}_{mn} = \Gamma^k{}_{mn}(G) - \tfrac12 H^k{}_{mn}$ and $\alpha'$ is understood as a loop-counting parameter that we will set to one.

Let us now compute the Riemann tensor and H-flux corresponding to the Lagrangian~\eqref{tr'}.
To account for the gauge symmetry, we first introduce a ``regulator'' explicitly breaking it.
We do this by simply taking the matrix $\tilde a_{ij}$ in eq.~\eqref{tr'} to be a generic $2\times 2$ matrix.
We then compute the Riemann tensor and H-flux for the regulated theory before projecting out the ``pure gauge'' direction and taking the regulator to zero, i.e., setting $\tilde a_{ij}$ to its value in eq.~\eqref{tr'}.
This is equivalent to the proper gauge-fixing procedure explained in foot. 38 of \cite{Levine:2021fof}.

Choosing a convenient frame to diagonalise the metric $G_{mn}$
\begin{equation}
\begin{gathered}
ds^2 = G_{mn} d\varphi^m d\varphi^n = \frac{\tilde k_I}{2} \Tr[ E^I E^I ] ~, \qquad I = (i, \bar \imath ) ~, \quad i = L,R ~, \quad \bar\imath = \bar L, \bar R ~,\\
E^i = J_i +c_{i j} \bar {K}_{j} ~, \quad E^{\bar\imath} = d_{ i j} \bar K_{ j} ~,\qquad J_i = h_i^{-1} d h_i ~, \quad \bar K_{ i} = d \bar h_i \bar h_i^{-1} ~, \\
c_{ij} = \frac{\tilde a_{i j}}{\tilde k_i} ~, \qquad \tilde k_i \, d_{ij} d_{ik} = \tilde k_j \delta_{jk} - \tilde k_i c_{ij} c_{ik} ~, \label{ceq}
\end{gathered}\end{equation}
where we take $d_{ij}$ to be symmetric and denote its inverse by $q = d^{-1}$, and expanding the frame field in terms of generators $T_A$ of the Lie algebra $\alg{g}$, $E^I = E^{AI} T_A$, the H-flux $H=dB = \tfrac{1}{6} H_{AI,BK,CL} E^{AI} \wedge E^{BK} \wedge E^{CL}$ is given by
\begin{equation}\begin{gathered}
\begin{aligned}
& H_{Ai,Bk,Cl} = -\frac{i}{2} f_{ABC} \, \tilde k_i \, \delta_{ikl} ~, \qquad H_{A \bar\imath,B\bar k,C\bar l} = -\frac i2 f_{ABC} \, \sum_j \alpha_{ijkl} ~, \\
& H_{Ai,Bk,C\bar l} = 0 ~, \qquad H_{Ai,B\bar k,C\bar l} = -\frac i2 f_{ABC} \, \tilde k_i \, \Big( (cq)_{ik} (cq)_{il} + \sum_j c_{ij}q_{jk} q_{jl} \Big) ~,
\end{aligned}
\\
\alpha_{ijkl} = \tilde k_j \Big(q_{ji} q_{jk} q_{jl} + 2(cq)_{ji} (cq)_{jk} (cq)_{jl} - \sum_{m, \sigma} [ (cq)_{j\sigma(i)} c_{jm} q_{m\sigma(k)} q_{m\sigma(l)} ] \Big) ~.
\end{gathered}\end{equation}
Here we have indicated sums over indices explicitly, $\sigma$ are the three cyclic permutations of $\{i,k,l\}$, $(cq) = c\cdot q$ denotes the usual matrix product and $\delta_{ikl}$ is $1$ when $i=k=l$ and $0$ otherwise.

From Cartan's structure equation $dE^M+\widehat{\omega}^M{}_N E^N = T^M$ with torsion $T^M = \tfrac 12 H^M{}_{NP} E^N\wedge E^P$ (here with $M=A,I$) we can compute the torsionful spin connection
\begin{gather}
\nonumber
\widehat{\omega}^{Ai}{}_{Cl} = \sum_{Bk} A_{ikl} f_{ABC} E^{B\bar k} ~, \quad \widehat{\omega}^{Ai}{}_{C\bar l} =\sum_{Bk} B_{ikl} f_{ABC} E^{B\bar k} ~, \quad \widehat{\omega}^{A\bar\imath}{}_{C\bar l} = \sum_{Bk} C_{ikl} f_{ABC} E^{B\bar k} ~, \\
A_{ikl} = -i \delta_{il}(cq)_{ik} ~, \qquad B_{ikl} = i \, [ (cq)_{ik}(cq)_{il} - \sum_j c_{ij} q_{jk} q_{jl} ] ~, \\
\nonumber
C_{ikl} = -\frac{i}{2 \tilde k_i } \, \sum_j [ \tilde k_i d_{ij} q_{jk} q_{jl} - \tilde k_k d_{kj} q_{ji} q_{jl} +\tilde k_l d_{lj} q_{jk} q_{ji} + \alpha_{ijkl} ] ~.
\end{gather}
The torsionful Riemann curvature tensor is defined by $\tfrac 12 \widehat R^M{}_{NPQ} E^P \wedge E^Q = d\widehat \omega^M{}_{N} + \widehat\omega ^M{}_P \wedge \widehat \omega ^P{}_N$.
Rotating from $E^i, E^{\bar\imath}$ back to the basis $J_i, \bar K_i$, fixing the gauge $\bar h_\ind{R}=1$, projecting out the corresponding directions, and setting $\tilde{a}_{ij}$ to its value in eq.~\eqref{tr'}, we obtain the following non-zero components
\begin{equation}\begin{aligned}
&\widehat R^{Ai}{}_{Cl,D\bar{\ind{L}}, E \bar{\ind{L}}} = f^A{}_{BC} f^B{}_{DE} \, \frac{(-1)^{l-1} h^{i-1} \tilde k_\ind{L}(h+2\tilde k_\ind{L})(h+2\tilde k_\ind{R})^{3-i}}{2(4 \tilde k_\ind{L} \tilde k_\ind{R} + h \tilde k_\ind{L} +h \tilde k_\ind{R})^2}\ ,
\quad \ \ \qquad i,l=L,R \ , \\
&\widehat R^{A\bar{\ind{L}}}{}_{Cl,D\bar{\ind{L}}, E \bar{\ind{L}}} = f^A{}_{BC} f^B{}_{DE} \, \frac{(-1)^{l} (h+2\tilde k_\ind{L})(h+2\tilde k_\ind{R})(2\tilde k_\ind{L} \tilde k_\ind{R} + h \tilde k_\ind{L} \tilde k_\ind{R})}{2(4 \tilde k_\ind{L} \tilde k_\ind{R} + h \tilde k_\ind{L} +h \tilde k_\ind{R})^2}\ , \qquad l=L,R \ , \label{Req}
\end{aligned}\end{equation}
and the remaining components of the H-flux become
\begin{equation}\begin{aligned}
&H_{Ai,Bk,Cl} = -\frac{i}{2} f_{ABC} \, \tilde k_i \, \delta_{ikl} \ , \qquad \quad \quad H_{A \bar{\ind{L}}, B \bar{\ind{L}}, C \bar{\ind{L}}} = -\frac{i}{2} f_{ABC} \, \tilde k_\ind{L} \ , \\
&H_{Ai, Bk, C\bar{\ind{L}}} = -\frac{i}{2} f_{ABC} \, \tilde k_i \, \delta_{ik} \, c_{i\ind{L}} \ , \ \ \qquad H_{Ai, B\bar{\ind{L}}, C\bar{\ind{L}}} = -\frac{i}{2} f_{ABC} \, \tilde k_i \, c_{i\ind{L}} \ . \label{Heq}
\end{aligned}\end{equation}

Substituting eqs.~\eqref{Req}, \eqref{Heq} into the beta function~\eqref{genb}, we find that the Lagrangian $\Lag'$ is indeed renormalisable with only the coupling $\h$ running according to
\begin{equation}\begin{aligned}
\ddt \h = \, & 2 \cG \frac{(\h + 2 \tilde{\kk}_\ind{L}) (\h + 2 \tilde{\kk}_\ind{R}) (2 \tilde{\kk}_\ind{L} \tilde{\kk}_\ind{R} + \h (\tilde{\kk}_\ind{L} + \tilde{\kk}_\ind{R}))}{(4 \tilde{\kk}_\ind{L} \tilde{\kk}_\ind{R} +
\h (\tilde{\kk}_\ind{L} + \tilde{\kk}_\ind{R}))^2}
\\ & \ \Big( 1 + \frac{\cG}{\h(4 \tilde{\kk}_\ind{L} \tilde{\kk}_\ind{R} +
\h (\tilde{\kk}_\ind{L} + \tilde{\kk}_\ind{R}))^3} \Big( 32 \tilde{\kk}_\ind{L}^3 \tilde{\kk}_\ind{R}^3 + 64 \h \tilde{\kk}_\ind{L}^2 \tilde{\kk}_\ind{R}^2 (\tilde{\kk}_\ind{L} + \tilde{\kk}_\ind{R})
+ 3 \h^4 (\tilde{\kk}_\ind{L} + \tilde{\kk}_\ind{R})^2
\\ & \hspace{35pt} + 4 \h^3 (\tilde{\kk}_\ind{L} + \tilde{\kk}_\ind{R}) (2 \tilde{\kk}_\ind{L} + \tilde{\kk}_\ind{R}) (\tilde{\kk}_\ind{L} + 2 \tilde{\kk}_\ind{R}) + 8 \h^2 \tilde{\kk}_\ind{L} \tilde{\kk}_\ind{R} (5 \tilde{\kk}_\ind{L}^2 + 9 \tilde{\kk}_\ind{L} \tilde{\kk}_\ind{R} + 5 \tilde{\kk}_\ind{R}^2)\Big)\Big) ~.\label{bf}
\end{aligned}\end{equation}
The leading term in eq.~\eqref{bf} agrees with the one-loop result of \cite{Sfetsos:2017sep}, with the coupling of that paper identified as $\lambda_\ind{\text{there}} = (\tilde k_\ind{L}+\tilde k_\ind{R})/(2h+\tilde k_\ind{L} + \tilde k_\ind{R})$.

The two-loop fixed points $\h=-2\tilde\kk_\ind{L},-2\tilde\kk_\ind{R},-2\tilde\kk_\ind{L}\tilde\kk_\ind{R}/(\tilde\kk_\ind{L}+\tilde\kk_\ind{R})$ and $\h \to \infty$ are the same in this scheme as those at one loop up to the correction $\kk_i\to \tilde \kk_i$.
At the fixed points $\h=-2\tilde{\kk}_\ind{L}$ and $\h=-2\tilde{\kk}_\ind{R}$ (related by the $\Integer_2$ transformation \eqref{eq:z22}), the bi-$\l$ model becomes the $(\grp{G}_{\kk_\ind{L}-\kk_\ind{R}}\times \grp{G}_{\kk_\ind{R}})/\grp{G}_{\kk_\ind{L}}$ and $(\grp{G}_{\kk_\ind{L}}\times \grp{G}_{\kk_\ind{R}-\kk_\ind{L}})/\grp{G}_{\kk_\ind{R}}$ gauged WZW models respectively \cite{Sfetsos:2017sep}.
This may be seen by explicitly integrating out the gauge fields and substituting $\h=-2\tilde{\kk}_{1,2}$.
The fixed point $\h = -2 \tfrac{\tilde\kk_\ind{L} \tilde\kk_\ind{R}}{\tilde\kk_\ind{L}+\tilde\kk_\ind{R}}$ is related by the $\Integer_2$ transformation~\eqref{eq:z22} to the fixed point $\h \to \infty$, which, as discussed in \secref{sec2}, gives the $(\grp{G}_{k_\ind{L}} \times \grp{G}_{k_\ind{R}})/\grp{G}_{k_\ind{L}+k_\ind{R}}$ gauged WZW model.

In the symmetric limit $\kk_\ind{L}=\kk_\ind{R} \equiv \kk$ when the bi-$\l$ deformation reduces to the standard $\l$-deformation of the coset $\tfrac{\grp{G}\times \grp{G}}{\grp{G}}$, the two-loop beta function \eqref{bf} matches the known one in the same scheme \cite{Hoare:2019mcc},
\unskip\footnote{As explained in app.~A of \cite{Hoare:2019mcc}, the correct result \eqref{cres} for the $\l$ model on the coset $\tfrac{\grp{F}}{\grp{G}}=\tfrac{\grp{G}\times \grp{G}}{\grp{G}}$ is obtained by substituting $\cF=c_2(\grp{G})$ and $\cG=\tfrac{1}{2} c_2(\grp{G})$ in the general formulae of that paper, where $c_2(\grp{G})$ is the dual Coxeter number of $\grp{G}$.}
\begin{equation}\begin{aligned}
\ddt \kk = 0 ~, \qquad \quad \ddt \l = - \frac{\cG}{\kk_\ind{L}} \l \Big[ 1 - \frac{ \cG(1-3\l^2)}{2 \kk (1-\l^2)}\Big] ~, \qquad \l^{-1} \equiv \frac{\h}{\kk_\ind{L}}+1 ~.
\label{cres}
\end{aligned}\end{equation}

In the limit $\kk_\ind{R} \to \infty$, which corresponds to the NATD of the $\l$-model, the result \eqref{bf} reproduces the two-loop beta function of the $\l$-model in the same scheme as \cite{Hoare:2019mcc} (see also \cite{Georgiou:2019nbz})
\begin{equation}\begin{aligned}
\ddt \kk_\ind{L} = 0 ~, \qquad \ddt \l = - \frac{2\cG}{\kk_\ind{L}} \Big( \frac{\l}{1+\l}\Big)^2\Big[ 1 - \frac{2 \cG\l^2(1-2\l)}{\kk_\ind{L} (1-\l)(1+\l)^3}\Big] ~, \qquad \l^{-1} \equiv \frac{\h}{2\kk_\ind{L}}+1 ~.
\end{aligned}\end{equation}
Further taking the limit $\kk_\ind{L} \to \infty$ (holding $\h = 2\kk_\ind{L}(\l-1)$ fixed), which corresponds to the bi-NATD of the PCM, reproduces the scheme-invariant two-loop beta function of the PCM
\begin{equation}
\ddt \h = \cG + \half \cG^2 \h^{-1} ~.
\end{equation}

Finally, let us note that there is a curious weak-coupling limit
\begin{equation}
\kk_\ind{L}, \kk_\ind{R} \to +\infty ~, \quad \h \to -\infty ~, \qquad \quad \bar{\h}\equiv - \h - \frac{4\kk_\ind{L}\kk_\ind{R}}{\kk_\ind{L}+\kk_\ind{R}} \text{ fixed} ~, \qquad \bar \kk \equiv \kk_\ind{R}-\kk_\ind{L} \text{ fixed} ~,
\end{equation}
in which the two-loop beta function \eqref{bf} becomes that of the PCM with WZ term in the same subtraction scheme (extending the one-loop observation of \cite{Sfetsos:2017sep})
\begin{equation}
\ddt \bar \h = \cG\Big(1- \frac{\bar \kk^2}{\bar \h^2}\Big)\Big[ 1+ \frac12 \cG \bar \h^{-1} \Big(1-\frac{3\bar \kk^2}{\bar \h^2}\Big)\Big] ~, \qquad \ddt \bar \kk = 0~.
\end{equation}
It remains to be understood if there is a first-principles explanation of this fact.

\section{Conclusions}\label{sec6}

In this paper we have constructed and investigated integrable bi-deformations of $\Integer_4$ permutation supercoset sigma models with superisometry group $\grp{G} \times \grp{G}$.
These are expected to define integrable deformations of type II superstrings on $\AdS_3 \times \Sp^3 \times \To^4$ and $\AdS_3 \times \Sp^3 \times \Sp^3 \times \Sp^1$.
Starting from the classically integrable GS and PS sigma models~\eqref{eq:z4coset}, with $P_- = P_1 + 2P_2 - P_3$ and $P_- = P_1 + 2 P_2 + 3P_3$ respectively, we constructed their bi-$\eta$ and bi-$\lambda$ deformations in \secref{sec1} and \secref{sec2}.
In \secref{sec:pl} we showed that these models are related by Poisson-Lie duality, with each obtained by integrating out different degrees of freedom from the same $\mathcal E$ model on the Drinfel'd double.
This also allowed us to construct an $\eta$-$\lambda$ deformation, with one copy of $\grp{G}$ $\eta$ deformed and the other copy $\lambda$ deformed.

The particular form of the bi-$\eta$ and bi-$\lambda$ models suggests an underlying pattern in the deformations.
It would be interesting to uncover this by extending the construction to general $\Integer_{2N}$ permutation (super)cosets, generalising the results of~\cite{Hoare:2021dix}.
Moreover, in addition to PS and GS type models, it is known that for $N>2$ there are other choices of $P_-$ that define classically integrable sigma models~\cite{Osten:2021opf} and these should also admit bi-$\eta$ and bi-$\lambda$ deformations.

\medskip

An important open problem is the explicit construction of the type II supergravity backgrounds for the bi-deformations of $\AdS_3\times \Sp^3 \times \To^4$ and $\AdS_3\times \Sp^3 \times \Sp^3 \times \Sp^1$.
The type II supergravity equations of motion imply one-loop Weyl invariance, a basic consistency condition for string sigma models.
Below we summarise what is known in the literature for deformations of $\AdS_3\times \Sp^3 \times \To^4$ and its non-abelian T-duals.

\medskip

The bi-$\eta$ deformation has been studied in detail~\cite{Seibold:2019dvf} in the case where the operator $R$ is built from two $\alg{psu}(1,1|2)$ Drinfel'd-Jimbo solutions of the non-split inhomogeneous Yang-Baxter equation. 
When both Drinfel'd-Jimbo solutions are associated to the fermionic Dynkin diagram of $\alg{psu}(1,1|2)$, the background solves the type II supergravity equations.
For other Dynkin diagrams, the background instead solves the generalised equations of~\cite{Arutyunov:2015mqj,Wulff:2016tju} (see also \cite{Araujo:2018rbc}). 
This follows the general pattern that supergravity backgrounds are associated to unimodular operators $R$ \cite{Borsato:2016ose}, which in turn define unimodular Lie (super)algebras $\tilde{\alg{f}}$~\eqref{eq:dd} through the R-bracket.
It would be interesting to understand the precise form of the unimodularity condition for the bi-$\eta$ deformation, in particular for operators $R$ that mix the left and right copies of the symmetry algebra.

\medskip

Less is known for the bi-$\lambda$ deformation.
It is expected that the corresponding background will solve the type II supergravity equations since the lack of isometries means that the generalised supergravity equations are equivalent to the standard supergravity equations.
Moreover, the degrees of freedom that are integrated out in the $\mathcal{E}$ model are associated to the unimodular Lie (super)algebra $\alg{f}_{\diag}$~\eqref{eq:dd}.
For the symmetric $\lambda$ deformation, a candidate supergravity background has been written down in~\cite{Chervonyi:2016ajp}.
An alternative dilaton and set of R-R fluxes supporting the same metric and B-field was given earlier in \cite{Sfetsos:2014cea}.
This second background is expected to be the bosonic Poisson-Lie dual
\unskip\footnote{By this we mean Poisson-Lie dualising the symmetric $\eta$ deformation ($\eta_\ind{L} = \eta_\ind{R}$) with the operator $R$ built from the Drinfel'd-Jimbo solution associated to the distinguished Dynkin diagram with respect to the bosonic subalgebra.
For a discussion of Poisson-Lie dualities with respect to subalgebras see~\cite{Hoare:2017ukq}.}
of the symmetric $\eta$ deformation.
This is in contrast to the first background, corresponding to the symmetric $\lambda$ deformation~\cite{Hollowood:2014qma}, which is the Poisson-Lie dual with respect to the full superisometry algebra.
It would be interesting to construct the generalisation of both these backgrounds for the bi-deformed $\lambda$ models.
The two bi-deformed backgrounds would have the same metric 
(and no B-field)~\cite{Sfetsos:2017sep} as each other, but would be supported by different dilatons and R-R fluxes.

To gain a better understanding of the bi-$\lambda$ deformations, and $\lambda$ deformations more generally, it is informative to take the $\lambda \to 0$ limit.
As recalled in \secref{sec2}, taking this limit in the bosonic truncation gives the $(\grp{G}_{k_\ind{L}} \times \grp{G}_{k_\ind{R}})/\grp{G}_{k_\ind{L}+k_\ind{R}}$ gauged WZW model.
This is a CFT, hence the associated metric and B-field can be completed, with the requisite flat directions and a non-trivial dilaton, to a supergravity background, i.e., there are no R-R fluxes.
By analogy with the $\AdS_2 \times \Sp^2 \times \To^6$ case~\cite{Sfetsos:2014cea,Hoare:2015gda,Borsato:2016zcf,Hoare:2018ebg}, we expect this NS-NS background to be the $\lambda \to 0$ limit of the bosonic Poisson-Lie dual background, while for the bi-$\lambda$ deformation we expect R-R fluxes and a more complicated dilaton.
In addition to taking the $\lambda\to0$ limit at the level of the supergravity background, it would be important to understand it abstractly in the sigma model~\eqref{eq:ansatz}, particularly in the GS case~\eqref{GSsol} given the simple form of the $\kappa$-symmetry transformations~\eqref{eq:ks1}, \eqref{eq:ks2} in this limit.

\medskip

Similarly, very little is known about the $\eta$-$\lambda$ deformation.
We again expect that the backgrounds will solve the type II supergravity equations assuming that the operator $R$ satisfies a unimodularity condition, or equivalently the Lie (super)algebra $\alg{f}_2$~\eqref{eq:bhybrid} is unimodular.
The $\eta$-$\lambda$ deformation can be understood as the single-sided Poisson-Lie dual of the bi-$\eta$ deformation, and in the $\eta \to 0$ limit becomes the single-sided $\lambda$ deformation.
On general grounds, it is expected that this single-sided $\lambda$ deformation is the same as~\eqref{single} (up to interchanging the two copies of $\grp{G}$), which is found by first taking $k_\ind{R} \to \infty$ in the bi-$\lambda$ model to give the NATD-$\lambda$ model~\eqref{actrnatd} and then undoing the non-abelian T-duality.
However, this remains to be confirmed.

A supergravity background embedding the $\lambda$ deformation of the PCM, i.e., the bosonic truncation of the single-sided $\lambda$ deformation, is given in~\cite{Sfetsos:2014cea}.
By a similar logic to before, this should correspond to the single-sided bosonic Poisson-Lie dual of the single-sided $\eta$ deformation.
We note however, that this background has imaginary R-R fluxes.
Supergravity backgrounds for a different type of $\eta$-$\lambda$ deformation were discussed in \cite{Chervonyi:2016bfl}.
These are based on (super)cosets of the form $\grp{F}/\grp{F}_0$ and formally both $\eta$ and $\lambda$ deforming the (super)isometry group $\grp{F}$ at the same time~\cite{Sfetsos:2015nya}.
Trying to do this puts a strong constraint on the operator $R$ defining the $\eta$ deformation, which implies that the extra deformation acts trivially in many cases of interest.

The single-sided $\lambda$ deformation is particularly interesting since it has global $\grp{G}$ symmetry, so describes supersymmetric string backgrounds.
The presence of supersymmetry may mean the resulting supergravity backgrounds have certain ``nicer'' properties, as happened for the single-sided $\eta$ deformation in \cite{Hoare:2022asa}.
Moreover, the WZW model appears as the bosonic truncation in the further limit $\lambda\to 0$.

\medskip

Beyond one-loop Weyl invariance, it would also be interesting to investigate the higher-loop properties of these models.
In \secref{sec:rg} we investigated the higher-loop renormalisability of the bosonic truncation of the bi-$\lambda$ model.
We showed that in a ``tripled'' formulation certain fields decouple and the bosonic bi-$\lambda$ model becomes manifestly renormalizable to all orders due to the symmetries.
Therefore, it could be insightful to try to use a similar approach to study the bi-$\lambda$ deformations of $\Integer_4$ permutation supercosets.

\paragraph{Acknowledgments}

We would like to thank R.~Borsato, S.~Lacroix and A.~Tseytlin for related discussions and A.~Tseytlin for comments on the draft.
The work of BH was supported by a UKRI Future Leaders Fellowship (grant number MR/T018909/1).
FS was supported by the Swiss National Science Foundation via the Early Postdoc.Mobility fellowship ``q-deforming AdS/CFT'' and by the European Union's Horizon 2020 research and innovation programme under the Marie Sklodowska-Curie grant agreement number 1010272.
FS also thanks Ben Hoare and the Department of Mathematical Sciences at Durham University for hospitality during the final stage of this work.
This research was supported in part by the National Science Foundation under Grant No.~NSF PHY-1748958.
NL was supported by the Institut Philippe Meyer at the {\'E}cole Normale Sup{\'e}rieure in Paris.

\appendix

\section{Action of the \texorpdfstring{$\eta$}{eta}-\texorpdfstring{$\lambda$}{lambda} deformation}\label{app:hybrid}

Using the definition of the bilinear form $\ls \cdot, \cdot \rs$ as in \eqref{eq:bilin}, together with the projectors $\mathcal E \mathcal P (\mathcal P \pm 1)$ of \eqref{eq:epehybrid}, the action \eqref{eq:EmodB} can be expanded into
\unskip\footnote{In this appendix we use the shorthand notation $\int \equiv \int \extder^2 x$.}
\begin{equation} \begin{aligned}
\Act =\,& \frac{1}{2} \int \tSTr \left[ (J_{\ind{L}+} - \cA_{\ind{L}+}, J_{\ind{R}+} - \cA_{\ind{R}+}) \begin{pmatrix}
R_{\ind{L}, g_{\ind{L}}} + 1 & 0 \\
0 & \Ad_{g_\ind{R}}^{-1}
\end{pmatrix} f_+ \widetilde{K}_+ \right] \\
& - \frac{1}{2} \int \tSTr \left[ (J_{\ind{L}+} - \cA_{\ind{L}+}, - \cA_{\ind{R}+}) \begin{pmatrix}
R_{\ind{L}, g_{\ind{L}}} - 1 & 0 \\
0 & 1
\end{pmatrix} f_+ \widetilde{K}_+ \right] \\
&- \frac{1}{2} \int \tSTr \left[ (J_{\ind{L}-} - \cA_{\ind{L}-}, J_{\ind{R}-} - \cA_{\ind{R}-}) \begin{pmatrix}
R_{\ind{L}, g_{\ind{L}}} + 1 & 0 \\
0 & \Ad_{g_{\ind{R}}}^{-1}
\end{pmatrix} f_- \widetilde{K}_- \right] \\
& +\frac{1}{2} \int \tSTr \left[ (J_{\ind{L}-} - \cA_{\ind{L}-}, - \cA_{\ind{R}-}) \begin{pmatrix}
R_{\ind{L}, g_{\ind{L}}} - 1 & 0 \\
0 & 1
\end{pmatrix} f_- \widetilde{K}_- \right] \\
&- \frac{1}{2} \int \tSTr \left[ (J_{\ind{L}+}, J_{\ind{R}+}) (\cA_{\ind{L}-}, \cA_{\ind{R}-})^t \right] + \frac{1}{2} \int \tSTr \left[ (J_{\ind{L}-} , J_{\ind{R}-}) (\cA_{\ind{L}+}, \cA_{\ind{R}+})^t \right] \\
&+ \frac{1}{2} \int \tSTr \left[ (J_{\ind{L}+} , 0) (\cA_{\ind{L}-}, \cA_{\ind{R}-})^t \right] -\frac{1}{2} \int \tSTr \left[ (J_{\ind{L}-}, 0) (\cA_{\ind{L}+}, \cA_{\ind{R}+})^t \right] \\
&+ \Act_{\textrm{WZ}}(g_{\ind{R}}; \tSTr)~,
\end{aligned}
\end{equation}
where
\begin{equation} \label{eq:Kt}
\begin{aligned}
\widetilde{K}_+ &= (P_\ind{L} - P_-^\lambda) J_- - (1-P_-^\lambda) \cA_-~,\\
\widetilde{K}_- &= (P_\ind{L} - (P_+^\lambda)^{-1}) J_+ - (1-(P_+^\lambda)^{-1}) \cA_+~.
\end{aligned}
\end{equation}
Adding the first two lines together, as well as the third and fourth line, gives
\begin{equation} \begin{aligned}
\Act =\,& \frac{1}{2} \int \tSTr \left[ (J_{\ind{L}+} - \cA_{\ind{L}+}, J_{\ind{R}+} - \cA_{\ind{R}+} +\Ad_{g_{\ind{R}}}^{-1} \cA_{\ind{R}+}) \begin{pmatrix}
2 & 0 \\
0 & \Ad_{g_{\ind{R}}}^{-1}
\end{pmatrix} f_+ \tilde{K}_+ \right] \\
&- \frac{1}{2} \int \tSTr \left[ (J_{\ind{L}-} - \cA_{\ind{L}-}, J_{\ind{R}-} - \cA_{\ind{R}-} + \Ad_{g_{\ind{R}}}^{-1} \cA_{\ind{R}-}) \begin{pmatrix}
2 & 0 \\
0 & \Ad_{g_{\ind{R}}}^{-1}
\end{pmatrix} f_- \tilde{K}_- \right] \\
&- \frac{1}{2} \int \tSTr \left[ (0, J_{\ind{R}+} ) (\cA_{\ind{L}-}, \cA_{\ind{R}-})^t \right] + \frac{1}{2} \int \tSTr \left[ (0, J_{\ind{R}-} ) (\cA_{\ind{L}+}, \cA_{\ind{R}+})^t \right]+ \Act_{\textrm{WZ}}(g_{\ind{R}}; \tSTr)~.
\end{aligned}
\end{equation}
Using the explicit expression of $\tilde{K}_\pm$ in \eqref{eq:Kt} and defining
\begin{equation}
\mathcal M_\pm = \begin{pmatrix}
2 & 0 \\
0 & \Ad_{g_{\ind{R}}}^{-1}
\end{pmatrix} f_\pm~,
\end{equation}
this becomes
\begin{equation} \begin{aligned}
\Act =\,& \frac{1}{2} \int \tSTr \left[ (J_+-\cA_+ + P_{\ind{R}} \Ad_{g_{\ind{R}}}^{-1} \cA_+) \mathcal M_+
\left((P_{\ind{L}} - P_-^\lambda)J_- - (1-P_-^\lambda)\cA_- \right)
\right] \\
&- \frac{1}{2} \int \tSTr \left[ (J_- -\cA_- + P_{\ind{R}} \Ad_{g_{\ind{R}}}^{-1} \cA_-) \mathcal M_- \left((P_{\ind{L}} - (P_+^\lambda)^{-1})J_+ - (1-(P_+^\lambda)^{-1})\cA_+ \right) \right] \\
&- \frac{1}{2} \int \tSTr \left[ J_+ P_{\ind{R}} \cA_- \right] + \frac{1}{2} \int \tSTr \left[ J_- P_{\ind{R}} \cA_+ \right]+ \Act_{\textrm{WZ}}(g_{\ind{R}}; \tSTr)~.
\end{aligned}
\end{equation}
Grouping the terms of the form $JJ$, $J\cA$, $\cA J$ and $\cA \cA$ together gives
\begin{equation} \begin{aligned}
\Act =\,& \frac{1}{2} \int \tSTr \left[ J_+ \mathcal M_+
(P_{\ind{L}} - P_-^\lambda)J_- \right] - \frac{1}{2} \int \tSTr \left[ J_- \mathcal M_- (P_{\ind{L}} - (P_+^\lambda)^{-1})J_+ \right] \\
&- \frac{1}{2} \int \tSTr \left[ J_+ \mathcal M_+
(1-P_-^\lambda)\cA_- \right] - \frac{1}{2} \int \tSTr \left[ J_+ P_{\ind{R}} \cA_- \right] \\
&+ \frac{1}{2} \int \tSTr \left[ J_- \mathcal M_- (1-(P_+^\lambda)^{-1})\cA_+ \right]+ \frac{1}{2} \int \tSTr \left[ J_- P_{\ind{R}} \cA_+ \right]\\
&- \frac{1}{2} \int \tSTr \left[ \cA_+ (1- P_{\ind{R}} \Ad_{g_{\ind{R}}}) \mathcal M_+
\left((P_{\ind{L}} - P_-^\lambda)J_- - (1-P_-^\lambda)\cA_- \right)
\right] \\
&+ \frac{1}{2} \int \tSTr \left[ \cA_- (1- P_{\ind{R}} \Ad_{g_{\ind{R}}}) \mathcal M_- \left((P_{\ind{L}} - (P_+^\lambda)^{-1})J_+ - (1-(P_+^\lambda)^{-1})\cA_+ \right) \right] \\
& + \Act_{\textrm{WZ}}(g_{\ind{R}}; \tSTr)~.
\end{aligned}
\end{equation}
Now, recall that $P_+^\lambda$ and $P_-^\lambda$ are transpose to each other with respect to $\tSTr$ so one can write
\begin{equation} \begin{aligned}
\Act =\,& \frac{1}{2} \int \tSTr \left[ J_+ \left(\mathcal M_+
(P_{\ind{L}} - P_-^\lambda)- (P_{\ind{L}} - (P_-^\lambda)^{-1}) \mathcal M_-^T \right)J_- \right] \\
&+ \frac{1}{2} \int \tSTr \left[ J_+ \left( -\mathcal M_+
(1-P_-^\lambda) - P_{\ind{R}} + (P_{\ind{L}} - (P_-^\lambda)^{-1})\mathcal M_-^T (1- P_{\ind{R}} \Ad_{g_{\ind{R}}}^{-1})\right)\cA_- \right] \\
&+ \frac{1}{2} \int \tSTr \left[ \cA_+ \left( (1-(P_-^\lambda)^{-1}) \mathcal M_-^T + P_{\ind{R}} - (1- P_{\ind{R}} \Ad_{g_{\ind{R}}}) \mathcal M_+
(P_{\ind{L}} - P_-^\lambda)\right) J_- \right] \\
&+\frac{1}{2} \int \tSTr \left[ \cA_+ \left( (1- P_{\ind{R}} \Ad_{g_{\ind{R}}}) \mathcal M_+
( 1-P_-^\lambda) - (1-(P_-^\lambda)^{-1}) \mathcal M_-^T (1- P_{\ind{R}} \Ad_{g_{\ind{R}}}^{-1}) \right) \cA_-
\right] \\
& + \Act_{\textrm{WZ}}(g_{\ind{R}}; \tSTr)~.
\end{aligned}
\end{equation}
To make progress we note the useful identities
\begin{equation}
\mathcal M_+ (P_{\ind{L}} - P_-^\lambda) + (1+\Ad_{g_{\ind{R}}}^{-1}) (1-\Ad_{g_{\ind{R}}}^{-1})^{-1} P_{\ind{R}} = \left( \mathcal M_+ (1-P_-^\lambda) + P_{\ind{R}} \right)(P_{\ind{L}} + P_{\ind{R}}(1-\Ad_{g_{\ind{R}}}^{-1})^{-1})~,
\end{equation}
and
\begin{equation} \begin{aligned}
(1- P_{\ind{R}} \Ad_{g_{\ind{R}}}) \mathcal M_+
(P_{\ind{L}} - P_-^\lambda) -P_{\ind{R}} &= (1- P_{\ind{R}} \Ad_{g_{\ind{R}}}) \mathcal M_+
( 1-P_-^\lambda)((1- P_{\ind{R}} \Ad_{g_{\ind{R}}}^{-1}))^{-1} \\
&=- (1-(P_-^\lambda)^{-1}) \mathcal M_-^T \\
&= - (P_{\ind{L}}-(P_-^\lambda)^{-1}) \mathcal M_-^T - P_{\ind{R}} \mathcal M_-^T~.
\end{aligned}
\end{equation}
From this we deduce that
\begin{equation} \begin{aligned}
&\mathcal M_+
(P_{\ind{L}} - P_-^\lambda)- (P_{\ind{L}} - (P_-^\lambda)^{-1}) \mathcal M_-^T \\
& = \left( -2 P_{\ind{L}} - P_{\ind{R}} (1+\Ad_{g_{\ind{R}}}^{-1} P_-^\lambda) (1-P_-^\lambda)^{-1} \right) (1-(P_-^\lambda)^{-1})\mathcal M_-^T \end{aligned}
\end{equation}
Finally, the action becomes
\begin{equation} \begin{aligned}
\Act =\,& - \frac{1}{2} \int \tSTr \left[ J_+ \left( 2 P_{\ind{L}} + P_{\ind{R}} (1+\Ad_{g_{\ind{R}}}^{-1} P_-^\lambda) (1-P_-^\lambda)^{-1} \right) \mathcal O^{-1} \left(J_- - \cA_- + P_{\ind{R}} \Ad_{g_{\ind{R}}}^{-1}\cA_- \right)\right] \\
&-\frac{1}{2} \int \tSTr \left[ J_+ (1+\Ad_{g_{\ind{R}}}^{-1}) P_{\ind{R}} \cA_-\right] + \Act_{\textrm{WZ}}(g_{\ind{R}}; \tSTr) \\
&+ \int \tSTr \left[ \cA_+ \mathcal O^{-1}\left( J_- - \cA_- + P_{\ind{R}} \Ad_{g_{\ind{R}}}^{-1}\cA_- \right) \right]~,
\end{aligned}
\end{equation}
where
\begin{equation}
\mathcal O = \frac{1}{2} P_{\ind{L}} (P_E^{-1}-R_{\ind{L},g_{\ind{L}}}) + P_{\ind{R}} (1-\Ad_{g_{\ind{R}}}^{-1} P_-^\lambda) (1-P_-^\lambda)^{-1} ~.
\end{equation}
Using the identities
\begin{align}
&\nn \left( 2 P_{\ind{L}} + P_{\ind{R}} (1+\Ad_{g_{\ind{R}}}^{-1} P_-^\lambda) (1-P_-^\lambda)^{-1} \right) \mathcal O^{-1}
\\
&\quad = 2 \left( P_\ind{L} + P_\ind{R} (1-P_-^\lambda)^{-1} \right) \mathcal O^{-1} - P_\ind{R} \\
&\quad = 2 \left( P_\ind{L} + P_\ind{R} \Ad_{g_\ind{R}}^{-1} P_-^\lambda (1-P_-^\lambda)^{-1} \right) \mathcal O^{-1} + P_\ind{R}~,
\end{align}
we provide two alternative formulations of the action.
The first is
\begin{equation} \begin{aligned}
\Act = &- \int \tSTr \left[ J_+ \left(P_{\ind{L}} + P_{\ind{R}} (1-P_-^\lambda)^{-1} \right) \mathcal O^{-1} \left(J_- - \cA_- + P_{\ind{R}} \Ad_{g_{\ind{R}}}^{-1}\cA_- \right)\right] \\
&+ \frac{1}{2} \int \tSTr \left[ J_+P_{\ind{R}} \left(J_- - \cA_- + \Ad_{g_{\ind{R}}}^{-1}\cA_- \right)\right] \\
&-\frac{1}{2} \int \tSTr \left[ J_+ (1+\Ad_{g_{\ind{R}}}^{-1}) P_{\ind{R}} \cA_-\right] + \Act_{\textrm{WZ}}(g_{\ind{R}}; \tSTr) \\
&+ \int \tSTr \left[ \cA_+ \mathcal O^{-1}\left( J_- - \cA_- + P_{\ind{R}} \Ad_{g_{\ind{R}}}^{-1}\cA_- \right) \right]~,
\end{aligned}
\end{equation}
while the second is
\begin{equation} \begin{aligned}
\Act =\,& - \int \tSTr \left[ J_+ \left(P_{\ind{L}} + P_{\ind{R}} \Ad_{g_\ind{R}}^{-1} P_-^\lambda (1-P_-^\lambda)^{-1} \right) \mathcal O^{-1} \left(J_- - \cA_- + P_{\ind{R}} \Ad_{g_{\ind{R}}}^{-1}\cA_- \right)\right] \\
&- \frac{1}{2} \int \tSTr \left[ J_+P_{\ind{R}} \left(J_- - \cA_- + \Ad_{g_{\ind{R}}}^{-1}\cA_- \right)\right] \\
&-\frac{1}{2} \int \tSTr \left[ J_+ (1+\Ad_{g_{\ind{R}}}^{-1}) P_{\ind{R}} \cA_-\right] + \Act_{\textrm{WZ}}(g_{\ind{R}}; \tSTr) \\
&+ \int \tSTr \left[ \cA_+ \mathcal O^{-1}\left( J_- - \cA_- + P_{\ind{R}} \Ad_{g_{\ind{R}}}^{-1}\cA_- \right) \right]~.
\end{aligned}
\end{equation}

\begin{bibtex}[\jobname]

@article{Georgiou:2019nbz,
author = "Georgiou, George and Sagkrioti, Eftychia and Sfetsos, Konstantinos and Siampos, Konstantinos",
title = "{An exact symmetry in $\lambda$-deformed CFTs}",
eprint = "1911.02027",
archivePrefix = "arXiv",
primaryClass = "hep-th",
reportNumber = "CERN-TH-2019-053",
doi = "10.1007/JHEP01(2020)083",
journal = "JHEP",
volume = "01",
pages = "083",
year = "2020"
}

@article{Polyakov:1983tt,
author = "Polyakov, Alexander M. and Wiegmann, P. B.",
editor = "Stone, M.",
title = "{Theory of Nonabelian Goldstone Bosons}",
doi = "10.1016/0370-2693(83)91104-8",
journal = "Phys. Lett. B",
volume = "131",
pages = "121--126",
year = "1983"
}

@article{Rahmfeld:1998zn,
author = "Rahmfeld, J. and Rajaraman, Arvind",
title = "{The GS string action on $AdS_3 \times S^3$ with Ramond-Ramond charge}",
eprint = "hep-th/9809164",
archivePrefix = "arXiv",
reportNumber = "SU-ITP-98-55, RU-98-41",
doi = "10.1103/PhysRevD.60.064014",
journal = "Phys. Rev. D",
volume = "60",
pages = "064014",
year = "1999"
}

@article{Park:1998un,
author = "Park, Jaemo and Rey, Soo-Jong",
title = "{Green-Schwarz superstring on $AdS_3 \times S^3$}",
eprint = "hep-th/9812062",
archivePrefix = "arXiv",
reportNumber = "IASSNS-HEP-98-84, SNUTP-98-121",
journal = "JHEP",
volume = "01",
pages = "001",
year = "1999"
}

@article{Metsaev:2000mv,
author = "Metsaev, R. R. and Tseytlin, Arkady A.",
title = "{Superparticle and superstring in $AdS_3 \times S^3$ Ramond-Ramond background in light cone gauge}",
eprint = "hep-th/0011191",
archivePrefix = "arXiv",
reportNumber = "FIAN-TD-00-18, OHSTPY-HEP-T-00-029",
doi = "10.1063/1.1377274",
journal = "J. Math. Phys.",
volume = "42",
pages = "2987--3014",
year = "2001"
}

@article{Babichenko:2009dk,
author = "Babichenko, A. and Stefanski, Jr., B. and Zarembo, K.",
title = "{Integrability and the AdS(3)/CFT(2) correspondence}",
eprint = "0912.1723",
archivePrefix = "arXiv",
primaryClass = "hep-th",
reportNumber = "ITEP-TH-59-09, LPTENS-09-36, UUITP-25-09",
doi = "10.1007/JHEP03(2010)058",
journal = "JHEP",
volume = "03",
pages = "058",
year = "2010"
}

@article{Berkovits:1999im,
author = "Berkovits, Nathan and Vafa, Cumrun and Witten, Edward",
title = "{Conformal field theory of AdS background with Ramond-Ramond flux}",
eprint = "hep-th/9902098",
archivePrefix = "arXiv",
reportNumber = "IFT-P-012-99, HUTP-99-A004, IASSNS-HEP-99-5",
doi = "10.1088/1126-6708/1999/03/018",
journal = "JHEP",
volume = "03",
pages = "018",
year = "1999"
}

@article{Berkovits:1999du,
author = "Berkovits, Nathan",
title = "{Quantization of the type II superstring in a curved six-dimensional background}",
eprint = "hep-th/9908041",
archivePrefix = "arXiv",
reportNumber = "IFT-P-059-99",
doi = "10.1016/S0550-3213(99)00690-2",
journal = "Nucl. Phys. B",
volume = "565",
pages = "333--344",
year = "2000"
}

@article{Metsaev:1998it,
author = "Metsaev, R. R. and Tseytlin, Arkady A.",
title = "{Type IIB superstring action in $AdS_5 \times S^5$ background}",
eprint = "hep-th/9805028",
archivePrefix = "arXiv",
reportNumber = "FIAN-TD-98-21, IMPERIAL-TP-97-98-44, NSF-ITP-98-055",
doi = "10.1016/S0550-3213(98)00570-7",
journal = "Nucl. Phys. B",
volume = "533",
pages = "109--126",
year = "1998"
}

@article{Berkovits:1999zq,
author = "Berkovits, N. and Bershadsky, M. and Hauer, T. and Zhukov, S. and Zwiebach, B.",
title = "{Superstring theory on $AdS_2 \times S^2$ as a coset supermanifold}",
eprint = "hep-th/9907200",
archivePrefix = "arXiv",
reportNumber = "IFT-P-060-99, HUTP-99-A044, MIT-CTP-2878, CTP-MIT-2878",
doi = "10.1016/S0550-3213(99)00683-5",
journal = "Nucl. Phys. B",
volume = "567",
pages = "61--86",
year = "2000"
}

@article{Henneaux:1984mh,
author = "Henneaux, Marc and Mezincescu, Luca",
title = "{A Sigma Model Interpretation of Green-Schwarz Covariant Superstring Action}",
reportNumber = "UTTG-26-84",
doi = "10.1016/0370-2693(85)90507-6",
journal = "Phys. Lett. B",
volume = "152",
pages = "340--342",
year = "1985"
}

@article{Berkovits:2000fe,
author = "Berkovits, Nathan",
title = "{Super Poincare covariant quantization of the superstring}",
eprint = "hep-th/0001035",
archivePrefix = "arXiv",
reportNumber = "IFT-P-005-2000",
doi = "10.1088/1126-6708/2000/04/018",
journal = "JHEP",
volume = "04",
pages = "018",
year = "2000"
}

@article{Bena:2003wd,
author = "Bena, Iosif and Polchinski, Joseph and Roiban, Radu",
title = "{Hidden symmetries of the $AdS_5 \times S^5$ superstring}",
eprint = "hep-th/0305116",
archivePrefix = "arXiv",
reportNumber = "NSF-KITP-03-34, UCLA-03-TEP-14",
doi = "10.1103/PhysRevD.69.046002",
journal = "Phys. Rev. D",
volume = "69",
pages = "046002",
year = "2004"
}

@article{Vallilo:2003nx,
author = "Vallilo, Brenno Carlini",
title = "{Flat currents in the classical $AdS_5 \times S^5$ pure spinor superstring}",
eprint = "hep-th/0307018",
archivePrefix = "arXiv",
reportNumber = "IFT-P-029-2003",
doi = "10.1088/1126-6708/2004/03/037",
journal = "JHEP",
volume = "03",
pages = "037",
year = "2004"
}

@article{Mikhailov:2007eg,
author = "Mikhailov, Andrei and Schafer-Nameki, Sakura",
title = "{Algebra of transfer-matrices and Yang-Baxter equations on the string worldsheet in $AdS_5 \times S^5$}",
eprint = "0712.4278",
archivePrefix = "arXiv",
primaryClass = "hep-th",
reportNumber = "CALT-68-2666, NI-07085",
doi = "10.1016/j.nuclphysb.2008.04.029",
journal = "Nucl. Phys. B",
volume = "802",
pages = "1--39",
year = "2008"
}

@article{Magro:2008dv,
author = "Magro, Marc",
title = "{The Classical Exchange Algebra of $AdS_5 \times S^5$}",
eprint = "0810.4136",
archivePrefix = "arXiv",
primaryClass = "hep-th",
reportNumber = "AEI-2008-085",
doi = "10.1088/1126-6708/2009/01/021",
journal = "JHEP",
volume = "01",
pages = "021",
year = "2009"
}

@article{Vicedo:2010qd,
author = "Vicedo, Benoit",
title = "{The classical R-matrix of AdS/CFT and its Lie dialgebra structure}",
eprint = "1003.1192",
archivePrefix = "arXiv",
primaryClass = "hep-th",
reportNumber = "IPHT-T10-026",
doi = "10.1007/s11005-010-0446-9",
journal = "Lett. Math. Phys.",
volume = "95",
pages = "249--274",
year = "2011"
}

@article{Maillet:1985ek,
author = "Maillet, Jean Michel",
title = "{New Integrable Canonical Structures in Two-dimensional Models}",
reportNumber = "PAR/LPTHE-85-32",
doi = "10.1016/0550-3213(86)90365-2",
journal = "Nucl. Phys. B",
volume = "269",
pages = "54--76",
year = "1986"
}

@article{Maillet:1985ec,
author = "Maillet, Jean Michel",
title = "{Hamiltonian Structures for Integrable Classical Theories From Graded Kac-moody Algebras}",
reportNumber = "PAR-LPTHE-85/40",
doi = "10.1016/0370-2693(86)91289-X",
journal = "Phys. Lett. B",
volume = "167",
pages = "401--405",
year = "1986"
}

@article{Sevostyanov:1995hd,
author = "Sevostyanov, Alexei",
title = "{The Classical R matrix method for nonlinear sigma model}",
eprint = "hep-th/9509030",
archivePrefix = "arXiv",
doi = "10.1142/S0217751X96001978",
journal = "Int. J. Mod. Phys. A",
volume = "11",
pages = "4241--4254",
year = "1996"
}

@article{Lacroix:2018njs,
author = "Lacroix, Sylvain",
title = "{Integrable models with twist function and affine Gaudin models}",
eprint = "1809.06811",
archivePrefix = "arXiv",
primaryClass = "hep-th",
reportNumber = "tel-01900498, 2018LYSEN014",
school = "Lyon, Ecole Normale Superieure",
year = "2018"
}

@article{Delduc:2013fga,
author = "Delduc, Francois and Magro, Marc and Vicedo, Benoit",
title = "{On classical $q$-deformations of integrable sigma-models}",
eprint = "1308.3581",
archivePrefix = "arXiv",
primaryClass = "hep-th",
doi = "10.1007/JHEP11(2013)192",
journal = "JHEP",
volume = "11",
pages = "192",
year = "2013"
}

@article{Benitez:2018xnh,
author = "Ben\'\i{}tez, H\'ector A. and Rivelles, Victor O.",
title = "{Yang-Baxter deformations of the $AdS_{5}\times S^{5}$ pure spinor superstring}",
eprint = "1807.10432",
archivePrefix = "arXiv",
primaryClass = "hep-th",
doi = "10.1007/JHEP02(2019)056",
journal = "JHEP",
volume = "02",
pages = "056",
year = "2019"
}

@article{Kawaguchi:2014qwa,
author = "Kawaguchi, Io and Matsumoto, Takuya and Yoshida, Kentaroh",
title = "{Jordanian deformations of the $AdS_5 \times S^5$ superstring}",
eprint = "1401.4855",
archivePrefix = "arXiv",
primaryClass = "hep-th",
reportNumber = "KUNS-2477, ITP-UU-14-05, SPIN-14-05",
doi = "10.1007/JHEP04(2014)153",
journal = "JHEP",
volume = "04",
pages = "153",
year = "2014"
}

@article{Delduc:2013qra,
author = "Delduc, Francois and Magro, Marc and Vicedo, Benoit",
title = "{An integrable deformation of the $AdS_5 \times S^5$ superstring action}",
eprint = "1309.5850",
archivePrefix = "arXiv",
primaryClass = "hep-th",
doi = "10.1103/PhysRevLett.112.051601",
journal = "Phys. Rev. Lett.",
volume = "112",
number = "5",
pages = "051601",
year = "2014"
}

@article{Hoare:2014oua,
author = "Hoare, Ben",
title = "{Towards a two-parameter q-deformation of $AdS_3 \times S^3 \times M^4$ superstrings}",
eprint = "1411.1266",
archivePrefix = "arXiv",
primaryClass = "hep-th",
reportNumber = "HU-EP-14-44",
doi = "10.1016/j.nuclphysb.2014.12.012",
journal = "Nucl. Phys. B",
volume = "891",
pages = "259--295",
year = "2015"
}

@article{Vicedo:2015pna,
author = "Vicedo, Benoit",
title = "{Deformed integrable $\sigma$-models, classical R-matrices and classical exchange algebra on Drinfel'd doubles}",
eprint = "1504.06303",
archivePrefix = "arXiv",
primaryClass = "hep-th",
doi = "10.1088/1751-8113/48/35/355203",
journal = "J. Phys. A",
volume = "48",
number = "35",
pages = "355203",
year = "2015"
}

@article{Hoare:2021dix,
author = "Hoare, Ben",
title = "{Integrable deformations of sigma models}",
eprint = "2109.14284",
archivePrefix = "arXiv",
primaryClass = "hep-th",
doi = "10.1088/1751-8121/ac4a1e",
journal = "J. Phys. A",
volume = "55",
number = "9",
pages = "093001",
year = "2022"
}

@article{Klimcik:2008eq,
author = "Klimcik, Ctirad",
title = "{On integrability of the Yang-Baxter sigma-model}",
eprint = "0802.3518",
archivePrefix = "arXiv",
primaryClass = "hep-th",
doi = "10.1063/1.3116242",
journal = "J. Math. Phys.",
volume = "50",
pages = "043508",
year = "2009"
}

@article{Novikov:1982ei,
author = "Novikov, S.P.",
title = "{The Hamiltonian formalism and a many-valued analogue of Morse theory}",
doi = "10.1070/RM1982v037n05ABEH004020",
journal = "Usp. Mat. Nauk",
volume = "37N5",
number = "5",
pages = "3--49",
year = "1982"
}

@article{Witten:1983ar,
author = "Witten, Edward",
title = "{Non-abelian bosonization in two dimensions}",
doi = "10.1007/BF01215276",
journal = "Commun. Math. Phys.",
volume = "92",
pages = "455--472",
year = "1984"
}

@article{Witten:1983tw,
author = "Witten, Edward",
title = "{Global aspects of current algebra}",
doi = "10.1016/0550-3213(83)90063-9",
journal = "Nucl. Phys. B",
volume = "223",
pages = "422--432",
year = "1983"
}

@article{Cagnazzo:2012se,
author = "Cagnazzo, A. and Zarembo, K.",
title = "{B-field in AdS(3)/CFT(2) Correspondence and Integrability}",
eprint = "1209.4049",
archivePrefix = "arXiv",
primaryClass = "hep-th",
reportNumber = "NORDITA-2012-67, UUITP-24-12",
doi = "10.1007/JHEP11(2012)133",
journal = "JHEP",
volume = "11",
pages = "133",
year = "2012",
note = "[Erratum: JHEP 04, 003 (2013)]"
}

@article{Abdalla:1984gm,
author = "Abdalla, M. C. B.",
title = "{Integrability of Chiral Nonlinear $\sigma$ Models Summed to a {Wess-Zumino} Term}",
reportNumber = "NBI-HE-84-38",
doi = "10.1016/0370-2693(85)91172-4",
journal = "Phys. Lett. B",
volume = "152",
pages = "215--217",
year = "1985"
}

@article{Veselov:1984,
author = "Veselov, A. P. and Takhtajan, L. A.",
title = "{Integrability of the Novikov equations for principal chiral fields with a multivalued Lagrangian}",
journal = "Sov. Phys. Dokl.",
volume = "29",
pages = "994",
year = "1984"
}

@article{Delduc:2017fib,
author = "Delduc, Francois and Hoare, Ben and Kameyama, Takashi and Magro, Marc",
title = "{Combining the bi-Yang-Baxter deformation, the Wess-Zumino term and TsT transformations in one integrable $\sigma$-model}",
eprint = "1707.08371",
archivePrefix = "arXiv",
primaryClass = "hep-th",
doi = "10.1007/JHEP10(2017)212",
journal = "JHEP",
volume = "10",
pages = "212",
year = "2017"
}

@article{Hoare:2020mpv,
author = "Hoare, B. and Lacroix, S.",
title = "{Yang\textendash{}Baxter deformations of the principal chiral model plus Wess\textendash{}Zumino term}",
eprint = "2009.00341",
archivePrefix = "arXiv",
primaryClass = "hep-th",
reportNumber = "ZMP-HH/20-17",
doi = "10.1088/1751-8121/abc43d",
journal = "J. Phys. A",
volume = "53",
number = "50",
pages = "505401",
year = "2020"
}

@article{Hoare:2015gda,
author = "Hoare, B. and Tseytlin, A. A.",
title = "{On integrable deformations of superstring sigma models related to $AdS_n \times S^n$ supercosets}",
eprint = "1504.07213",
archivePrefix = "arXiv",
primaryClass = "hep-th",
reportNumber = "IMPERIAL-TP-AT-2015-02, HU-EP-15-21",
doi = "10.1016/j.nuclphysb.2015.06.001",
journal = "Nucl. Phys. B",
volume = "897",
pages = "448--478",
year = "2015"
}

@article{Sfetsos:2015nya,
author = "Sfetsos, Konstantinos and Siampos, Konstantinos and Thompson, Daniel C.",
title = "{Generalised integrable $\lambda$- and $\eta$-deformations and their relation}",
eprint = "1506.05784",
archivePrefix = "arXiv",
primaryClass = "hep-th",
doi = "10.1016/j.nuclphysb.2015.08.015",
journal = "Nucl. Phys. B",
volume = "899",
pages = "489--512",
year = "2015"
}

@article{Hoare:2017ukq,
author = "Hoare, Ben and Seibold, Fiona K.",
title = "{Poisson-Lie duals of the $\eta$ deformed symmetric space sigma model}",
eprint = "1709.01448",
archivePrefix = "arXiv",
primaryClass = "hep-th",
doi = "10.1007/JHEP11(2017)014",
journal = "JHEP",
volume = "11",
pages = "014",
year = "2017"
}

@article{Klimcik:2015gba,
author = "Klimcik, Ctirad",
title = "{$\eta$ and $\lambda$ deformations as $\mathcal{E}$-models}",
eprint = "1508.05832",
archivePrefix = "arXiv",
primaryClass = "hep-th",
doi = "10.1016/j.nuclphysb.2015.09.011",
journal = "Nucl. Phys. B",
volume = "900",
pages = "259--272",
year = "2015"
}

@article{Sfetsos:2017sep,
author = "Sfetsos, Konstantinos and Siampos, Konstantinos",
title = "{Integrable deformations of the $G_{k_1} \times G_{k_2}/G_{k_1+k_2}$ coset CFTs}",
eprint = "1710.02515",
archivePrefix = "arXiv",
primaryClass = "hep-th",
reportNumber = "CERN-TH-2017-199",
doi = "10.1016/j.nuclphysb.2017.12.011",
journal = "Nucl. Phys. B",
volume = "927",
pages = "124--139",
year = "2018"
}

@article{Hoare:2019mcc,
author = "Hoare, Ben and Levine, Nat and Tseytlin, Arkady A.",
title = "{Integrable sigma models and 2-loop RG flow}",
eprint = "1910.00397",
archivePrefix = "arXiv",
primaryClass = "hep-th",
reportNumber = "Imperial-TP-AT-2019-06",
doi = "10.1007/JHEP12(2019)146",
journal = "JHEP",
volume = "12",
pages = "146",
year = "2019"
}

@article{Levine:2021fof,
author = "Levine, Nat and Tseytlin, Arkady A.",
title = "{Integrability vs. RG flow in $G \times G$ and $G \times G /H$ sigma models}",
eprint = "2103.10513",
archivePrefix = "arXiv",
primaryClass = "hep-th",
reportNumber = "Imperial-TP-NL-2021-01",
doi = "10.1007/JHEP05(2021)076",
journal = "JHEP",
volume = "05",
pages = "076",
year = "2021"
}

@article{Metsaev:1987zx,
author = "Metsaev, R. R. and Tseytlin, Arkady A.",
title = "{Order alpha-prime (Two Loop) Equivalence of the String Equations of Motion and the Sigma Model Weyl Invariance Conditions: Dependence on the Dilaton and the Antisymmetric Tensor}",
reportNumber = "PRINT-87-0184 (LEBEDEV)",
doi = "10.1016/0550-3213(87)90077-0",
journal = "Nucl. Phys. B",
volume = "293",
pages = "385--419",
year = "1987"
}

@article{Klimcik:2002zj,
author = "Klimcik, Ctirad",
title = "{Yang-Baxter sigma models and dS/AdS T duality}",
eprint = "hep-th/0210095",
archivePrefix = "arXiv",
reportNumber = "IML-02-XY",
doi = "10.1088/1126-6708/2002/12/051",
journal = "JHEP",
volume = "12",
pages = "051",
year = "2002"
}

@article{Delduc:2018xug,
author = "Delduc, F. and Hoare, B. and Kameyama, T. and Lacroix, S. and Magro, M.",
title = "{Three-parameter integrable deformation of $\mathbb{Z}_4$ permutation supercosets}",
eprint = "1811.00453",
archivePrefix = "arXiv",
primaryClass = "hep-th",
reportNumber = "ZMP-HH/18-22",
doi = "10.1007/JHEP01(2019)109",
journal = "JHEP",
volume = "01",
pages = "109",
year = "2019"
}

@article{Sfetsos:2013wia,
author = "Sfetsos, Konstadinos",
title = "{Integrable interpolations: From exact CFTs to non-Abelian T-duals}",
eprint = "1312.4560",
archivePrefix = "arXiv",
primaryClass = "hep-th",
reportNumber = "DMUS-MP-13-23, DMUS--MP--13-23",
doi = "10.1016/j.nuclphysb.2014.01.004",
journal = "Nucl. Phys. B",
volume = "880",
pages = "225--246",
year = "2014"
}

@article{Klimcik:2019kkf,
author = "Klim\v{c}\'\i{}k, Ctirad",
title = "{Dressing cosets and multi-parametric integrable deformations}",
eprint = "1903.00439",
archivePrefix = "arXiv",
primaryClass = "hep-th",
doi = "10.1007/JHEP07(2019)176",
journal = "JHEP",
volume = "07",
pages = "176",
year = "2019"
}

@article{Hollowood:2014qma,
author = "Hollowood, Timothy J. and Miramontes, J. Luis and Schmidtt, David M.",
title = "{An Integrable Deformation of the $AdS_5 \times S^5$ Superstring}",
eprint = "1409.1538",
archivePrefix = "arXiv",
primaryClass = "hep-th",
doi = "10.1088/1751-8113/47/49/495402",
journal = "J. Phys. A",
volume = "47",
number = "49",
pages = "495402",
year = "2014"
}

@article{Hollowood:2014rla,
author = "Hollowood, Timothy J. and Miramontes, J. Luis and Schmidtt, David M.",
title = "{Integrable Deformations of Strings on Symmetric Spaces}",
eprint = "1407.2840",
archivePrefix = "arXiv",
primaryClass = "hep-th",
doi = "10.1007/JHEP11(2014)009",
journal = "JHEP",
volume = "11",
pages = "009",
year = "2014"
}

@article{delaOssa:1992vci,
author = "de la Ossa, Xenia C. and Quevedo, Fernando",
title = "{Duality symmetries from non-Abelian isometries in string theory}",
eprint = "hep-th/9210021",
archivePrefix = "arXiv",
reportNumber = "NEIP-92-004",
doi = "10.1016/0550-3213(93)90041-M",
journal = "Nucl. Phys. B",
volume = "403",
pages = "377--394",
year = "1993"
}

@article{Sfetsos:2014cea,
author = "Sfetsos, Konstantinos and Thompson, Daniel C.",
title = "{Spacetimes for $\lambda$-deformations}",
eprint = "1410.1886",
archivePrefix = "arXiv",
primaryClass = "hep-th",
doi = "10.1007/JHEP12(2014)164",
journal = "JHEP",
volume = "12",
pages = "164",
year = "2014"
}

@article{Fateev:1992tk,
author = "Fateev, V. A. and Onofri, E. and Zamolodchikov, Alexei B.",
title = "{Integrable deformations of the $O(3)$ sigma model. The sausage model}",
reportNumber = "PAR-LPTHE-92-46, LPTHE-92-46",
doi = "10.1016/0550-3213(93)90001-6",
journal = "Nucl. Phys. B",
volume = "406",
pages = "521--565",
year = "1993"
}

@article{Hoare:2019ark,
author = "Hoare, Ben and Levine, Nat and Tseytlin, Arkady A.",
title = "{Integrable 2d sigma models: quantum corrections to geometry from RG flow}",
eprint = "1907.04737",
archivePrefix = "arXiv",
primaryClass = "hep-th",
reportNumber = "Imperial-TP-AT-2019-05",
doi = "10.1016/j.nuclphysb.2019.114798",
journal = "Nucl. Phys. B",
volume = "949",
pages = "114798",
year = "2019"
}

@article{Delduc:2018hty,
author = "Delduc, F. and Lacroix, S. and Magro, M. and Vicedo, B.",
title = "{Integrable Coupled $\sigma$ Models}",
eprint = "1811.12316",
archivePrefix = "arXiv",
primaryClass = "hep-th",
reportNumber = "ZMP-HH/18-26",
doi = "10.1103/PhysRevLett.122.041601",
journal = "Phys. Rev. Lett.",
volume = "122",
number = "4",
pages = "041601",
year = "2019"
}

@article{Seibold:2019dvf,
author = "Seibold, Fiona K.",
title = "{Two-parameter integrable deformations of the $AdS_3 \times S^3 \times T^4$ superstring}",
eprint = "1907.05430",
archivePrefix = "arXiv",
primaryClass = "hep-th",
doi = "10.1007/JHEP10(2019)049",
journal = "JHEP",
volume = "10",
pages = "049",
year = "2019"
}

@article{Hoare:2022asa,
author = "Hoare, Ben and Seibold, Fiona K. and Tseytlin, Arkady A.",
title = "{Integrable supersymmetric deformations of $AdS_3 \times S^3 \times T^4$}",
eprint = "2206.12347",
archivePrefix = "arXiv",
primaryClass = "hep-th",
reportNumber = "Imperial-TP-AT-2022-02",
doi = "10.1007/JHEP09(2022)018",
journal = "JHEP",
volume = "09",
pages = "018",
year = "2022"
}

@article{Witten:1991mm,
author = "Witten, Edward",
title = "{On Holomorphic factorization of WZW and coset models}",
reportNumber = "IASSNS-HEP-91-25",
doi = "10.1007/BF02099196",
journal = "Commun. Math. Phys.",
volume = "144",
pages = "189--212",
year = "1992"
}

@article{Appadu:2015nfa,
author = "Appadu, Calan and Hollowood, Timothy J.",
title = "{Beta function of k deformed $AdS_5 \times S^5$ string theory}",
eprint = "1507.05420",
archivePrefix = "arXiv",
primaryClass = "hep-th",
doi = "10.1007/JHEP11(2015)095",
journal = "JHEP",
volume = "11",
pages = "095",
year = "2015"
}

@article{Borsato:2016ose,
author = "Borsato, Riccardo and Wulff, Linus",
title = "{Target space supergeometry of $\eta$ and $\lambda$-deformed strings}",
eprint = "1608.03570",
archivePrefix = "arXiv",
primaryClass = "hep-th",
reportNumber = "IMPERIAL-TP-LW-2016-03",
doi = "10.1007/JHEP10(2016)045",
journal = "JHEP",
volume = "10",
pages = "045",
year = "2016"
}

@article{Klimcik:1995dy,
author = "Klimcik, C. and Severa, P.",
title = "{Poisson-Lie T duality and loop groups of Drinfeld doubles}",
eprint = "hep-th/9512040",
archivePrefix = "arXiv",
reportNumber = "CERN-TH-95-330",
doi = "10.1016/0370-2693(96)00025-1",
journal = "Phys. Lett. B",
volume = "372",
pages = "65--71",
year = "1996"
}

@article{Klimcik:1996nq,
author = "Klimcik, C. and Severa, P.",
title = "{Non-Abelian momentum winding exchange}",
eprint = "hep-th/9605212",
archivePrefix = "arXiv",
reportNumber = "CERN-TH-96-142",
doi = "10.1016/0370-2693(96)00755-1",
journal = "Phys. Lett. B",
volume = "383",
pages = "281--286",
year = "1996"
}

@article{Klimcik:1995ux,
author = "Klimcik, C. and Severa, P.",
title = "{Dual non-Abelian duality and the Drinfeld double}",
eprint = "hep-th/9502122",
archivePrefix = "arXiv",
reportNumber = "CERN-TH-95-39, CERN-TH-95-039",
doi = "10.1016/0370-2693(95)00451-P",
journal = "Phys. Lett. B",
volume = "351",
pages = "455--462",
year = "1995"
}

@article{Klimcik:1995jn,
author = "Klimcik, C.",
editor = "Gava, E. and Narain, K. S. and Vafa, C.",
title = "{Poisson-Lie T duality}",
eprint = "hep-th/9509095",
archivePrefix = "arXiv",
reportNumber = "CERN-TH-95-248",
doi = "10.1016/0920-5632(96)00013-8",
journal = "Nucl. Phys. B Proc. Suppl.",
volume = "46",
pages = "116--121",
year = "1996"
}

@article{Tseytlin:1990nb,
author = "Tseytlin, Arkady A.",
title = "{Duality Symmetric Formulation of String World Sheet Dynamics}",
reportNumber = "KCL-TP-1990-2",
doi = "10.1016/0370-2693(90)91454-J",
journal = "Phys. Lett. B",
volume = "242",
pages = "163--174",
year = "1990"
}

@article{Tseytlin:1990va,
author = "Tseytlin, Arkady A.",
title = "{Duality symmetric closed string theory and interacting chiral scalars}",
reportNumber = "KCL-TP-1990-3",
doi = "10.1016/0550-3213(91)90266-Z",
journal = "Nucl. Phys. B",
volume = "350",
pages = "395--440",
year = "1991"
}

@article{vanTongeren:2019dlq,
author = "van Tongeren, Stijn J.",
title = "{Unimodular jordanian deformations of integrable superstrings}",
eprint = "1904.08892",
archivePrefix = "arXiv",
primaryClass = "hep-th",
doi = "10.21468/SciPostPhys.7.1.011",
journal = "SciPost Phys.",
volume = "7",
pages = "011",
year = "2019"
}

@article{Hoare:2018ngg,
author = "Hoare, Ben and Seibold, Fiona K.",
title = "{Supergravity backgrounds of the $\eta$-deformed AdS$_2 \times S^2 \times T^6 $ and AdS$_5 \times S^5$ superstrings}",
eprint = "1811.07841",
archivePrefix = "arXiv",
primaryClass = "hep-th",
doi = "10.1007/JHEP01(2019)125",
journal = "JHEP",
volume = "01",
pages = "125",
year = "2019"
}

@article{Chervonyi:2016ajp,
author = "Chervonyi, Yuri and Lunin, Oleg",
title = "{Supergravity background of the $\lambda$-deformed AdS$_3 \times$ S$^3$ supercoset}",
eprint = "1606.00394",
archivePrefix = "arXiv",
primaryClass = "hep-th",
doi = "10.1016/j.nuclphysb.2016.07.023",
journal = "Nucl. Phys. B",
volume = "910",
pages = "685--711",
year = "2016"
}

@article{Chervonyi:2016bfl,
author = "Chervonyi, Yuri and Lunin, Oleg",
title = "{Generalized $\lambda$-deformations of AdS$_p \times$ S$^p$}",
eprint = "1608.06641",
archivePrefix = "arXiv",
primaryClass = "hep-th",
doi = "10.1016/j.nuclphysb.2016.10.014",
journal = "Nucl. Phys. B",
volume = "913",
pages = "912--941",
year = "2016"
}

@article{Arutyunov:2015mqj,
author = "Arutyunov, G. and Frolov, S. and Hoare, B. and Roiban, R. and Tseytlin, A. A.",
title = "{Scale invariance of the $\eta$-deformed $AdS_5\times S^5$ superstring, T-duality and modified type II equations}",
eprint = "1511.05795",
archivePrefix = "arXiv",
primaryClass = "hep-th",
reportNumber = "ZMP-HH-15-27, TCDMATH-15-12, IMPERIAL-TP-AT-2015-08",
doi = "10.1016/j.nuclphysb.2015.12.012",
journal = "Nucl. Phys. B",
volume = "903",
pages = "262--303",
year = "2016"
}

@article{Balog:1993es,
author = "Balog, J. and Forgacs, P. and Horvath, Z. and Palla, L.",
editor = "Lust, D. and Weigt, G.",
title = "{A New family of SU(2) symmetric integrable sigma models}",
eprint = "hep-th/9307030",
archivePrefix = "arXiv",
reportNumber = "ITP-502-BUDAPEST",
doi = "10.1016/0370-2693(94)90213-5",
journal = "Phys. Lett. B",
volume = "324",
pages = "403--408",
year = "1994"
}

@article{Itsios:2014lca,
author = "Itsios, Georgios and Sfetsos, Konstadinos and Siampos, Konstantinos",
title = "{The all-loop non-Abelian Thirring model and its RG flow}",
eprint = "1404.3748",
archivePrefix = "arXiv",
primaryClass = "hep-th",
doi = "10.1016/j.physletb.2014.04.061",
journal = "Phys. Lett. B",
volume = "733",
pages = "265--269",
year = "2014"
}

@article{Fateev:1996ea,
author = "Fateev, V. A.",
title = "{The sigma model (dual) representation for a two-parameter family of integrable quantum field theories}",
doi = "10.1016/0550-3213(96)00256-8",
journal = "Nucl. Phys. B",
volume = "473",
pages = "509--538",
year = "1996"
}

@article{Hoare:2014pna,
author = "Hoare, B. and Roiban, R. and Tseytlin, A. A.",
title = "{On deformations of $AdS_n \times S^n$ supercosets}",
eprint = "1403.5517",
archivePrefix = "arXiv",
primaryClass = "hep-th",
reportNumber = "IMPERIAL-TP-AT-2014-02, HU-EP-14-10",
doi = "10.1007/JHEP06(2014)002",
journal = "JHEP",
volume = "06",
pages = "002",
year = "2014"
}

@article{Wulff:2016tju,
author = "Tseytlin, A. A. and Wulff, L.",
title = "{Kappa-symmetry of superstring sigma model and generalized 10d supergravity equations}",
eprint = "1605.04884",
archivePrefix = "arXiv",
primaryClass = "hep-th",
reportNumber = "IMPERIAL-TP-LW-2016-02",
doi = "10.1007/JHEP06(2016)174",
journal = "JHEP",
volume = "06",
pages = "174",
year = "2016",
}

@article{Levine:2022hpv,
author = "Levine, Nat",
title = "{Universal 1-loop divergences for integrable sigma models}",
eprint = "2209.05502",
archivePrefix = "arXiv",
primaryClass = "hep-th",
month = "9",
year = "2022"
}

@article{Berkovits:2001ue,
author = "Berkovits, Nathan and Howe, Paul S.",
title = "{Ten-dimensional supergravity constraints from the pure spinor formalism for the superstring}",
eprint = "hep-th/0112160",
archivePrefix = "arXiv",
reportNumber = "IFT-P-072-2001, KCL-TH-01-49, IFT-P.072-2001",
doi = "10.1016/S0550-3213(02)00352-8",
journal = "Nucl. Phys. B",
volume = "635",
pages = "75--105",
year = "2002"
}

@article{Borsato:2016zcf,
author = "Borsato, R. and Tseytlin, A. A. and Wulff, L.",
title = "{Supergravity background of $\lambda$-deformed model for AdS$_2 \times$ S$^2$ supercoset}",
eprint = "1601.08192",
archivePrefix = "arXiv",
primaryClass = "hep-th",
reportNumber = "IMPERIAL-TP-RB-2016-01",
doi = "10.1016/j.nuclphysb.2016.02.018",
journal = "Nucl. Phys. B",
volume = "905",
pages = "264--292",
year = "2016"
}

@article{Witten:1985nt,
author = "Witten, Edward",
title = "{Twistor-Like Transform in Ten-Dimensions}",
reportNumber = "PRINT-85-0458 (PRINCETON)",
doi = "10.1016/0550-3213(86)90090-8",
journal = "Nucl. Phys. B",
volume = "266",
pages = "245--264",
year = "1986"
}

@article{Grisaru:1985fv,
author = "Grisaru, Marcus T. and Howe, Paul S. and Mezincescu, L. and Nilsson, B. and Townsend, P. K.",
title = "{N=2 Superstrings in a Supergravity Background}",
reportNumber = "Print-85-0603 (CAMBRIDGE)",
doi = "10.1016/0370-2693(85)91071-8",
journal = "Phys. Lett. B",
volume = "162",
pages = "116--120",
year = "1985"
}

@article{Howe:1983sra,
author = "Howe, Paul S. and West, Peter C.",
title = "{The Complete N=2, D=10 Supergravity}",
reportNumber = "Print-83-0565 (KING'S COLL)",
doi = "10.1016/0550-3213(84)90472-3",
journal = "Nucl. Phys. B",
volume = "238",
pages = "181--220",
year = "1984"
}

@article{Arutyunov:2009ga,
author = "Arutyunov, Gleb and Frolov, Sergey",
title = "{Foundations of the AdS$_{5} \times S^{5}$ Superstring. Part I}",
eprint = "0901.4937",
archivePrefix = "arXiv",
primaryClass = "hep-th",
reportNumber = "ITP-UU-09-05, SPIN-09-05, TCD-MATH-09-06, HMI-09-03",
doi = "10.1088/1751-8113/42/25/254003",
journal = "J. Phys. A",
volume = "42",
pages = "254003",
year = "2009"
}

@article{Osten:2021opf,
author = "Osten, David",
title = "{Lax pairs for new ZN-symmetric coset \ensuremath{\sigma}-models and their Yang-Baxter deformations}",
eprint = "2112.07438",
archivePrefix = "arXiv",
primaryClass = "hep-th",
doi = "10.1016/j.nuclphysb.2022.115856",
journal = "Nucl. Phys. B",
volume = "981",
pages = "115856",
year = "2022"
}

@article{Hoare:2018ebg,
author = "Hoare, Ben and Seibold, Fiona K.",
title = "{Poisson-Lie duals of the $\eta$-deformed $\mathrm{AdS}_2 \times \mathrm{S}^2 \times \mathrm{T}^6$ superstring}",
eprint = "1807.04608",
archivePrefix = "arXiv",
primaryClass = "hep-th",
doi = "10.1007/JHEP08(2018)107",
journal = "JHEP",
volume = "08",
pages = "107",
year = "2018"
}

@article{Frolov:2021fmj,
author = "Frolov, Sergey and Sfondrini, Alessandro",
title = "{New dressing factors for AdS3/CFT2}",
eprint = "2112.08896",
archivePrefix = "arXiv",
primaryClass = "hep-th",
doi = "10.1007/JHEP04(2022)162",
journal = "JHEP",
volume = "04",
pages = "162",
year = "2022"
}

@article{Seibold:2022mgg,
author = "Seibold, Fiona K. and Sfondrini, Alessandro",
title = "{Transfer matrices for AdS3/CFT2}",
eprint = "2202.11058",
archivePrefix = "arXiv",
primaryClass = "hep-th",
reportNumber = "Imperial-TP-FS-2022-01",
doi = "10.1007/JHEP05(2022)089",
journal = "JHEP",
volume = "05",
pages = "089",
year = "2022"
}

@article{Frolov:2021bwp,
author = "Frolov, Sergey and Sfondrini, Alessandro",
title = "{Mirror thermodynamic Bethe ansatz for AdS3/CFT2}",
eprint = "2112.08898",
archivePrefix = "arXiv",
primaryClass = "hep-th",
doi = "10.1007/JHEP03(2022)138",
journal = "JHEP",
volume = "03",
pages = "138",
year = "2022"
}

@article{Cavaglia:2021eqr,
author = "Cavagli\`a, Andrea and Gromov, Nikolay and Stefa\'nski, Jr., Bogdan and Jr. and Torrielli, Alessandro",
title = "{Quantum Spectral Curve for AdS$_{3}$/CFT$_{2}$: a proposal}",
eprint = "2109.05500",
archivePrefix = "arXiv",
primaryClass = "hep-th",
reportNumber = "DMUS-MP/21-14, DMUS-MP-21/14",
doi = "10.1007/JHEP12(2021)048",
journal = "JHEP",
volume = "12",
pages = "048",
year = "2021"
}

@article{Ekhammar:2021pys,
author = "Ekhammar, Simon and Volin, Dmytro",
title = "{Monodromy bootstrap for SU(2|2) quantum spectral curves: from Hubbard model to AdS$_{3}$/CFT$_{2}$}",
eprint = "2109.06164",
archivePrefix = "arXiv",
primaryClass = "math-ph",
reportNumber = "UUITP-44/21, NORDITA 2021-090",
doi = "10.1007/JHEP03(2022)192",
journal = "JHEP",
volume = "03",
pages = "192",
year = "2022"
}

article{Ekhammar:2021pys,
author = "Ekhammar, Simon and Volin, Dmytro",
title = "{Monodromy bootstrap for SU(2|2) quantum spectral curves: from Hubbard model to AdS$_{3}$/CFT$_{2}$}",
eprint = "2109.06164",
archivePrefix = "arXiv",
primaryClass = "math-ph",
reportNumber = "UUITP-44/21, NORDITA 2021-090",
doi = "10.1007/JHEP03(2022)192",
journal = "JHEP",
volume = "03",
pages = "192",
year = "2022"
}

@article{Cavaglia:2022xld,
author = "Cavagli\`a, Andrea and Ekhammar, Simon and Gromov, Nikolay and Ryan, Paul",
title = "{Exploring the Quantum Spectral Curve for $AdS_3$/CFT${}_2$}",
eprint = "2211.07810",
archivePrefix = "arXiv",
primaryClass = "hep-th",
month = "11",
year = "2022"
}

@article{Araujo:2018rbc,
author = "Araujo, Thiago and Colg\'ain, Eoin \'O. and Yavartanoo, Hossein",
title = "{Embedding the modified CYBE in Supergravity}",
eprint = "1806.02602",
archivePrefix = "arXiv",
primaryClass = "hep-th",
reportNumber = "APCTP Pre2018-004, APCTP-PRE2018-004",
doi = "10.1140/epjc/s10052-018-6335-6",
journal = "Eur. Phys. J. C",
volume = "78",
number = "10",
pages = "854",
year = "2018"
}

\end{bibtex}

\bibliographystyle{nb}
\bibliography{\jobname}

\end{document}